\documentclass[12pt,preprint,onecolumn]{emulateapj}
\usepackage{latexsym}
\usepackage{url}
\usepackage{color,ulem}
\usepackage[hypertex]{hyperref}
\usepackage{amssymb}
\usepackage{amsmath}
\usepackage{epsfig}
\usepackage{appendix}
\usepackage{wrapfig}
\singlespace
\def\HI{\hbox{H~$\scriptstyle\rm I$}}
\def\HII{\hbox{H~$\scriptstyle\rm II$}}
\def\nHI{{\rm HI}}
\def\nH{{\rm H}}
\def\nHII{{\rm HII}}
\def\nHe{{\rm He}}
\def\nHeI{{\rm HeI}}
\def\nHeII{{\rm HeII}}
\def\nHeIII{{\rm HeIII}}

\def\HeI{\hbox{He~$\scriptstyle\rm I$}}
\def\HeII{\hbox{He~$\scriptstyle\rm II$}}
\def\HeIII{\hbox{He~$\scriptstyle\rm III$}}

\def\HeIII{\hbox{He~$\scriptstyle\rm III$}}
\def\CIV{\hbox{C~$\scriptstyle\rm IV$}}
\def\CIII{\hbox{C~$\scriptstyle\rm III$}}

\def\OVI{\hbox{O~$\scriptstyle\rm VI$}}

\def\OIII{\hbox{O~$\scriptstyle\rm III$}}

\def\SiIII{\hbox{Si~$\scriptstyle\rm III$}}
\def\SiIV{\hbox{Si~$\scriptstyle\rm IV$}}
\def\kms{\,{\rm km\,s^{-1}}}

\def\cmm{\,{\rm cm^{-2}}}

\def\lumunits{\,{\rm ergs\,s^{-1}}}
\def\uvunits{\,{\rm ergs\,cm^{-2}\,s^{-1}\,Hz^{-1}\,sr^{-1}}}

\def\lumdens{\,{\rm ergs\,s^{-1}\,Mpc^{-3}\,Hz^{-1}}}

\def\msun{\,{\rm M_\odot}}

\def\sfrd{\,{\rm M_\odot\,yr^{-1}\,Mpc^{-3}}}

\def\Lya{Ly$\alpha$}
\def\Lyb{Ly$\beta$}
\def\Lyg{Ly$\gamma$}
\def\Lyn{Ly$n$}
\def\etal{{et al.\ }}

\def\spose#1{\hbox to 0pt{#1\hss}}
\def\lta{\mathrel{\spose{\lower 3pt\hbox{$\mathchar"218$}}
     \raise 2.0pt\hbox{$\mathchar"13C$}}}
\def\gta{\mathrel{\spose{\lower 3pt\hbox{$\mathchar"218$}}
     \raise 2.0pt\hbox{$\mathchar"13E$}}}
\def\ni{\noindent}

\lefthead{Haardt \& Madau}
\righthead{New synthesis models of the cosmic UV/X-ray background}
\submitted{Accepted for publication in The Astrophysical Journal}


\begin{document}

\title{Radiative transfer in a clumpy universe: IV. New synthesis models of the 
cosmic UV/X-ray background}

\author{Francesco Haardt\altaffilmark{1,2} and Piero Madau\altaffilmark{2}}

\altaffiltext{1}{Dipartimento di Scienza e alta Tecnologia, Universit\`a dell'Insubria, 
via Valleggio 11, 22100 Como, Italy; haardt@uninsubria.it.}
\altaffiltext{2}{Department of Astronomy \& Astrophysics, University of 
California, 1156 High Street, Santa Cruz, CA 95064; pmadau@ucolick.org.}

\begin{abstract}
We present improved synthesis models of the evolving spectrum of the UV/X-ray diffuse 
background, updating and extending our previous results. Five new main components are added 
to our radiative transfer code CUBA: (1) the sawtooth modulation of the background intensity
from resonant line absorption in the Lyman series of cosmic hydrogen and helium; (2) the X-ray 
emission from the obscured and unobscured quasars that gives origin to the X-ray background; (3) 
a piecewise parameterization of the distribution in redshift and column density of intergalactic
absorbers that fits recent measurements of the mean free path of 1 ryd photons; (4) an accurate 
treatment of the photoionization structure of absorbers, which enters in the calculation of the 
helium continuum opacity and recombination emissivity; and (5) the UV emission from star-forming 
galaxies at all redshifts. We provide tables of the predicted \HI\ and \HeII\ 
photoionization and photoheating rates for use, e.g., in cosmological hydrodynamics simulations of the \Lya\ forest,
and a new metallicity-dependent calibration to the UV luminosity density-star formation rate density relation.
A ``minimal cosmic reionization model" is also presented in which the galaxy UV emissivity traces recent determinations of the 
cosmic history of star formation, the luminosity-weighted escape fraction of hydrogen-ionizing radiation increases 
rapidly with lookback time, the clumping factor of the high-redshift intergalactic medium evolves 
following the results of hydrodynamic simulations, and Population III stars and miniquasars make 
a negligible contribution to the metagalactic flux. The model provides a good fit to the 
hydrogen-ionization rates inferred from flux decrement and proximity effect measurements, predicts that cosmological \HII\ (\HeIII) regions 
overlap at redshift 6.7 (2.8), and yields an optical depth to Thomson scattering, $\tau_{\rm es}=0.084$ that is in 
agreement with {\it WMAP} results. Our new background intensities and spectra are sensitive to a 
number of poorly determined input parameters and suffer from various degeneracies. Their predictive power should be constantly 
tested against new observations. We are therefore making our redshift-dependent UV/X emissivities 
and CUBA outputs freely available for public use at \url{http://www.ucolick.org/~pmadau/CUBA}.
\end{abstract}

\keywords{cosmology: theory -- diffuse radiation -- intergalactic medium -- galaxies: evolution -- quasars: general}

\section{Introduction}

The reionization of the all-pervading intergalactic medium (IGM) is a landmark event in the history 
of cosmological structure formation. Studies of Gunn-Peterson absorption in the spectra of distant quasars 
show that hydrogen is highly photoionized out to redshift $z\gta 6$ (e.g., Fan, Carilli, \& Keating 2006a; 
Songaila 2004), while polarization data from the {\it Wilkinson Microwave Anisotropy Probe (WMAP)} constrain the 
redshift of a sudden reionization event to be significantly higher, $z= 10.5 \pm 1.2$ (Jarosik \etal 2011). 
It is generally thought that the IGM is kept ionized by the integrated UV emission from active nuclei and 
star-forming galaxies, but the relative contributions of these sources as a function of epoch are poorly 
known. Because of the high ionization threshold (54.4 eV) and small photoionization cross section of \HeII, and of 
the rapid recombination rate of \HeIII, the double ionization of helium is expected 
to be completed by hard UV-emitting quasars around the peak of their activity at $z\approx 2.5$ 
(e.g., Madau \& Meiksin 1994; Sokasian, Abel, \& Hernquist 2002; McQuinn \etal 2009), much later 
than the reionization of \HI\ and \HeI.  At $z>3$, the declining population of bright quasars 
appears to make an increasingly small contribution to the 1 ryd radiation background, and it is believed 
that massive stars in galactic and subgalactic systems may provide the additional ionizing flux needed at early times 
(e.g., Madau, Haardt, \& Rees 1999; Gnedin 2000; Haehnelt \etal 2001; Wyithe \& Loeb 2003; Meiksin 2005; Trac \& Cen 2007; 
Faucher-Gigu\`ere \etal 2008a; Gilmore \etal 2009; Robertson \etal 2010). This idea may be supported by the detection of escaping 
ionizing radiation from individual Lyman-break galaxies at $z\sim 3$ (e.g., Shapley \etal 2006). 

Despite much recent progress, a coherent description of the thermal state and ionization degree 
of the IGM remains elusive. The intensity and spectrum of the cosmic ultraviolet background remain 
one of the most uncertain yet critically important astrophysical input parameters for cosmological
simulations of the IGM and for interpreting quasar absorption-line data and derive information on the 
distribution of primordial baryons (traced by \HI, \HeI, \HeII\ transitions) and of the 
nucleosynthetic products of star formation (\CIII, \CIV, \SiIII, \SiIV, \OVI, etc.). This is the 
fourth paper in a series aimed at a detailed study of the generation and reprocessing of photoionizing 
radiation in a clumpy universe, and of the transfer of energy from this diffuse background flux to the IGM. 
In Paper I (Madau 1995) we showed how the stochastic attenuation produced by neutral hydrogen along 
the line of sight affects the colors of distant galaxies. In Paper II (Haardt \& Madau 1996) we developed CUBA, a radiative 
transfer code that followed the propagation of Lyman-continuum (LyC) photons through a partially ionized inhomogeneous IGM. 
CUBA outputs have been extensively used to model the \Lya\ forest in large cosmological 
simulations (e.g., Tytler \etal 2004; Theuns \etal 1998; Dav\'e \etal 1997; Zhang 
\etal 1997). In Paper III (Madau \etal 1999) we focused on the candidate sources 
of photoionization at early times and on the history of the transition from 
a neutral IGM to one that is almost fully ionized. In this paper we describe a new version of CUBA and use it 
to compute improved synthesis models of the UV/X-ray cosmic background spectrum and evolution, combining, 
updating, and extending many of our previous results in this field. The five main upgrade to CUBA are: (1) the sawtooth modulation 
from resonant line absorption in the Lyman series of intergalactic helium as well as hydrogen; 
(2) the X-ray emissivity from the obscured and unobscured populations of active galactic nuclei (AGNs) that gives 
origin to the X-ray background; (3) an up-to-date piecewise parameterization of the distribution 
in column density of intervening absorbers, which establishes the ``super Lyman-limit systems" as the dominant contributors
to the hydrogen LyC intergalactic opacity; (4) an accurate treatment of the absorber photoionization structure, 
entering in the calculation of the helium continuum opacity and recombination emissivity of the clumpy IGM; and (5)
the UV flux from star-forming galaxies at all redshifts. 

The plan is as follows. In \S\,2 we review the basic theory of cosmological radiative transfer in a clumpy universe.
\S\,3 and \S\,4 discuss the distribution of absorbers along the line of sight and their photoionization structure.
The recombination radiation from the clumpy IGM is calculated in \S\,5. In \S\,6 and \S\,7 we compute the UV and X-ray emissivity
from quasars, and in \S\,8 the UV emissivity from star-forming galaxies. An overview of the main results generated by the
updated CUBA radiative transfer code is given in \S\,9. Finally, we summarize our findings in \S\,10.       
Unless otherwise stated, all results shown below will assume a $(\Omega_M, \Omega_\Lambda, \Omega_b, h)=(0.3, 0.7, 0.045, 0.7)$ 
cosmology. Note that, while the source volume emissivities must be
evaluated in a given cosmological model, the resulting background intensity does not explicitly depend 
on the choice of cosmological parameters.    

\section{Cosmological radiative transfer}

We start by summarizing the basic theory describing the propagation of ionizing radiation 
in a clumpy, primordial IGM (e.g., Paper I; Paper II; Madau \& Haardt 2009). The equation 
of cosmological radiative transfer describing the time evolution of the 
space- and angle-averaged monochromatic intensity $J_\nu$ is
\begin{equation}
\left({\partial \over \partial t}-\nu H {\partial \over \partial \nu}\right)J_\nu+3HJ_\nu=
- c\kappa_\nu J_\nu + {c\over 4\pi}\epsilon_\nu, \label{eq:rad}
\end{equation}
where $H(z)$ is the Hubble parameter, $c$ the speed of the light, $\kappa_\nu$ is the 
absorption coefficient, and $\epsilon_\nu$ the proper volume emissivity. The integration of 
equation (\ref{eq:rad}) gives the background intensity at the observed frequency
$\nu_o$, as seen by an observer at redshift $z_o$, 
\begin{equation}
J_{\nu_o}(z_o)={c\over 4\pi}\int_{z_o}^{\infty}\, |dt/dz| dz
{(1+z_o)^3 \over (1+z)^3} \epsilon_\nu(z) e^{-\bar\tau},
\label{Jnu}
\end{equation}
where $\nu=\nu_o(1+z)/(1+z_o)$, $|dt/dz|=H^{-1}(1+z)^{-1}$, $\bar\tau\equiv -\ln \langle e^{-\tau}\rangle$ 
is the effective absorption optical depth of a clumpy IGM, and $\epsilon_\nu$ is the proper volume emissivity. 

\subsection{Photoelectric absorption}

In the case of LyC absorption by Poisson-distributed systems, the effective opacity between $z_o$ and $z$ is
\begin{equation}
\bar\tau_c(\nu_o,z_o,z)=\int_{z_o}^z\,
dz'\int_0^{\infty}\, dN_\nHI\, f(N_\nHI,z') (1-e^{-\tau_c}), \label{tauC}
\end{equation}
where $f(N_\nHI,z')$ is the bivariate distribution of absorbers in redshift and
column density along the line of sight, $\tau_c$ is the continuum optical depth at frequency
$\nu'=\nu_o(1+z')/(1+z_o)$ through an individual absorber,
\begin{equation}
\tau_c(\nu')=N_\nHI\sigma_\nHI(\nu') +N_\nHeI\sigma_\nHeI(\nu') +N_\nHeII\sigma_\nHeII(\nu'),
\end{equation}
where $N_i$ and $\sigma_i$ are the column densities and photoionization cross sections of ion $i$.

\subsection{Resonant absorption}
\label{sec:sawtooth}

Besides photoelectric absorption, resonant absorption by the hydrogen and helium  Lyman series will produce 
a sawtooth modulation of the radiation spectrum (Madau \& Haardt 2009; Haiman, Rees, \& Loeb 1997). Continuum photons 
that are redshifted through the \Lya\ frequency, $\nu_\alpha$, are resonantly scattered until they redshift out of resonance: 
the only two \Lya\ line destruction mechanisms, two-photon decay and \OIII\ Bowen fluorescence (Kallman \& McCray 1980), 
can typically be neglected in the low metallicity, low density IGM. This is not true, however, for photons passing through a 
higher order Lyman-series resonance, which will be absorbed and degraded via a radiative cascade rather than escaping
by redshifting across the line width. Since the line absorption cross section is  a narrow, strongly 
peaked function, the effective line absorption optical depth for a photon observed at $(z_o, \nu_o<\nu_n)$
that passed through a resonance at redshift $z_n=(1+z_o)(\nu_n/\nu_o)-1$, can be written as
\begin{equation}
\bar\tau_n(z_n)=(1+z_n){\nu_n\over c} \int_0^\infty dN_\nHI\,f(N_\nHI,z_n)W_n,
\label{taui}
\end{equation}
where $\nu_n=\nu_\alpha \times 4(1-n^{-2})/3$ is the frequency of the $1s \rightarrow np$ Lyman-series 
transition ($n\ge 3$) and $W_n$ is the rest equivalent width of the line expressed in wavelength units.
This opacity is dominated by systems having line center optical depths of order unity, i.e., which lie at 
the transition between the linear and the flat part of the curve of growth. 

Consider, for example, radiation observed at frequency below the \Lyb\ of hydrogen or helium, $\nu_o<\nu_\beta$. 
Photons emitted between $z_o$ and $z_\beta=(1+z_o)(\nu_\beta/\nu_o)-1$ can reach the observer without undergoing
resonant absorption. Photons emitted between $z_\beta$ and $z_\gamma=(1+z_o)
(\nu_\gamma/\nu_o)-1$ pass instead through the \Lyb\ resonance at $z_\beta$ and are
absorbed. Photons emitted between $z_\gamma$ and $z_\delta=(1+z_o)(\nu_\delta/\nu_o)-1$
pass through both the \Lyb\ and the \Lyg\ resonances before reaching
the observer. The background intensity can then be written as (Madau \& Haardt 2009)
\begin{equation}
J_{\nu_o}(z_o)=O(z_o,z_\beta)+O(z_\beta,z_\gamma)e^{-\bar\tau_\beta}+
O(z_\gamma,z_\delta)e^{-\bar\tau_\beta-\bar\tau_\gamma}+....
+O(z_L,\infty)\exp(-\sum_{n=3}^\infty \bar\tau_n),
\label{Jbeta}
\end{equation}
where we have denoted with the symbol $O(z_i,z_j)$ the ``Olbers' integrals" on the right hand side of 
equation (\ref{Jnu}), calculated between redshifts $z_i$ and $z_j>z_i$ and with $\bar\tau_c$ equal to 
the relevant continuum opacity,
\begin{equation}
O(z_i,z_j)\equiv {c\over 4\pi}\int_{z_i}^{z_j}\, |dt/dz|dz
{(1+z_o)^3 \over (1+z)^3} \epsilon_\nu(z) e^{-\bar\tau_c}. 
\end{equation}
In equation (\ref{Jbeta}), $z_L=(1+z_o)(\nu_L/\nu_o)-1$,
$\nu_L$ is the frequency at the Lyman limit, and $\bar\tau_\beta, \bar\tau_\gamma, \bar\tau_\delta,...$ are 
the Lyman-series effective opacities at redshifts $z_\beta, z_\gamma, z_\delta,...$.
In the case of resonant absorption by \HI, the LyC optical depth $\bar\tau_c$ is zero in all $O$-integrals 
except the last, while in the case of \HeII\ all terms must include photoelectric absorption by \HI\ and \HeI\
(as well by \HeII\ in the last term). Equation (\ref{Jbeta}) is easily generalized to higher 
observed frequencies, $\nu_n<\nu_o<\nu_{n+1}$, to read 
\begin{equation}
J_{\nu_o}(z_o)=O(z_o,z_{n+1})+\sum_{k=n+1}^\infty O(z_k,z_{k+1})\,\exp(-\sum_{l=n+1}^k \bar\tau_l). 
\label{Jres}
\end{equation}
Note how, in the case of large resonant opacities, only sources between the observer and the ``screen" redshift 
$z_{n}=(1+z_o)(\nu_{n}/\nu_o)-1$ corresponding to the frequency of the nearest Lyman-series line above $\nu_o$ 
will not be blocked from view: the background energy spectrum will show a series of discontinuities, 
peaking at frequencies just above each resonance, as the 
first integral in equation (\ref{Jres}) extends over the largest redshift path, and going to zero at resonance. 

\subsection{Lyman series cascades}

Each photon absorbed through a Lyman series resonance causes a radiative 
cascade that ultimately terminates either in a Ly$\alpha$ photon or in two-photon $2s \rightarrow 1s$ continuum 
decay. In the former case the photon scatters until it is redshifted out of resonance, in the latter 
the photons escape to infinity without further interactions. Consider, for example, the
absorption of a \Lyb\ photon. The excited $3p$ level is depopulated via $3p\rightarrow 2s$ decay  
(H$\alpha$). In the low-density IGM, collisional $l$-mixing of the $2s-2p$ levels (Seaton 1959) 
is negligible, and the cascade can only terminate in two-photon $2s \rightarrow 1s$ emission (Hirata 2006). 
Without $l$-mixing, the quantum selection rules forbid a \Lyb\ photons from being converted
into \Lya: by contrast, excitation of the $4p$ level by absorption of \Lyg\ can    
decay via the $3s$ or $3d$ levels to the $2p$ and ultimately produce \Lya.  
More generally, the fraction, $f_n$, of decays from an $np$ state that generates \Lya\ photons can be
determined from the selection rules and the decay probabilities. This fraction is found to increase as 
$f_n=(0,0.2609,0.3078,0.3259,...)$ for $n=(3,4,5,6,...)$, and to asymptote to 0.359 
at large $n$ (Pritchard \& Furlanetto 2006). Note that this is valid in the approximation that the IGM 
is optically thick to higher-order Lyman-series transitions.

What is the \Lya\ diffuse flux produced by these Lyman-series cascades? Let $J_{\nu_n}(z)$ be
the background intensity measured just above the \HI\ or \HeII\ \Lyn\ resonance at redshift $z$.
The \Lyn\ flux that is absorbed and converted into \Lya\ is $f_n J_{\nu_n}(z)[1-e^{-\bar\tau_n(z)}]$, 
and the proper \Lya\ volume emissivity generated by this process can be written as
\begin{equation}
\epsilon^n_{\nu}(z)={4\pi \over c} f_n J_{\nu_n}(z)[1-e^{-\bar\tau_n(z)}]\, {\nu \delta(\nu-\nu_\alpha)\over (1+z)}\,|dz/dt|, 
\label{eq:epsia}
\end{equation}
where $\delta(x)$ is the Dirac delta function.  
The additional flux observed at frequency $\nu_o\le \nu_\alpha$ and redshift $z_o$ from this process is then
\begin{equation}
\Delta J_{\nu_o}^n(z_o)=\left(\frac{\nu_o}{\nu_\alpha}\right)^3\,e^{-\bar\tau_c(\nu_o,z_o,z_\alpha)}\,
\{f_n J_{\nu_n}(z_\alpha)[1-e^{-\bar\tau_n(z_\alpha)}]\}
\label{Jalpha}
\end{equation}
(Madau \& Haardt 2009), where $1+z_\alpha=(1+z_o)(\nu_\alpha/\nu_o)$ and the LyC optical depth $\bar\tau_c$ is 
zero in the case of resonant absorption by \HI. When summing up over all Lyman series lines, the term
in square brackets must be replaced by $\sum_{n>3}\{f_n J_{\nu_n}(z_\alpha)[1-e^{-\bar\tau_n(z_\alpha)}]\}$. 
The same \Lyn\ cascade also produces a two-photon continuum with emissivity given by
\begin{equation}
\epsilon^n_{\nu}(z)={4\pi \over c} (1-f_n) J_{\nu_n}(z)[1-e^{-\bar\tau_n(z)}]\, {\nu f_\nu\over (1+z)}\,|dz/dt|, 
\label{eq:epsi2}
\end{equation}
where the two-photon emission function $f_\nu$ is expressed in photons per unit frequency interval and is
symmetric about $\nu_\alpha/2$. 

We note that the underlying assumption in equations (\ref{eq:epsia}) and (\ref{eq:epsi2}) is that every absorber 
is a source of unprocessed \Lya\ line and two-photon continuum radiation, i.e., that these photons escape into intergalactic space
without appreciable local absorption. In the case of \HeII\ \Lya\ emission, this requires 
negligible ``in situ" destruction from dust, metals (\OIII\ Bowen fluorescence), and photoelectric absorption by \HI, so 
that the \Lya\ photons diffuses into the wings and eventually escape from the production site into the IGM. This is 
a good approximation at the low metallicities that characterize intergalactic absorbers (Kallman \& McCray 1980), 
even more so since the reprocessing of Lyman series photons occurs in a ``skin" layer at the surface of an absorption system.
In \S\,4 we will show that this is a poor approximation in the case of the reprocessing of LyC radiation, a proper 
treatment of which requires a numerical solution of the radiative transfer equation within individual absorbers. 

\begin{figure}[thb]
\centering
\includegraphics*[width=0.47\textwidth]{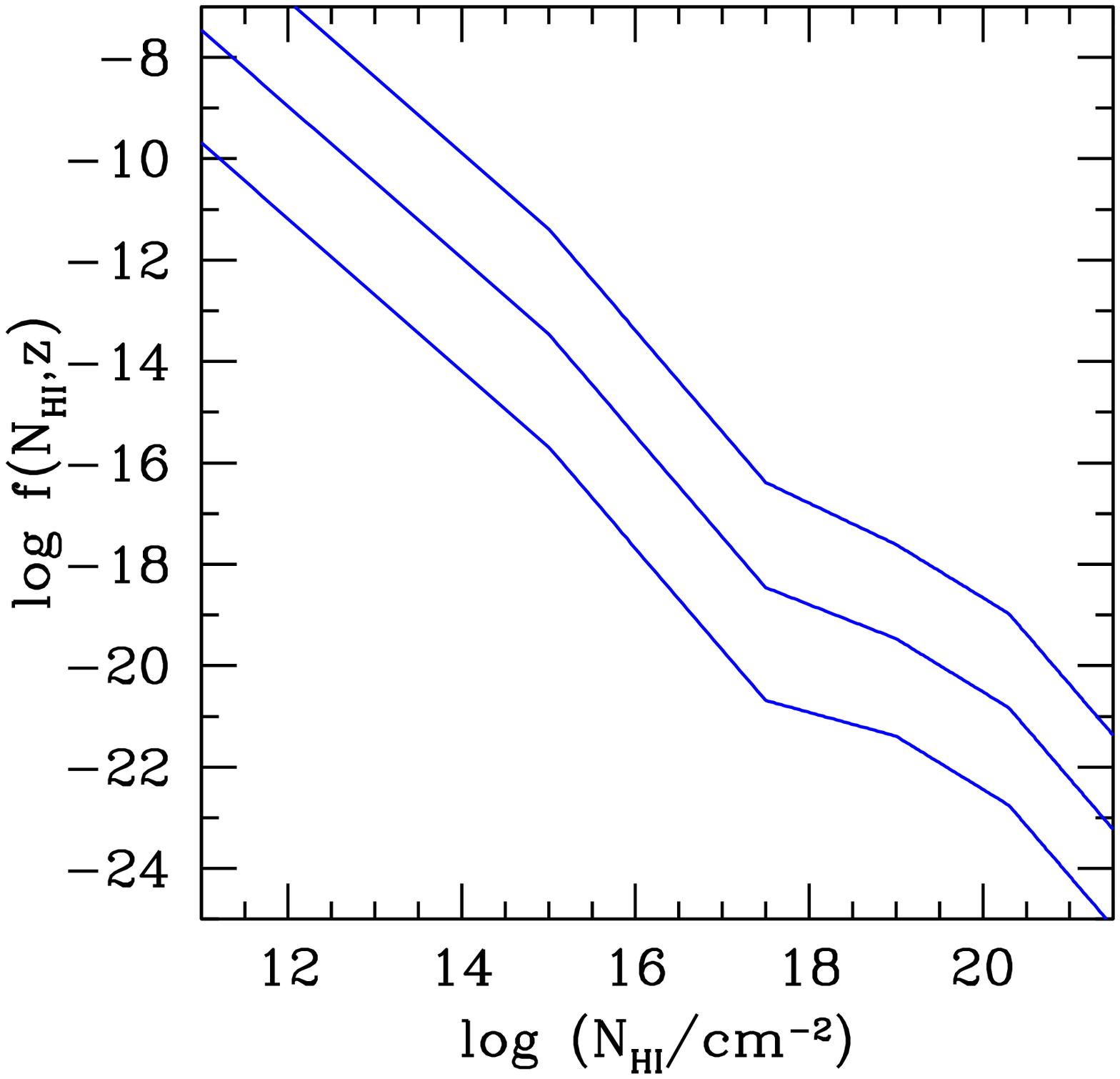}
\includegraphics*[width=0.47\textwidth]{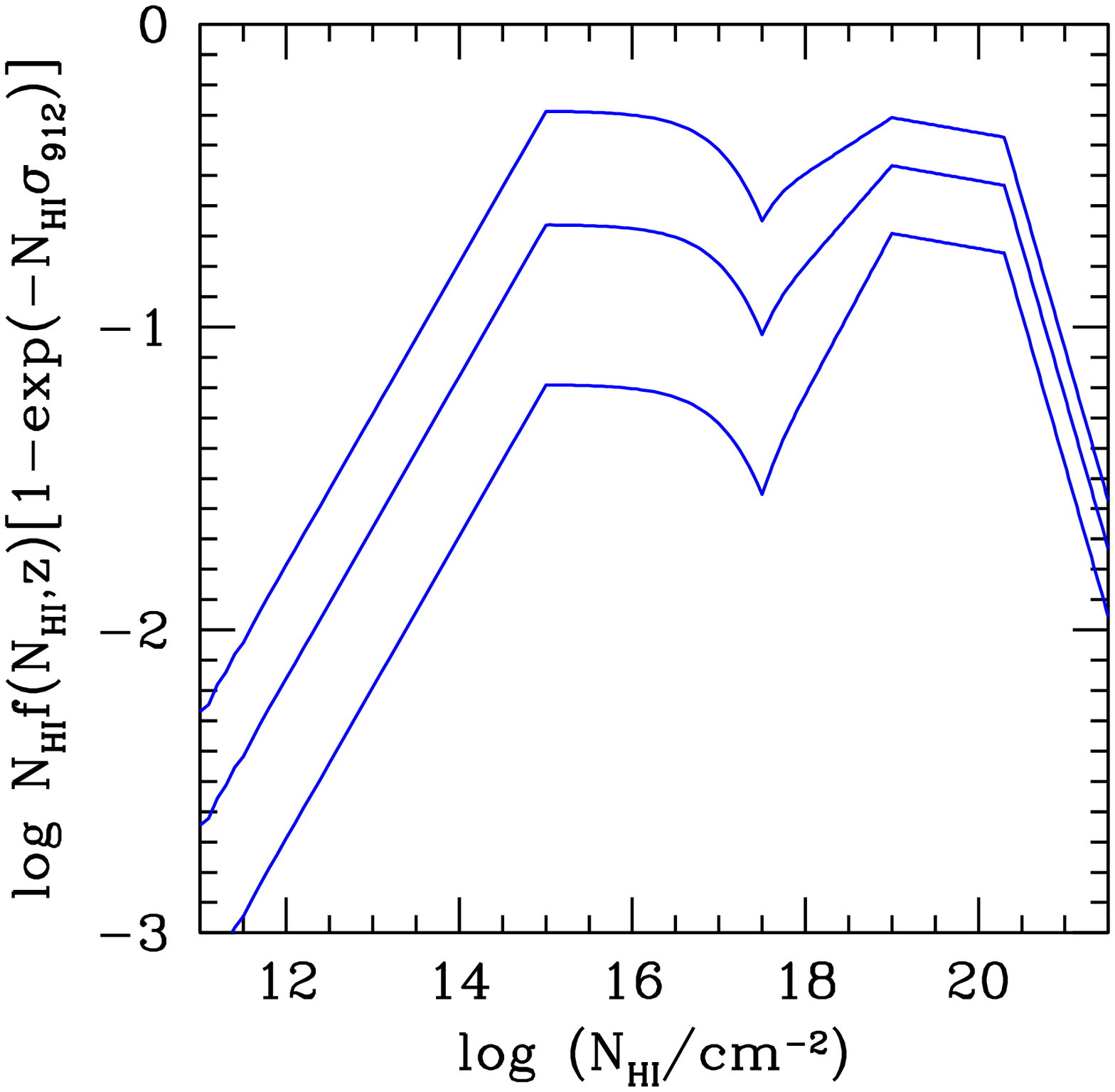}
\caption{\footnotesize {\it Left:} The assumed distribution of absorbers in \HI\ column density at 
redshift $z=2$ ({\it bottom curve}), $z=3.5$ ({\it middle curve}), and $z=5$ ({\it top curve}).
For clarity, we have multiplied the top and bottom curves by 50 and 1/50, respectively.  
{\it Right:} The quantity $N_\nHI f(N_\nHI,z) [1-\exp(-N_\nHI\sigma_{912}]$ at the same redshifts, 
showing the dominant contribution of the optically thick LLSs and SLLSs to the intergalactic opacity at 1 ryd. 
}
\label{fig1}
\vspace{+0.3cm}
\end{figure}

\begin{table}[h]
\caption{Parameters of the distribution of intergalactic absorbers}
\centering
\begin{tabular}{llllll}
\hline\hline
\\[-6pt]
{Absorbers class} & {log $(N_\nHI/$cm$^{-2})$} & {$\beta$} & {$A$ [cm$^{-2(\beta-1)}$]} & {$\gamma$} & {redshift} \\[3pt]
\hline
\\[-6pt]
\Lya\ forest & $11-15$ & $1.5$ & $10^{7.079}$ & 3.0 & $1.56<z<5.5$ \\
& $11-15$ & $1.5$ & $10^{8.238}$ & 0.16 & $z<1.56$ \\
& $11-15$ & $1.5$ & $10^{1.470}$ & 9.9 & $z>5.5$ \\
& $15-17.5$ & $2.0$ & $10^{14.580}$ & $3.0$ & $1.56<z<5.5$ \\
& $15-17.5$ & $2.0$ & $10^{15.740}$ & $0.16$ & $z<1.56$ \\
& $15-17.5$ & $2.0$ & $10^{8.970}$ & $9.9$ & $z>5.5$ \\[3pt]
\hline
\\[-6pt]
LLSs & $17.5-19$ & & & & \\[3pt]
\hline
\\[-6pt]
SLLSs & $19-20.3$ & $1.05$ & $10^{-0.347}$ & 1.27 & $z>1.56$ \\
& $19-20.3$ & $1.05$ & $10^{0.107}$ & 0.16 & $z<1.56$ \\[3pt]
\hline
\\[-6pt]
DLAs & $20.3-21.55$ & $2.0$ & $10^{18.940}$ & 1.27 & $z>1.56$ \\
& $20.3-21.55$ & $2.0$ & $10^{19.393}$ & 0.16 & $z<1.56$\\[3pt]
\hline
\\
\end{tabular}
\end{table}

\section{Distribution of absorbers along the line of sight}
 
The effective opacity of the IGM has traditionally been one of the main uncertainties affecting calculations
of the UV background. Our improved model uses a piecewise power-law parameterization for the distribution of 
absorbers along the line of sight, 
\begin{equation}
f(N_\nHI,z)=A\,N_\nHI^{-\beta}(1+z)^{\gamma},
\label{eq:ladis}
\end{equation}
and is designed to reproduce accurately a number of recent observations:

\begin{itemize}

\item Over the column density range 
$10^{11}<N_\nHI<10^{15}\,\cmm$, we use $(A,\beta,\gamma)=(1.2\times 10^7,1.5,3.0)$, where the normalization 
$A$ is expressed in units of cm$^{-2(\beta-1)}$, and $\beta=1.5$ is chosen following, e.g., Tytler (1987). 
As noted, e.g., by Meiksin \& Madau (1993), Petitjean et al. (1993), and Kim \etal (1997, 2002), $f(N_\nHI)$
starts to steepen from the empirical $-1.5$ power law at $N_\nHI>10^{14.5}\cmm$. Here, we assume 
$(A,\beta,\gamma)=(3.8\times 10^{14},2.0,3.0)$ for $10^{15}<N_\nHI<10^{17.5}\,\cmm$.
A ``curve of growth'' analysis (providing the relationship between equivalent width and column 
density) with Doppler parameter $b=32\,\kms$, together with equation (\ref{taui}) and the above 
distribution of \Lya-forest clouds, produces a \Lya\ effective opacity $\bar\tau_\alpha=0.0015(1+z)^4$, in agreement with the 
best-fits of Faucher-Gigu\`ere \etal (2008b) after metal correction. 

\item At the other end of the column density distribution, a recent survey of damped \Lya\ systems (DLAs) by Prochaska
\& Wolfe (2009) (see also Guimaraes \etal 2009) yields $dN/dz\equiv \int dN_\nHI f(N_\nHI,z)=0.294$ DLAs per unit redshift 
at $\langle z\rangle=3.7$ above $N_\nHI=10^{20.3}\,\cmm$. With a power-law exponent $\beta=2$ down to a break column
of $N_\nHI=10^{21.55}\,\cmm$ (Prochaska \& Wolfe 2009), and with an incidence per unit redshift $\propto (1+z)^{1.27}$
(Rao \etal 2006), the parameters for the DLAs becomes $(A,\beta,\gamma)=(8.7\times 10^{18},2,1.27)$. 

\item  For absorbers with $10^{19}<N_\nHI<10^{20.3}\,\cmm$ (the so-called ``super Lyman-limit systems", or SLLSs), we use 
O'Meara \etal (2007), who find $dN/dz=0.97$ SLLSs per unit redshift at $\langle z\rangle=3.5$ above $N_\nHI=10^{19}\,\cmm$. 
Matching with the DLAs abundance then requires $(A,\beta,\gamma)=(0.45,1.05,1.27)$ for the SLLSs.

\item There is obviously a significant mismatch between the power-law exponent for the \Lya\ clouds ($\gamma=3$) and the 
SLLSs ($\gamma=1.27$). Continuity then requires the shape of $f(N_\nHI,z)$ to change with redshift over the column 
density range of the Lyman-limit systems (LLSs), $10^{17.5}<N_\nHI<10^{19}\,\cmm$. In this interval of column densities 
we match the distribution function with a power law of redshift-dependent slope. The procedure yields the slopes 
$\beta=0.47,0.61,0.72,0.82$ at redshifts $z=2,3,4,5$, respectively, in agreement with Prochaska \etal (2010) who
find for the LSSs $\beta=0.8^{+0.4}_{-0.2}$ at $z\approx 3.5$. 

\item The ensuing $f(N_\nHI,z)$ distribution is shown in the left panel of Figure \ref{fig1} for $z=2,3.5,5$ where, for clarity,
we have multiplied the values at the highest and lowest redshift by 50 and 1/50, respectively. Its shape is similar to the 
distribution inferred by Prochaska \etal (2010). In the right panel of the same figure we have plotted
the quantity $N_\nHI d\bar\tau_c/(dz dN_\nHI)\vert_{\nu=\nu_{912}}=N_\nHI f(N_\nHI,z) [1-\exp(-N_\nHI \sigma_{912}]$, i.e., the effective
optical depth at 1 ryd per unit redshift per unit logarithmic interval of hydrogen column. This shows the dominant contribution 
of the LLSs and SLLSs to the LyC opacity.

\item The above parameterizations reproduce well the observations 
at $2\lta z \lta 5$. At low redshift, however, {\it Hubble Space Telescope} ({\it HST}) 
data show that the forest undergoes a much slower evolution. Following Weymann \etal (1998) 
we take $\gamma=0.16$ in the interval $0<z<z_{\rm low}$ and $dN/dz=34.7$ at $z=0$ above 
an equivalent width of 0.24 \AA\ (corresponding to a column of $10^{13.87}\,\cmm$ for $b=32\,\kms$).   
We derive $(A,\beta,\gamma)=(1.73\times 10^8,1.5,0.16)$ for $10^{11}<N_\nHI<10^{15}\,\cmm$ and
$(A,\beta,\gamma)=(5.49\times 10^{15},2,0.16)$ for $10^{15}<N_\nHI<10^{17.5}\,\cmm$ at all redshifts
below $z_{\rm low}=1.56$. We use a broken power-law for the redshift distribution of the SLLSs and DLAs as well; assuming that the 
same $\gamma=0.16$ slope and transition redshift $z_{\rm low}$ inferred for the forest also hold in the case of the thicker absorbers, we 
derive a normalization at $z<z_{\rm low}$ of $A=1.28$ for the SLLSs and $A=2.47\times 10^{19}$ for the DLAs. 
This yields $dN/dz=0.74$ absorbers above $N_\nHI=10^{17.2}\,\cmm$ at $\langle z\rangle =0.69$, in agreement 
with the value measured by Stengler-Larrea \etal (1995), $dN/dz=0.70\pm 0.2$.

\item Above $z=5.5$, the spectra of the highest redshift quasars known show an accelerated evolution in the \Lya\ 
opacity of the IGM, $\bar\tau_\alpha=2.68[(1+z)/6.5]^{10.9}$ (Fan \etal 2006b), indicating a sharp increase in  
in the average neutrality of the universe. This can be mimicked by assuming for the forest  
the values $(A,\beta,\gamma)=(29.5,1.5,9.9)$ ($10^{11}<N_\nHI<10^{15}\,\cmm$) and
$(A,\beta,\gamma)=(9.35\times 10^8,2,9.9)$ ($10^{15}<N_\nHI<10^{17.5}\,\cmm$) above redshift 5.5. 

\end{itemize}
\begin{figure}[thb]
\centering
\vspace{-0.8cm}
\includegraphics*[width=0.47\textwidth]{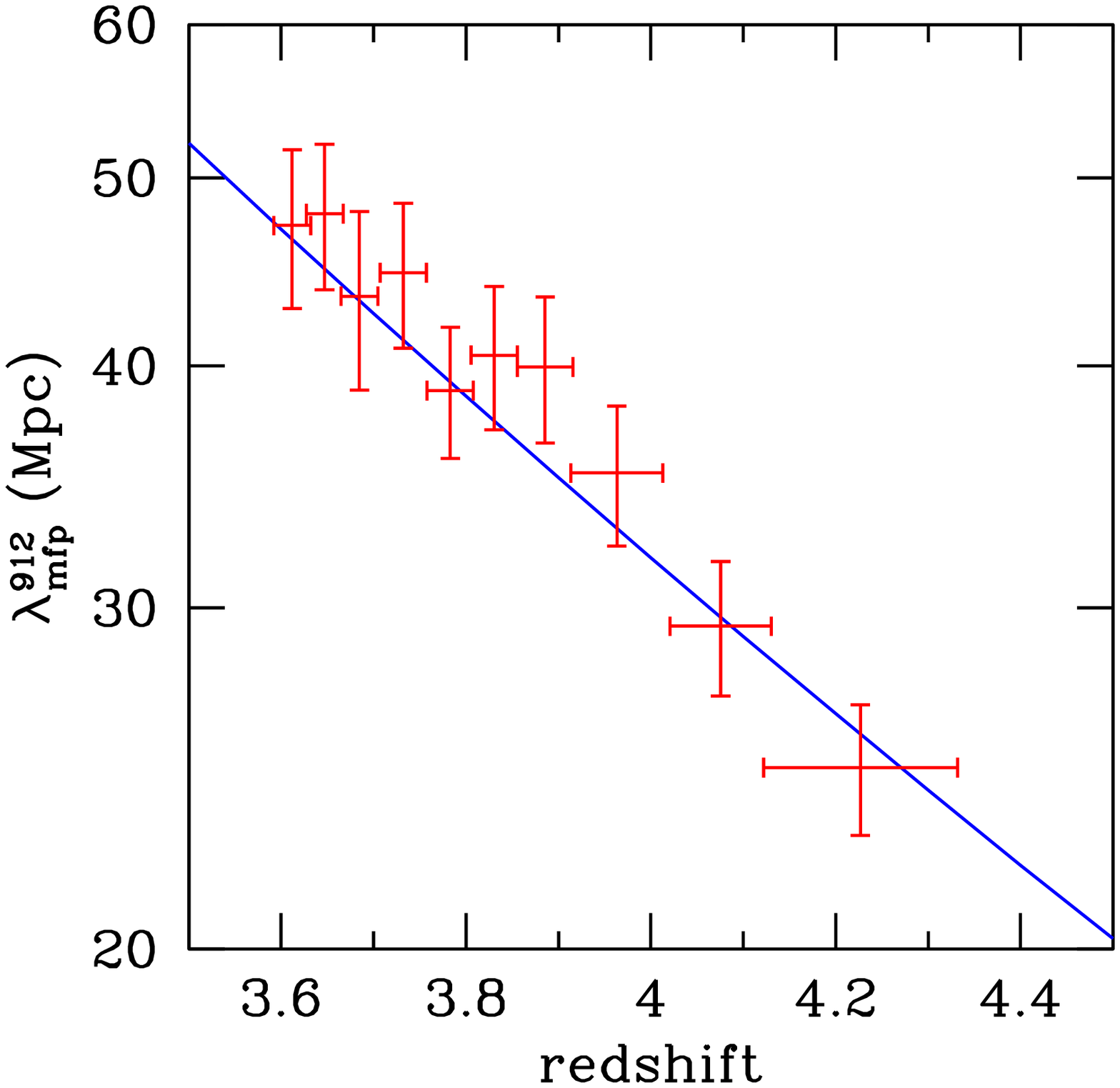}
\includegraphics*[width=0.47\textwidth]{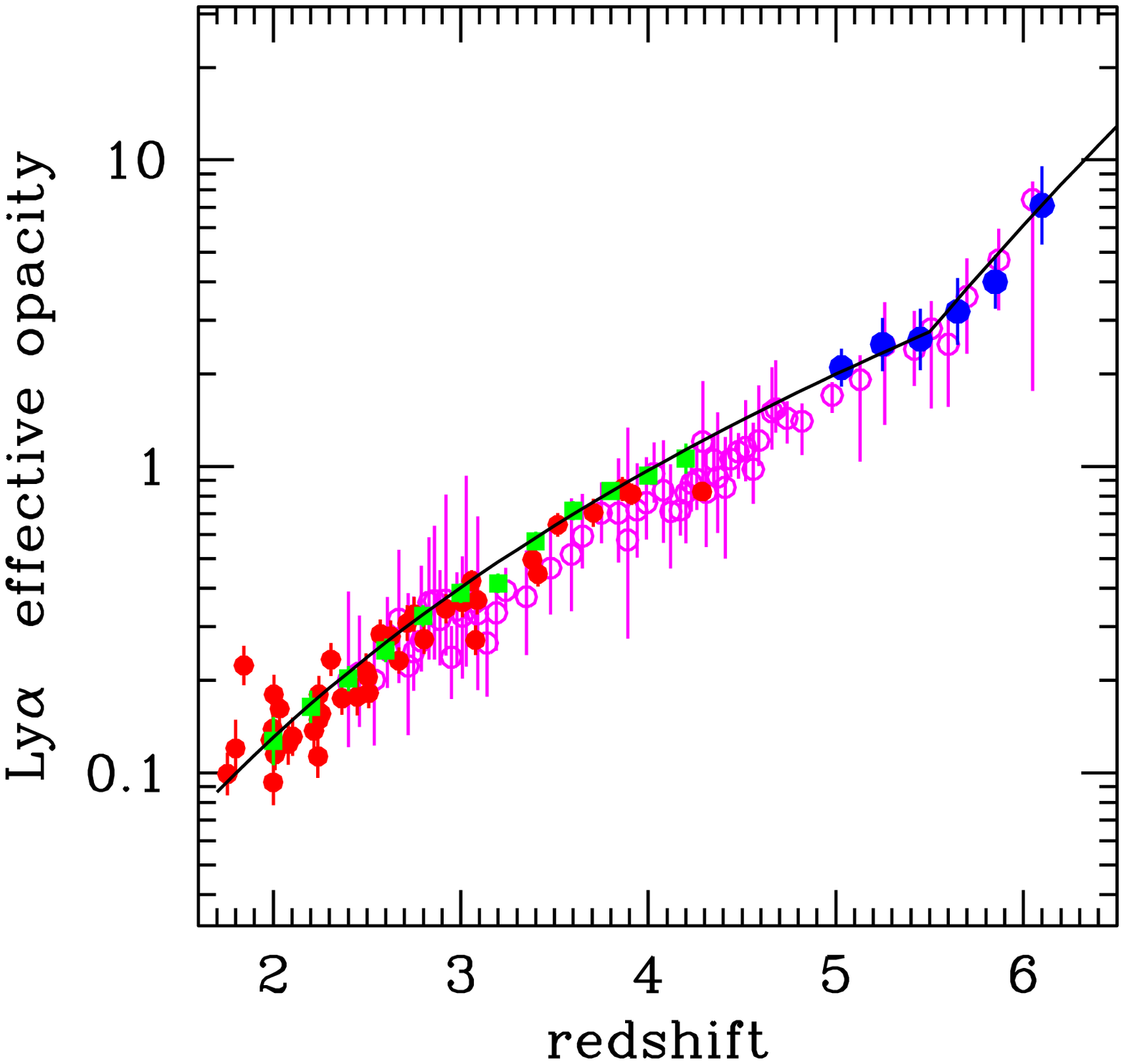}
\caption{\footnotesize {\it Left:} The predicted proper mean free path at 1 ryd ({\it solid line}) 
together with the measurements of Prochaska \etal (2009) ({\it crosses}). 
{\it Right:} Evolution of the observed effective \Lya\ optical depth,
$-\ln \langle \cal{T} \rangle$, where $\cal{T}$ is the transmitted flux ratio. Data points are from 
Schaye \etal (2003; {\it red filled circles}), Songaila (2004; {\it magenta empty squares}), 
Faucher-Gigu\`ere \etal (2008b; {\it green filled squares}), and Fan \etal (2006b; {\it blue filled circles}). 
The solid line shows the \Lya\ opacity, $\bar\tau_\alpha$, predicted by equations (\ref{taui}) (for 
$n=2$) and (\ref{eq:ladis}), and using a curve-of-growth analysis corresponding 
to a Doppler parameter $b=32\,\kms$. 
}
\vspace{+0.5cm}
\label{fig2}
\end{figure}
The parameters of the adopted distribution of intergalactic absorbers are summarized in Table 1. 

\subsection{Mean free path of hydrogen-ionizing radiation}

Inserting our $f(N_\nHI,z)$ in equation (\ref{tauC}), we can compute the (proper) LyC mean free 
path for 1 ryd photons as
\begin{equation}
\lambda^{912}_{\rm mfp}=c|dt/dz|\times {dz\over d\bar\tau_c}\vert_{\nu=\nu_{912}}. 
\end{equation}
This is plotted in the left panel of Figure \ref{fig2} in the redshift range 3.5-4.5.
At $z=3.5$, the major contributors to the LyC opacity are, in order of decreasing magnitude,
the high column-density \Lya\ forest ($10^{15}<N_\nHI<10^{17.5}\,\cmm$, 32\%), the SLLSs (28\%), the LLSs (20\%), the 
low column-density \Lya\ forest ($N_\nHI<10^{15}\,\cmm$, 12\%), and the DLAs (8\%). 
A new method to directly measure the IGM LyC opacity along quasar sight lines has been recently presented by 
Prochaska, Worseck, \& O'Meara (2009). The approach analyzes the ``stacked" spectrum of 1,800 quasars 
drawn from the {\it Sloan Digital Sky Survey (SDSS)}  to 
give an empirical determination of the mean free path $\lambda_{\rm mfp}^{912}$. Our new 
opacity model agrees very well with the measurements of Prochaska \etal (2009), and produces a continuum 
opacity that is approximately half of that adopted in Paper II. The right panel of the same figure 
shows how our model also provides a good fit to the \Lya\ quasar transmission data over 
the entire redshift range $2 \lta z \lta 6$. 

For a single population of absorbers described by equation (\ref{eq:ladis}), the mean free path scales with frequency and redshift
as $\lambda_{\rm mfp}(\nu,z)\propto (\nu/\nu_{912})^{3(\beta-1)} H^{-1}/(1+z)^{\gamma+1}$. Given the multi-component distribution  
summarized in Table 1, we can readily compute the mean free path of ionizing radiation in the range $13.6 \leq h\nu<48.4$ eV under 
the assumption that \HeI\ continuum absorption at energies above $24.2$ eV can be neglected (photons between 48.4 and 54.4 eV are 
reprocessed by \HeII\ Lyman series resonance absorption, see \S~\ref{sec:sawtooth}). 
For ease of use in analytical calculations, we fit our numerical results for the mean free path as
\begin{equation}
\lambda_{\rm mfp}(\nu,z)=c|dt/dz|\Delta z=c|dt/dz| A(s)(1+z)^{\gamma(s)}
\label{eq:deltazfit}
\end{equation}
where $s\equiv \nu/\nu_{912}$. Both the normalization $A(s)$ and the exponent $\gamma(s)$ are well fit by third order polynomials of the form
\begin{equation}
[A(s),\gamma(s)]=p_3(s^3-1)+p_2(s^2-1)+p_1(s-1)+p_0.
\end{equation}
Numerical values of the best-fit polynomial coefficients are given in Table 2: the fitting function is adjusted to be continuous in value 
at the redshifts where it changes slope. As discussed above, the fit is only valid in the
range $1\leq s \leq 3.56$. Close to the hydrogen Lyman edge, and at early 
enough epochs, only ``local" radiation sources -- sources within a mean free path of a few tens of Mpc -- contribute to the 
ionizing background intensity, and one can neglect cosmological effects such as source 
evolution and frequency shifts. In this ``source-function" approximation, $4\pi J_{912}(z)\approx \epsilon_{912}(z)\,
\lambda_{\rm mfp}^{912}(z)$.

\begin{table}[h]
\caption{Fitting parameters for the hydrogen LyC mean free path}
\centering
\begin{tabular}{lccccc}
\hline\hline
$$  & parameter & $p_3$ & $p_2$ & $p_1$ & $p_0$ \\
\hline
 $$&$A$& 0.0509 & -0.406 & 1.167 & 1.076\\
 $0<z<1.56$ & $\gamma$ &  0. &   0. & 0. &   -0.160\\
 $1.56<z<5.5$ & $\gamma$ & 0.0593&   -0.519& 1.586& -2.104\\
 $z>5.5$ & $\gamma$ & 0.122&   -1.356& 5.998&   -8.423\\
\hline
\end{tabular}
\label{tab:mfpHI}
\end{table}

\section{Photoionization structure of absorption systems} \label{sec:pho}

The ionization state of individual absorbers enters in calculations of the \HeI\ and \HeII\ opacities and of the 
continuum and line recombination radiation from hydrogen and helium. Under the assumption of photoionization equilibrium 
(generally accurate for quasar absorbers, see Paper II), in a pure H/He gas illuminated by 
a local radiation intensity ${\cal J}_\nu$, the ion fractions $Y_{\rm HI}$,  $Y_{\rm HeI}$, and $Y_{\rm HeII}$ can be written 
in implicit form as  
\begin{equation}
Y_{\rm HI}=(1+R_{\rm HI})^{-1};~~~~~Y_{\rm HeI}=(1+R_{\rm HeI}+R_{\rm HeI}R_{\rm HeII})^{-1};~~~~~
Y_{\rm HeII}=R_{\rm HeI}(1+R_{\rm HeI}+R_{\rm HeI}R_{\rm HeII})^{-1},
\label{eq:phot}
\end{equation}
where
\begin{equation}
R_i\equiv \frac{\Gamma_i}{n_e\alpha_i}, 
\end{equation}
$\Gamma_i$ is the photoionization rate of species $i\in \{$\HI, \HeI, \HeII$\}$,  
\begin{equation}
\Gamma_i\equiv \int{d\nu\,\frac{4\pi {\cal J}_{\nu}}{h\nu}\sigma_i(\nu)},
\end{equation}
and $\alpha_i$ is the (case A) recombination coefficient to all atomic levels of species $i$. The recombination 
rate of the next ionization state $i+1$ (e.g., if $i$ is \HI\ then $i+1$ is \HII) 
is $n_en_{i+1}\alpha_i$, where the electron number density $n_e$ is
\begin{equation}
n_e=n_{\rm H}(1-Y_{\rm HI})+n_{\rm He}Y_{\rm HeII}+2n_{\rm He}(1-Y_{\rm HeI}-Y_{\rm HeII}).
\end{equation} 
In the case of a highly ionized medium with $R_{\rm HI}$, $R_{\rm HeI}$, $R_{\rm HeII} \gg 1$, the densities of 
\HeI\ and \HeII\ can be expressed in terms of the \HI\ density as
\begin{equation}
\frac{n_{\rm HeI}}{n_{\rm HI}}\simeq \frac{n_{\rm He}}{n_{\rm H}}\,\frac{R_{\rm HI}}{R_{\rm HeI}R_{\rm HeII}}
\label{eq:zetathin}
\end{equation}
and
\begin{equation}
\frac{n_{\rm HeII}}{n_{\rm HI}} \simeq \frac{n_{\rm He}}{n_{\rm H}}\,\frac{R_{\rm HI}}{R_{\rm HeII}}.
\label{eq:etathin}
\end{equation}
For optically thin systems, the above relations with ${\cal J}_\nu=J_\nu$ clearly give the ratio between the column 
densities of different ions. Note how the quantity 
\begin{equation}
\eta\equiv N_\nHeII/ N_\nHI
\end{equation}
is independent on gas density only as long as the optically thin approximation holds, while the ratio
\begin{equation}
\zeta \equiv N_\nHeI /N_\nHI
\end{equation}
is always density dependent. 

\subsection{Slab approximation and fitting formulae}

An iterative solution to the equations of cosmological radiative transfer that included a detailed numerical calculation of the 
ionization and temperature structure of individual absorbers at every timestep would be a very computing-intensive task. 
To properly treat the self-shielding of LyC radiation, in Paper II we modeled absorbers as semi-infinite slabs, 
developed a ``steplike'' approximation to the function $\eta(N_{\rm HI})$, and used an analytical escape probability formalism 
to include the recombination emission from absorbers. Fardal, Giroux, \& Shull (1998) solved the local radiative transfer problem 
via an integral equation (the Milne solution 
for a gray atmosphere) for the number of photoionizations at any optical depth in a given slab. They also 
devised an approximation formula to $\eta$ that closely followed the numerical results. Faucher-Gigu\`ere \etal (2009) 
have recently generalized the treatment of Fardal \etal (1998) and applied a similar fitting formula to the 
results of a code that self-consistently solves the photoionization equilibrium balance, including the influence of recombination 
radiation. Here, we follow a similar method: under the assumption of Jeans length thickness for the absorbers, we solve the ionization 
and thermal structure of a slab of finite width illuminated by an external isotropic radiation field $J_\nu$, and derive analytical 
approximations for the ratios $\eta$ and $\zeta$ as a function of $N_\nHI$. Details of our calculations are provided in the Appendix. 
We parameterize the external 
background flux as a power-law, $J_\nu=J_{912}(\nu/\nu_{912})^{-\alpha}$, and as in Faucher-Gigu\`ere \etal (2009) divide 
the intensity above 54.4 eV by a factor of 10 to mimic a cosmological UV filtered spectrum. 

\begin{figure}[thb]
\centering
\includegraphics*[width=0.47\textwidth]{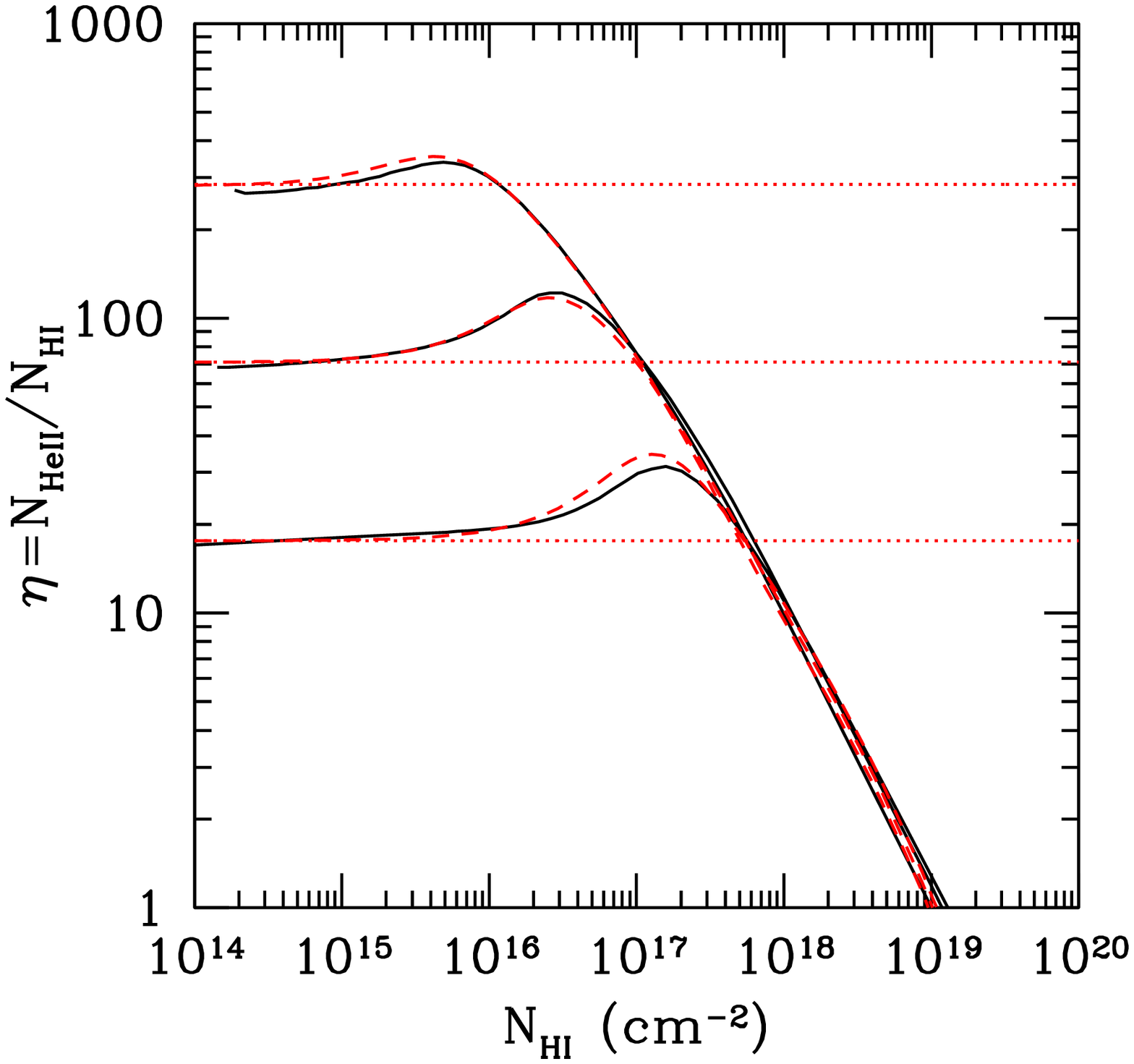}
\includegraphics*[width=0.47\textwidth]{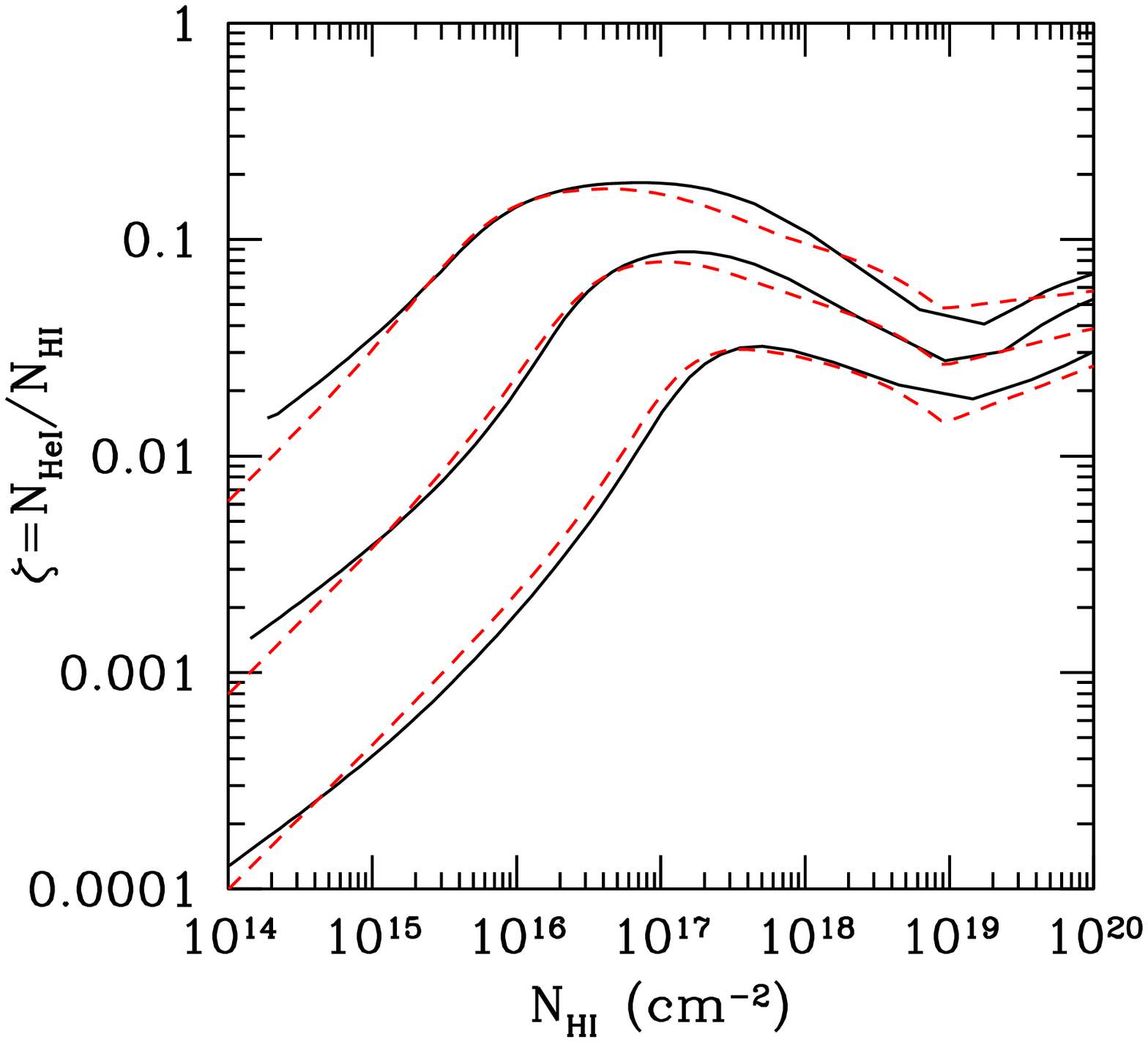}
\caption{\footnotesize {\it Left panel}: The ratio $\eta=N_\nHeII/N_\nHI$ as a function of $N_\nHI$
at redshift 3. The illuminating spectrum has intensity $J_\nu=J_{912}(\nu/\nu_{912})^{-\alpha}$, with $J_{912,-22}=1$ 
and, from bottom to top, spectral slopes $\alpha=0,\, 1,\, 2$, with a break of a factor of 10 at 54.4 eV. {\it Solid curves:} 
full numerical photoionization calculations. {\it Dashed curves:} our analytical approximations for $\eta$ 
based on equation (\ref{eq:fardal}). {\it Dotted curves:} optically thin limit.
{\it Right panel}: same for $\zeta=N_\nHeI/N_\nHI$.}
\label{fig:eta22}
\vspace{+0.5cm}
\end{figure}

Figure \ref{fig:eta22} shows the resulting ratios $\eta(N_\nHI)$ and $\zeta(N_\nHI)$ for a range of input spectra and for the 
representative intensity value at 1 ryd of $10^{-22}\,\uvunits$. The function $\eta$ remains constant at low \HI\ columns, as long as the 
optically thin approximation holds. As the neutral hydrogen column increases, the slab first becomes optically thick to \HeII-ionizing radiation,
and $\eta$ increases. Slabs with even larger columns become optically thick to \HI\ LyC: they are characterized by a highly ionized 
surface layer and an almost fully neutral core. This is the reason for the rapid decrease of 
$\eta$ after the peak, and the consequent trend of $\zeta$ toward the neutral limit, $\zeta \rightarrow n_\nHe/n_\nH$.
As in Fardal \etal (1998) and Faucher-Gigu\`ere et al. (2009), we calculate the column $N_\nHeII$ from the equation  
\begin{equation}
\frac{n_\nHe}{4 n_\nH}\,\frac{\tau_{912}}{1+A\tau_{912}}R_{\nHI}\,=\, \tau_{228}\,+\, \frac{\tau_{228}}{1+B\tau_{288}}R_{\nHeII}, 
\label{eq:fardal}
\end{equation}
where $\tau_{912}\equiv N_\nHI \sigma_{912}$, $\tau_{228}\equiv N_\nHeII \sigma_{228}$, and  $A$ and $B$ are constants fitted to 
our numerical results. To make use of the above expression, one must further specify the ionization rates $\Gamma_i$ to be used in the $R_i$ terms   
together with a relation between electron density and $N_\nHI$. It is in this second step that our approach differs from
that of Faucher-Gigu\`ere et al. (2009). These authors used the optically thin limit for $\Gamma_i$, which provides a poor
approximation to the numerical results.
Here, we first compute the ionization rates in the optically thin limit, 
\begin{equation}
\Gamma^{\rm thin}_i\equiv \int{d\nu\,\frac{4\pi J_{\nu}}{h\nu}\sigma_i(\nu)},
\end{equation}
and derive $\eta^{\rm thin}$ using equation (\ref{eq:etathin}). For a given (input value) of $\tau_\nHI$, we then
calculate $\tau^{\rm thin}_\nHeII$. We then write a first-order approximation to the \HeII\ ionization rate at the face of the slab,
\begin{equation}
\Gamma^{\rm abs}_\nHeII=\int{d\nu\, \frac{4\pi J_\nu}{h\nu} {\rm e}^{-\tau^{\rm thin}_\nHeII(\nu)}\, \sigma_\nHeII(\nu)}.
\end{equation}
The analogous expression for \HI\ is 
\begin{equation}
\Gamma^{\rm abs}_\nHI=\int{d\nu\, \frac{4\pi J_\nu}{h\nu} {\rm e}^{-\tau_\nHI(\nu)}\, \sigma_\nHI(\nu)}.
\end{equation}
Finally, we compute the $R_i$ factors for \HI\ and \HeII\ in equation (\ref{eq:fardal}) as
\begin{equation}
R_i=\frac{0.5\Gamma_i^{\rm thin}+0.5\Gamma_i^{\rm abs}}{n_e\alpha_i}.
\end{equation}
The recombination rates depend on the gas temperature. We found that our numerical results can be fit with the simple scaling   
\begin{equation}
T=(2\times 10^4\,~{\rm K})\, J_{912,-22}^{0.1},
\end{equation}
where $J_{912,-22}\equiv J_{912}/10^{-22}\,\uvunits$, is adequate for our purposes. The weak dependence on $J$ is 
related to the fact that, for $\log N_\nHI\gta 16$, cooling is largely provided by collisionally-excited line radiation rather than by 
recombinations. With this simplified treatment, we have been able to fit the numerically obtained values of $\eta$ for a broad range 
of input spectra. The best-fit curves shown in the left panel of Figure ~\ref{fig:eta22} have been obtained taking $A=0.02$ and $B=0.25$ 
in equation (\ref{eq:fardal}), and 
\begin{equation}
n_e=3.0\times 10^{-3}{\rm cm}^{-3}\,(N_{\nHI,17.2})^{2/3}\,(\Gamma^{\rm thin}_{\nHI,-12})^{2/3}
\end{equation}
for the electron density. Here, $N_{\nHI,17.2}\equiv N_\nHI/10^{17.2}\,{\rm cm}^{-2}$ and
$\Gamma^{\rm thin}_{\nHI,-12}\equiv \Gamma^{\rm thin}_{\nHI}/10^{-12}\,{\rm s}^{-1}$. The above relation can be derived assuming Jeans length thickness for the absorbers and optically thin photoionization equilibrium 
(Schaye 2001; see also Faucher-Gigu\`ere et al. 2009).
A simple approximation for $\zeta$ can be also derived, once $\eta$ is obtained. We use equation (\ref{eq:zetathin}), $\zeta=\eta/R_\nHeI$, with 
$\Gamma_\nHeI=\Gamma^{\rm thin}_\nHeI$ for $N_\nHI \lta 10^{19}$ cm$^{-2}$. At larger columns we apply a linear (in log space) 
extrapolation to the limiting vale $\zeta \rightarrow n_\nHe/n_\nH$ assumed to be reached at $N_\nHI=10^{22}$ cm$^{-2}$.

\section{Recombination emissivity}

\begin{figure}[thb]
\centering
\includegraphics*[width=0.47\textwidth]{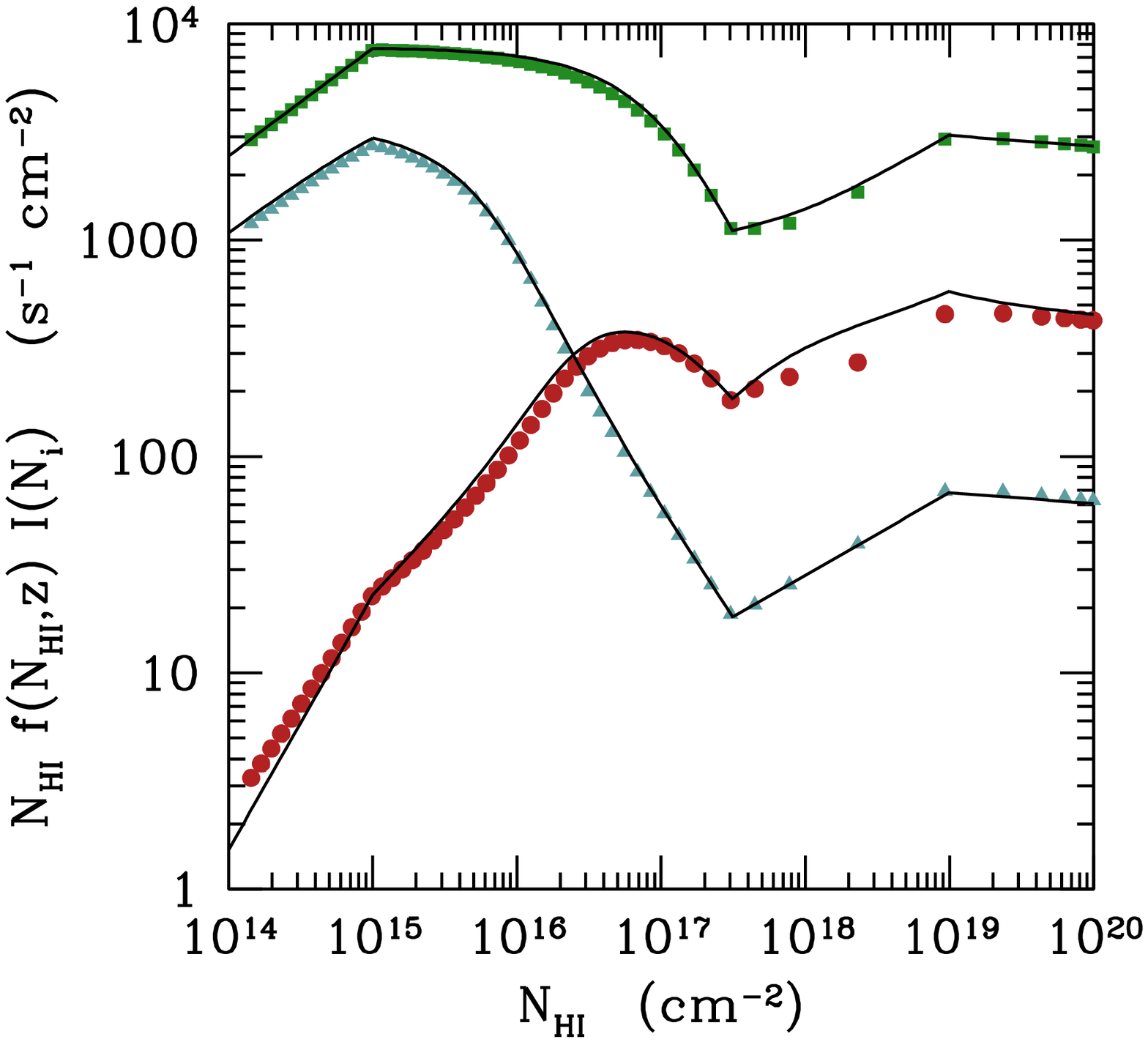}
\includegraphics*[width=0.47\textwidth]{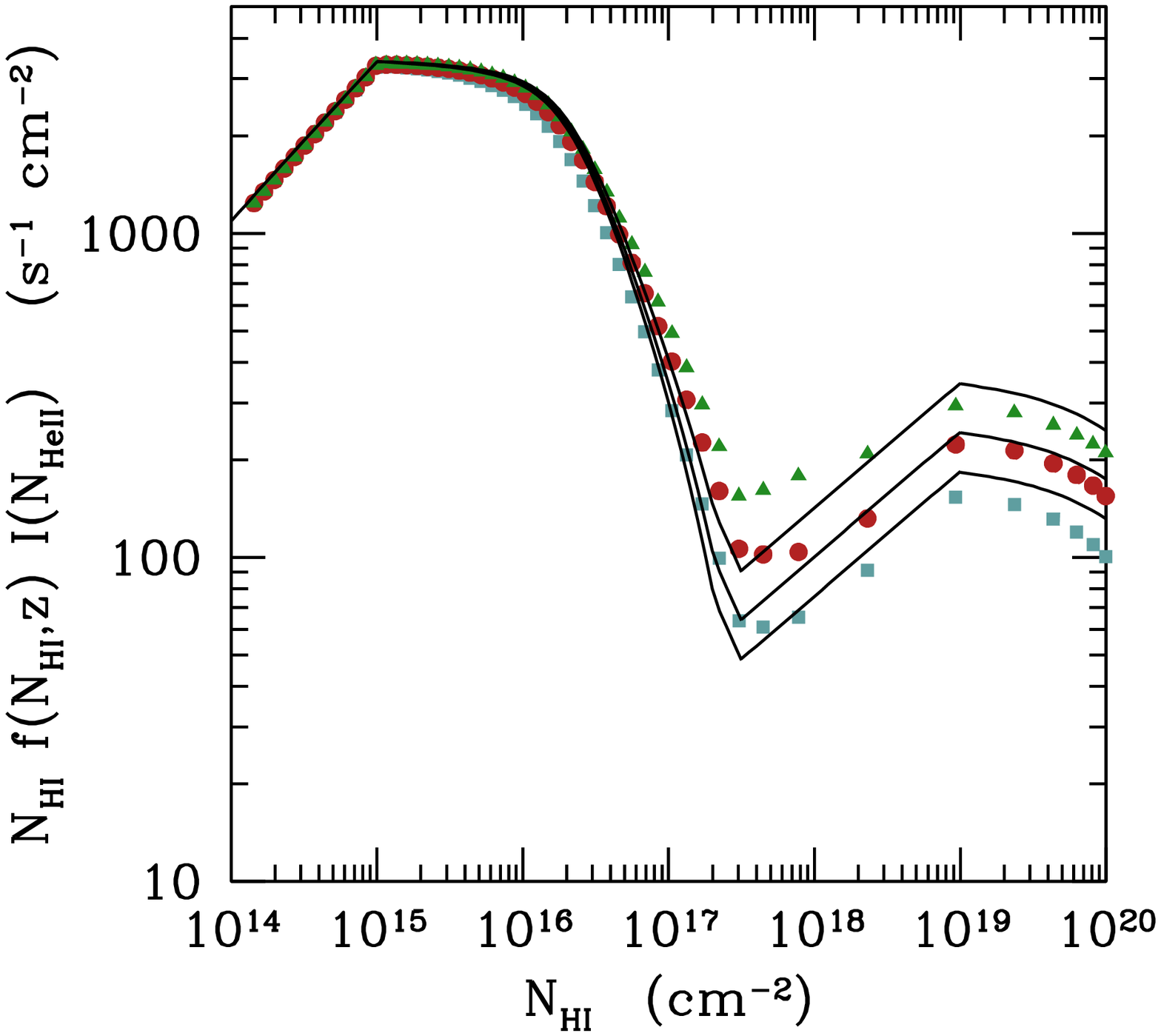}
\caption{\footnotesize {\it Left panel:} LyC recombination radiation from quasar absorbers as a function of $N_\nHI$. The points depict the results
of the full numerical integration of the local radiative transfer equation, while the lines show our analytical approximations (eq.~\ref{eq:recapx}).
{\it Green squares:} \HI\ LyC at 912 \AA.  {\it Blue triangles:} \HeII\ LyC at 228 \AA.  {\it Red circles:} \HeI\ LyC at 504 \AA.  
The quantity plotted is the integral $I(N_i)$ (eq.~\ref{eq:recapx}) multiplied by $N_\nHI  f(N_\nHI ,z)$ (eq.~\ref{eq:ladis}) at $z=3$, showing 
the contribution of optically thin and optically thick absorbers to the LyC emissivity. For this comparison we assumed an illuminating 
spectrum with $J_{912,-22}=1$ and spectral slope $\alpha=1$, with a break of a factor of 10 at 54.4 eV.
{\it Right panel}: same as left panel, but for recombination re-emission from \HeII\ BalC at 13.6 eV ({\it blue squares}), \Lya\ at 40.8 eV ({\it 
green triangles}), and two-photon continuum at 20.4 eV ({\it red circles}).
}
\label{fig:rec}
\vspace{+0.6cm}
\end{figure}

In \S~2 we have seen how background photons absorbed through a Lyman series resonance cause a radiative cascade that 
ultimately terminates either in a Ly$\alpha$ photon or in two-photon $2s \rightarrow 1s$ continuum decay. 
In this section we use the detailed photoionization structure of absorbing systems to    
calculate the reprocessing of background LyC radiation by the clumpy IGM via atomic recombination processes. We 
include recombinations from the continuum to the ground state of \HI, \HeI, and \HeII, as well as \HeII\ Balmer, two-photon, and 
\Lya\ emission. Using the formalism developed in the Appendix, the recombination flux at the slab surface, 
 \begin{equation}
F_\nu=\int{d\Omega\, \mu I_\nu (0,\mu)},
 \end{equation}
can be written as
\begin{equation}
F_\nu=\frac{1}{2}\int_0^L{dx j_\nu(x)}\,\int_0^1{d\mu\, {\rm e}^{-\tau_\nu(x)/\mu}}\,=\,
\frac{1}{2}\int_0^L{dx\,j_\nu(x)E_2(\tau_\nu(x))}.
\label{eq:recflux}
\end{equation}
The emission coefficient from a generic recombination process,  
\begin{equation}
j_\nu(x)\equiv h\nu\,\phi_\nu\, \alpha_r\, n_e(x)\,n_{i+1}(x)\,=h\nu\,\phi_\nu\,\frac{\alpha_r}{\alpha_i}\,n_i(x)\,\Gamma_i(x),
\label{eq:recrad}
\end{equation}
where $\phi_\nu$ is the normalized emission profile and $\alpha_r$ is the relevant recombination coefficient, is 
proportional to the density of species $i$, times the rate at which it absorbs ionizing photons ($\Gamma_i$), times 
the fraction of recombinations that lead to the radiative transition under consideration (the ratio $\alpha_r/\alpha_i$).
The emission profile of free-bound recombination radiation can be computed via the Milne detailed-balance relation, which relates the velocity-dependent 
recombination cross section to the photoionization cross section, while a delta-function profile is sufficient for bound-bound transitions.
The cosmological proper recombination emissivity for the relevant recombination process can then be computed by integrating over the distribution of absorbers,
\begin{equation}
\epsilon_\nu(z)=2|dz/cdt| \int_0^{\infty}\, dN_\nHI\, f(N_\nHI,z) F_\nu(N_\nHI),
\end{equation}
where the factor 2 accounts for the two surfaces of a slab. Using equations (\ref{eq:recflux}) 
and (\ref{eq:recrad}), and denoting with $N_i=\int_0^L n_idx$ the species $i$ column density of 
the absorber, the recombination emissivity becomes 
\begin{equation}
\epsilon_\nu(z)=|dz/cdt| h\nu\,\phi_\nu\,\frac{\alpha_r}{\alpha_i}\,\int_0^{\infty}\, dN_\nHI\, f(N_\nHI,z)
\int_0^{N_i(N_\nHI)}{dN'_i\,\Gamma_i(N'_i) E_2(\tau_\nu(N'_i))}.
\label{eq:recemiss}
\end{equation}
As with the ionization and thermal structure of individual absorbers, it is not practical to perform a self-consistent, iterative, 
numerical evaluation of the recombination emissivity at every timestep in the cosmological code. To derive a simple analytical formula
to the emergent radiation from absorbers, we make use of the fact the number of ionizing incident photons that are absorbed 
saturates in the optically thick regime, and approximate the second integral on the rhs of equation (\ref{eq:recemiss}) as (cf. Faucher-Gigu\`ere \etal 2009)
\begin{equation}
I(N_i)\equiv \int_0^{N_i}{dN'_i\, \Gamma(N'_i)E_2(\tau_\nu(N'_i))}\approx \left(0.5\Gamma_i^{\rm thin}+0.5\Gamma_i^{\rm abs}\right)\,N_T \, 
\left(1-{\rm e}^{-N_i/N_T}\right).
\label{eq:recapx}
\end{equation}
Here $N_T$ is the column density of ion $i$ above which the recombination emission saturates. As shown in Figure \ref{fig:rec}, the above formula 
works especially well in the case of LyC recombination re-emission from \HI\ and \HeII, where self-absorption 
by the emitting ion dominates the local reprocessing of recombination radiation. Our best-fit parameters to the full numerical results for \HI, \HeI, and 
\HeII\ LyC recombinations are $N_T=6.5\times 10^{16}\,(\nu/\nu_{912})^{1.5}\,\cmm$, $N_T=1.2\times 10^{16}\,(\nu/\nu_{504})^{1.5}\,\cmm$, 
and $N_T=2.3\times 10^{17}\,(\nu/\nu_{228})^{1.5}\,\cmm$, respectively. (In the case of non-ionizing \HI\ recombination \Lya\ and 
two-photon emission, we find $N_T=6.5\times 10^{17}\,\cmm$.)      

The emergent recombination flux from \HeII\ BalC, two-photon, and \Lya\ depends on the helium (emission) as well as hydrogen (absorption) ionization structure.  
With the adopted column density distribution, however, recombinations into \HeII\ are dominated by absorbers in the range of columns 
$10^{15}\lta N_\nHI \lta 10^{16}$ cm $^{-2}$: in these systems, \HI\  absorption can be neglected and a simple approximation can be 
found by setting $N_T=2.3\times 10^{18}$ cm$^{-2}$. The fit at large \HI\ columns is actually improved by multiplying the rhs of equation (\ref{eq:recapx}) by 
$\exp[-{\rm min}(\tau_{912},1.3)](\nu_{912}/\nu)^{0.6}$. A comparison between the results of the full numerical integration of the local radiative transfer 
equation and our analytical approximations to the recombination radiation from \HeII\ BalC, two-photon, and \Lya\ are shown in the right 
panel of Figure \ref{fig:rec}. Note that, in our calculations, we have again assumed that \HeII\ \Lya\ photons diffuse into the wings 
and then escape subject only to continuum absorption. 

\section{Quasar UV emissivity}

The only sources of ionizing radiation included in CUBA are star-forming galaxies and quasars. For the quasar comoving emissivity at 1 ryd, 
$\epsilon_{912}(z)/(1+z)^3$, we use the function
\begin{equation}
{\epsilon_{912}(z)\over (1+z)^3}=(10^{24.6}\,\lumdens)\,(1+z)^{4.68}\,{\exp(-0.28z)\over \exp(1.77z)+26.3},
\label{eqemiss}
\end{equation}
which closely fits the results of Hopkins, Richards, \& Hernquist (2007) in the redshift interval $1<z<5$. The UV SED is given by the 
broken power-law
\begin{equation}
L_\nu \propto \begin{cases} \nu^{-0.44} & (\lambda> 1300\,{\rm \AA});\\
\nu^{-1.57} & (\lambda < 1300\,{\rm \AA})
\end{cases}
\label{eq:LUV}
\end{equation}
(Vanden Berk \etal 2001; Telfer \etal 2002).

\begin{figure}[thb]
\centering
\includegraphics*[width=0.6\textwidth]{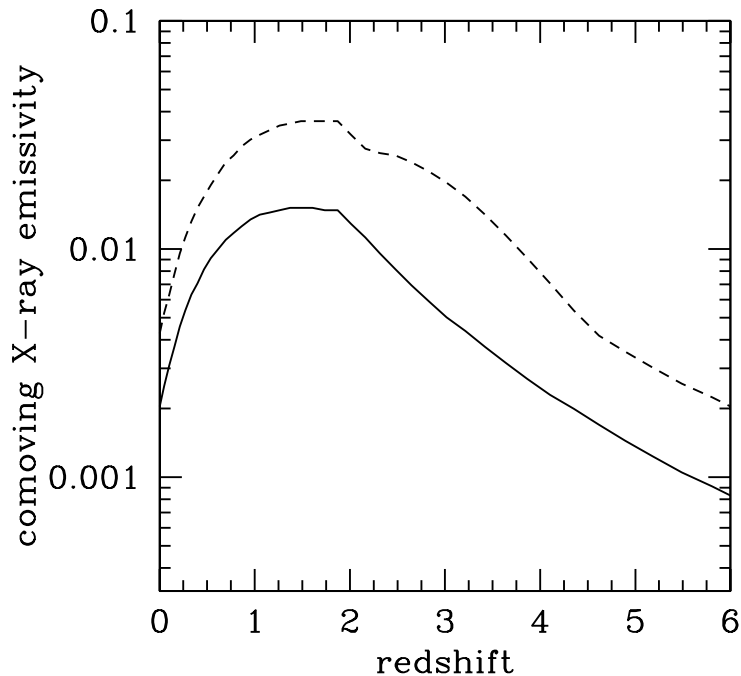}
\caption{\footnotesize Quasar comoving emissivity at 2 keV ({\it dashed line}) and 10 keV ({\it solid line}) 
in units of $10^{23}\,\lumdens$. The latter has been computed following Ueda \etal (2003) and Silverman \etal (2008), 
the former using the procedure outlined in Section \ref{sec:obscuration}.
}
\label{fig5}
\vspace{+0.5cm}
\end{figure}
 
\section{Quasar X-ray emissivity}

The extrapolation of the steep UV power-law in equation (\ref{eq:LUV}) to higher energies is unable to reproduce the 
X-ray properties of the quasar population as a whole, as recorded in the cosmic X-ray background (XRB). The XRB may play a unique role in 
regulating the thermodynamics and ionization degree of intergalactic absorbers. In a photoionized IGM,
soft X-rays between 0.5 and 0.9 keV are responsible for the highest ionization states of metals like carbon, nitrogen, and oxygen. 
At early epochs, X-rays penetrate regions that are optically thick to UV radiation, providing a source of heating and ionization. 
They could make the IGM warm and weakly ionized prior to the era of reionization breakthrough (e.g., Oh 2001; Venkatesan, Giroux, \& Shull 2001;
Madau \etal 2004; Ricotti \& Ostriker 2004; Kuhlen \& Madau 2005). Compton scattering of hard XRB photons may be a source of heating for 
highly ionized low-density intergalactic gas (Madau \& Efstathiou 1999).

Deep X-ray surveys aided by optical identification programs have shown that the bulk of the XRB is produced by a mixture of unobscured 
``Type 1" and obscured ``Type 2" AGNs (Mushotzky \etal 2000; Giacconi \etal 2001), as predicted by XRB synthesis models constructed 
within the framework of AGN unification schemes 
(e.g., Setti \& Woltjer 1989; Madau, Ghisellini, \& Fabian 1994; Comastri \etal 1995; Gilli, Comastri, \& Hasinger 2007). Here, we compute 
the total X-ray emissivity from Type 1 and Type 2 AGNs following a modern version of the original approach by Madau \etal (1994). 

\subsection{Intrinsic hard X-ray luminosity function}

According to Ueda \etal (2003), who combined various surveys from the {\it HEAO 1}, {\it ASCA} and {\it Chandra} satellites, the hard 2-10 keV quasar 
luminosity function (HXLF) follows a luminosity-dependent density evolution with a cutoff redshift (above which the evolution stops) that increases 
with luminosity. At the present epoch, the intrinsic (i.e., before absorption) HXLF of all AGNs (including both Type 1's and Type 2's) is best 
represented by 
\begin{equation}
\phi(L,0)={\phi_*/L_*\over (L/L_*)^{1.86}+(L/L_*)^{3.23}}
\label{eq:phiX}
\end{equation}
in the luminosity range $10^{41.5}-10^{46.5}\,\lumunits$, where $\phi_*=2190$ Gpc$^{-3}$ and 
$L_*=10^{43.94}\,\lumunits$. This changes with cosmic time (for redshift up to 3) as
\begin{equation}
\phi(L,z)=\phi(L,0)\,e(z,L),
\end{equation}
where the evolution factor is  
\begin{equation}
e(z,L)=\begin{cases} (1+z)^{e_1} & (z<z_c);\\
        e(z_c)\left(\frac{1+z}{1+z_c}\right)^{e_2} & (z\geq z_c)
\end{cases}
\end{equation}
and
\begin{equation}
z_c(L)=\begin{cases} z_c^* & (L\geq L_a);\\ 
        z_c^*(L/L_a)^{0.335} & (L<L_a).
\end{cases}
\end{equation}
Here, $e_1=4.23$, $e_2=-1.5$, $z_c^*=1.9$, and $L_a=10^{44.6}\,\lumunits$ (Ueda \etal 2003).  
An extension of the HXLF up to $z\sim 5$ by Silverman et al. (2008) shows a steeper decline in the number of
$z>3$ AGNs with an evolution rate similar to that found by studies of optically-selected QSOs. The new fit requires
a much stronger evolution above the cutoff redshift, $e_2=-3.27$, than previously found by Ueda \etal (2003, $e_2=-1.5$). In the 
following, we shall use Ueda \etal (2003) HXLF best fit parameters together with the Silverman \etal (2008) value for $e_2$. 

For the intrinsic spectrum before absorption, we assume the standard power-law multiplied by an exponential,
\begin{equation}
S_E\propto E^{-\alpha}\exp\left(-{E\over E_c}\right), 
\end{equation}
with $\alpha=0.9$ (Nandra \& Pounds 1994). The high-energy cutoff, $E_c=460\,$keV, is fixed by the shape of
the XRB turnover above 30 keV. These seed photons are then reflected towards the observer by a semi-infinite cold disk 
close to the primary emitter. This reflection component, commonly detected in the X-ray spectra of nearby Seyfert galaxies (Nandra \& Pounds 
1994), is comparable to the direct flux around 30 keV, decreases rapidly towards lower energies, and flattens the 
overall spectral slope above 10 keV (Lightman \& White 1988). 

\subsection{AGN emissivity after absorption}
\label{sec:obscuration}

According to the AGN unification scheme, obscuring matter at a distance of several parsecs
from the central powerhouse blocks our line of sight to the active nucleus. When our
view is unobscured, we see a Type 1 AGN; when our view is occulted, photons of all energies from the far IR to 
several keV are absorbed, and in these bands we  can only detect the nucleus in scattered light. Ueda \etal (2003) found
the following expression for the observed (normalized) distribution of absorbing $N_\nH$ columns:
\begin{equation}
f(L,N_\nH)= 
\begin{cases}
2-(5+2\epsilon)/(1+\epsilon)\,\psi & (20.0\leq \log{N_\nH} < 20.5),\\ 
1/(1+\epsilon)\,\psi & (20.5\leq \log{N_\nH}<23.0),\\
\epsilon/(1+\epsilon)\,\psi & (23.0\leq \log{N_\nH}<24),
\end{cases}
\label{eq:nhdistr}
\end{equation}
where the parameter
\begin{equation}
\psi(L)={\rm min}\{\psi_{\rm max},{\rm max}[\psi_{44}-\beta(\log{L}-44),\,0]\}
\end{equation}
accounts for the fact that the fraction of absorbed sources is smaller at higher luminosities.   
Here, $\psi_{\rm max}=(1+\epsilon)/(3+\epsilon)$, $\epsilon=1.7$, $\psi_{44}=0.47$, $\beta=0.1$, and 
$ \int_{20}^{24}{f(L,N_\nH) \, d\log{N_\nH}}=1$. Sources absorbed by a column larger (smaller) than $10^{22}\,\cmm$ are 
defined as X-ray Type 2 (Type 1) AGNs. It is assumed that ``Compton-thick" AGNs with columns $\log N_\nH>24$ are not present 
in samples detected below 10 keV. Such a population is added by extrapolating the $N_\nH$ function above $\log N_H>24$, keeping 
the same normalization up to $\log N_\nH=25$ as well as the same cosmological evolution of Compton-thin AGNs (Ueda \etal 2003).    

We then follow Madau, Ghisellini, \& Fabian (1993) and model the thick blocking material that covers most of the solid angle around the central X-ray source
as a homogeneous spherical cloud of cold material and column $N_\nH$. The radiation transfer is computed with a Monte Carlo code 
constructed using the photon-escape weighing method of Pozdnyakov, Sobol', \& Sunyaev (1983). We set the electron temperature equal to zero,
use the full Klein-Nishina scattering cross section, adopt the bound-free opacity associated with standard cosmic-abundance material from 
Morrison \& McCammon (1983), and ignore the iron K$\alpha$ emission line in the spectra. Each Monte Carlo run uses $10^6$ photons, and produces 
as output an absorbed spectrum, $S_E(N_\nH)$. After being reprocessed by cold material along the line of sight, the emergent specific intensity
forms a hump, whose position and width are determined by the competition of bound-free absorption at low energies, and Compton downscattering 
and exponential roll-off of the primary spectrum at high energies (Madau \etal 1993). A small spectral component, equal to 2.5\% 
of the primary incident power and representing the flux scattered into the line of sight by electrons in the warm ionized medium, 
is added to the transmitted Type 2 flux. The absorbed spectra are then averaged over the $N_\nH$-distribution corresponding to a given luminosity, and normalized 
to the unabsorbed 2-10 keV flux,
\begin{equation}
S_E(L)=\frac{\int_{20}^{24} {S_E(N_\nH)\,f(L,N_\nH)d\log{N_\nH}}}  {\int_{2-10\,{\rm keV}}{S_EdE}},
\end{equation}
to yield a flux normalized, luminosity-dependent, average AGN SED. The X-ray proper emissivity as a function of redshift is then obtained
by simply integrating over the HXLF,
\begin{equation}
\epsilon_E(z)=(1+z)^3 \int_{L_{\rm min}}^{L_{\rm max}} {S_E(L)\phi(L,z)L\, dL},
\end{equation}
where we set $L_{\rm min}=10^{41.5}\,\lumunits$, and $L_{\rm max}=10^{48}\,\lumunits$. 

\begin{figure}[thb]
\centering
\includegraphics*[width=0.47\textwidth]{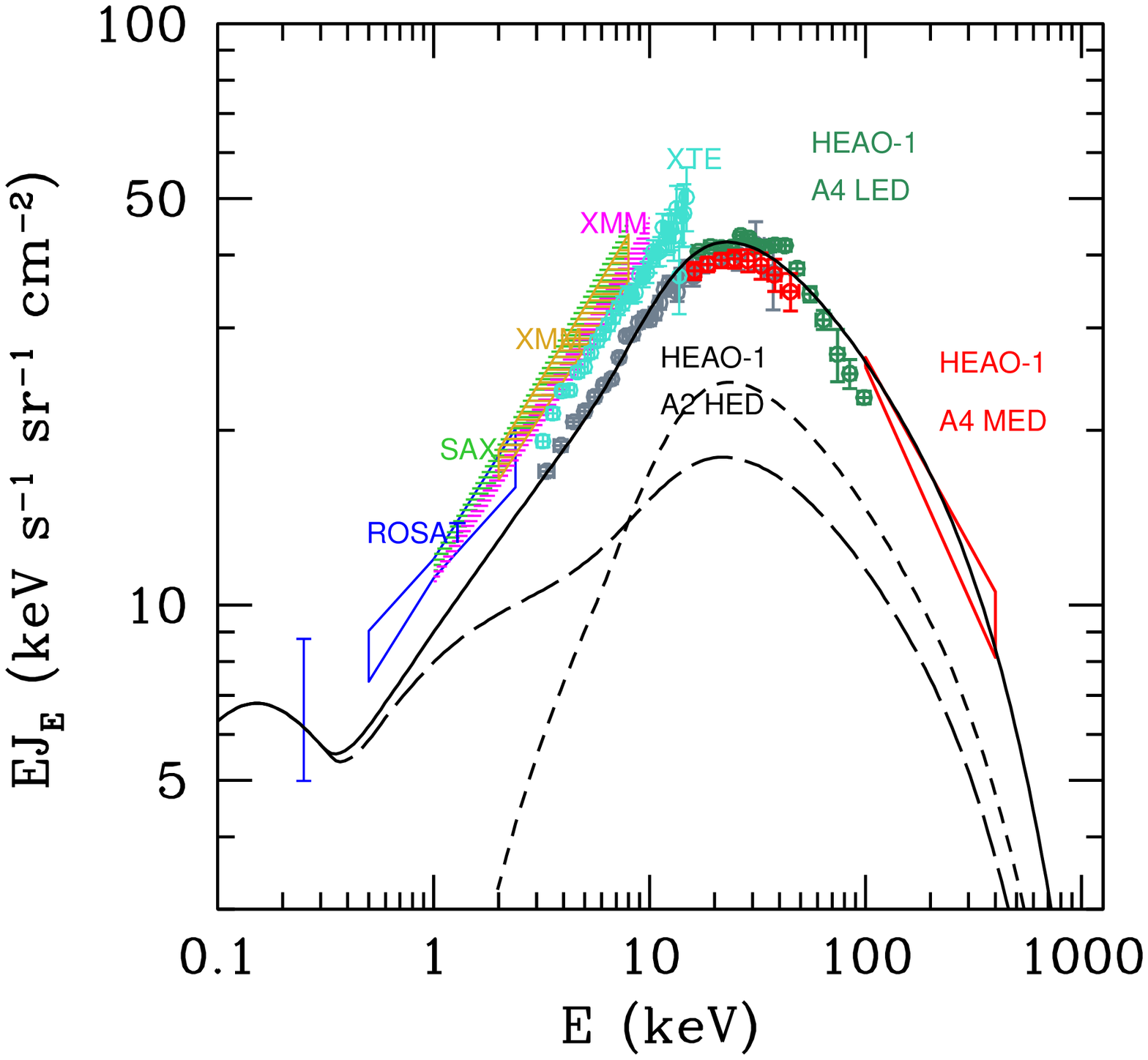}
\includegraphics*[width=0.47\textwidth]{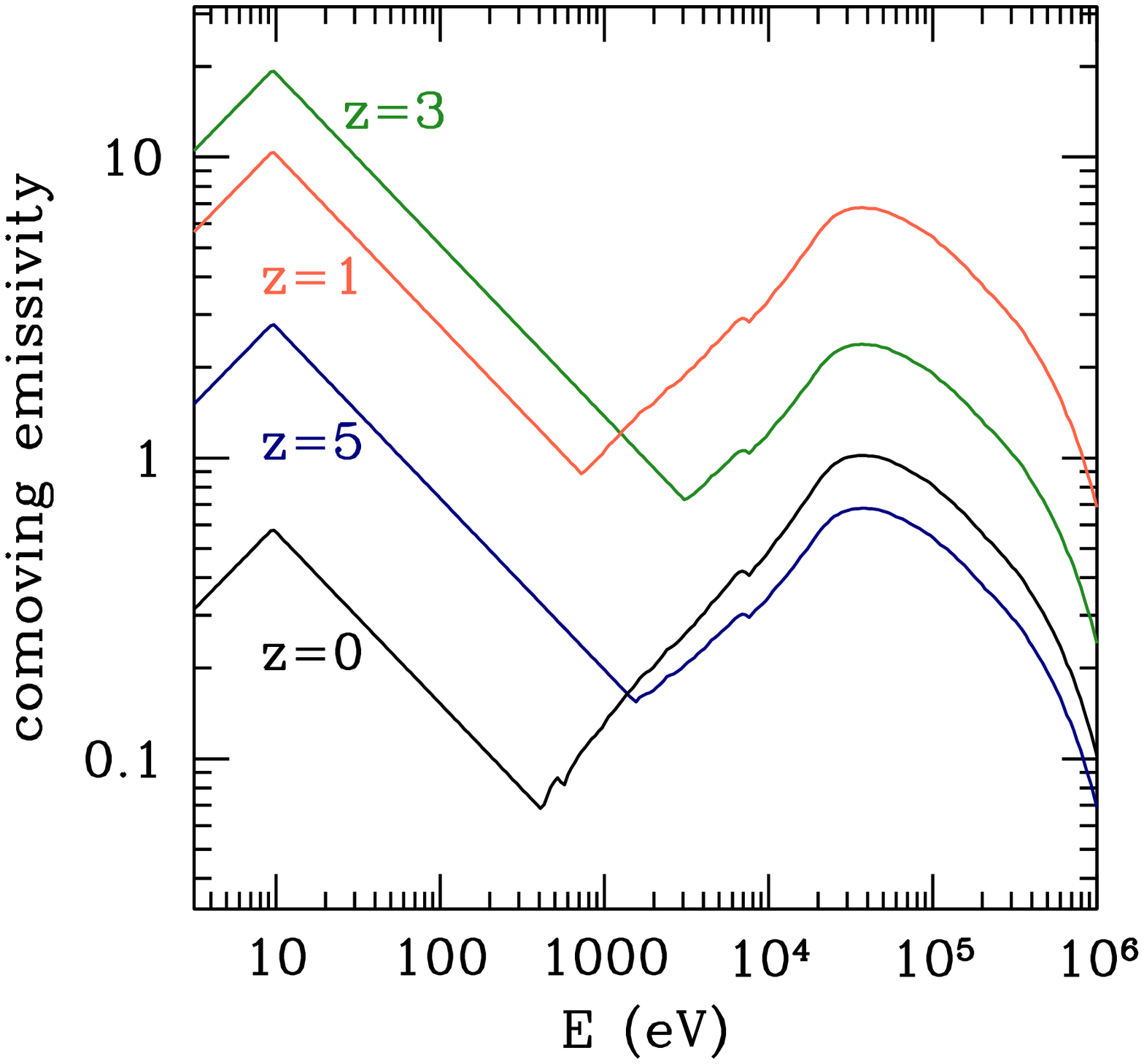}
\vspace{-0.3cm}
\caption{\footnotesize {\it Left panel:} The cosmic XRB spectrum and the predicted contribution from AGNs. {\it Grey points:} {\it HEAO}-1 A2 HED data 
(Gruber \etal 1999). {\it Dark green points:} {\it HEAO}-1 A4 LED (Gruber \etal 1999).  {\it Cyan points}: {\it Rossi-XTE} (Revnivtsev \etal 2003).
{\it Blue point:} 0.25 keV soft XRB intensity from {\it ROSAT} shadowing experiments (Warwick \& Roberts 1998).
{\it Red bowtie:} {\it HEAO}-1 A4 MED (Kinzer \etal 1997). 
{\it Blue bowtie}: {\it ROSAT} PSPC data (Georgantopoulos \etal 1996). 
{\it Light green bowtie:} {\it BeppoSAX} (Vecchi \etal 1999). {\it Purple and yellow bowties:} 
{\it Newton-XMM} (Lumb \etal 2002; De Luca \& Molendi 2004). {\it Solid line:} our synthesis model spectrum, produced by a mixture of 
absorbed ($\log N_\nH>22$, {\it short-dashed line}) and unabsorbed ($\log N_\nH<22$, {\it long-dashed line}) AGNs. See text for details. 
{\it Right panel:} The broadband quasar comoving emissivity per logarithmic bandwidth, $\nu\epsilon_\nu/(1+z)^3$ (in units of $10^{39}$ 
ergs s$^{-1}$ Mpc$^{-3}$), as a function of photon energy $E$ from the optical to hard X-rays. The composite spectrum is shown at redshifts $z=0,1,3,5$. 
}
\label{fig6}
\vspace{+0.3cm}
\end{figure}

The model described above is able to reproduce a number of X-ray observations, from the evolution of AGNs in the soft and hard X-ray bands, to the XRB. 
The quasar comoving emissivity at 2 keV and 10 keV is plotted in Figure \ref{fig5}, while a global fit to the XRB is shown in the left panel of 
Figure \ref{fig6}. The absolute XRB flux is still affected by rather large uncertainties: our model reproduces well the background intensity 
measured by {\it HEAO-1} and {\it BeppoSAX}, but the HEAO-1 A2 data are lower by about 20\% with respect to the determinations by, 
e.g., {\it XMM} and {\it RXTE} at energies below 10 keV. Figure \ref{fig6} (right panel) depicts the broadband quasar comoving emissivity 
per logarithmic bandwidth, $\nu\epsilon_\nu/(1+z)^3$, as a function of photon energy from the optical to hard X-rays. In terms of energy output, 
the composite spectrum for $\lambda<5000\,$\AA\ is characterized by two broad bumps, one in the UV at 10 eV and another in the X-ray region at 
30 keV (a third peak in the mid-infrared, see, e.g., Sazonov, Ostriker, \& Sunyaev 2004, can be neglected for the present purposes). While X-rays 
dominate the energy ouput at $z=0$, the peak of the emitted power moves increasingly towards the UV at redshifts above 1.       

\section{Galaxy emissivity}

Star-forming galaxies are expected to play a dominant role as sources of hydrogen-ionizing radiation at $z>3$ 
as the quasar population declines with lookback time. To compute the LyC emissivity from galaxies at all epochs, we start 
with an empirical determination of the star formation history of the universe following Madau \etal (1996).
We adopt the far-UV (FUV, 1500 \AA) luminosity functions of Schiminovich \etal (2005) in the redshift range $0\le z \le 2$, 
of Reddy \& Steidel (2009) at $z=2.3$ and 3.05, and of Bouwens \etal (2011) at redshifts 3.8, 5.0, 5.9, 6.8, and 
8.0. All were integrated down to $L_{\rm min}=0.01\,L_*$ using Schechter function fits with parameters 
$(\phi_*, L_*, \alpha)$ to compute the dust-reddened galaxy FUV luminosity density $\rho_{\rm FUV}$\footnote{In this section we use the notation 
$\rho_\nu(z)\equiv \epsilon_\nu(z)/(1+z)^3$, i.e., the term {\it luminosity density} is synonymous with {\it comoving specific emissivity}.},
\begin{equation}
\rho_{\rm FUV}(z)=\int_{0.01L_*}^\infty L\phi(L,z)dL=\Gamma(2+\alpha,0.01)\phi_*L_*.
\end{equation} 
Here $\alpha$ denotes the faint-end slope of the Schechter parameterization and $\Gamma$ is the incomplete gamma function. 
We used $\alpha=-1.6$ at $0<z<2$, $\alpha=-1.73$ at $z=2.3$ and $z=3.05$, and $\alpha=-1.73, -1.66, -1.74, -2.01, -1.91$ at 
$z=3.8, 5.0, 5.9, 6.8, 8.0$, respectively (see Schiminovich \etal 2005; Reddy \& Steidel 2009; Bouwens \etal 2011).

\begin{figure}[thb]
\centering
\includegraphics*[width=0.47\textwidth]{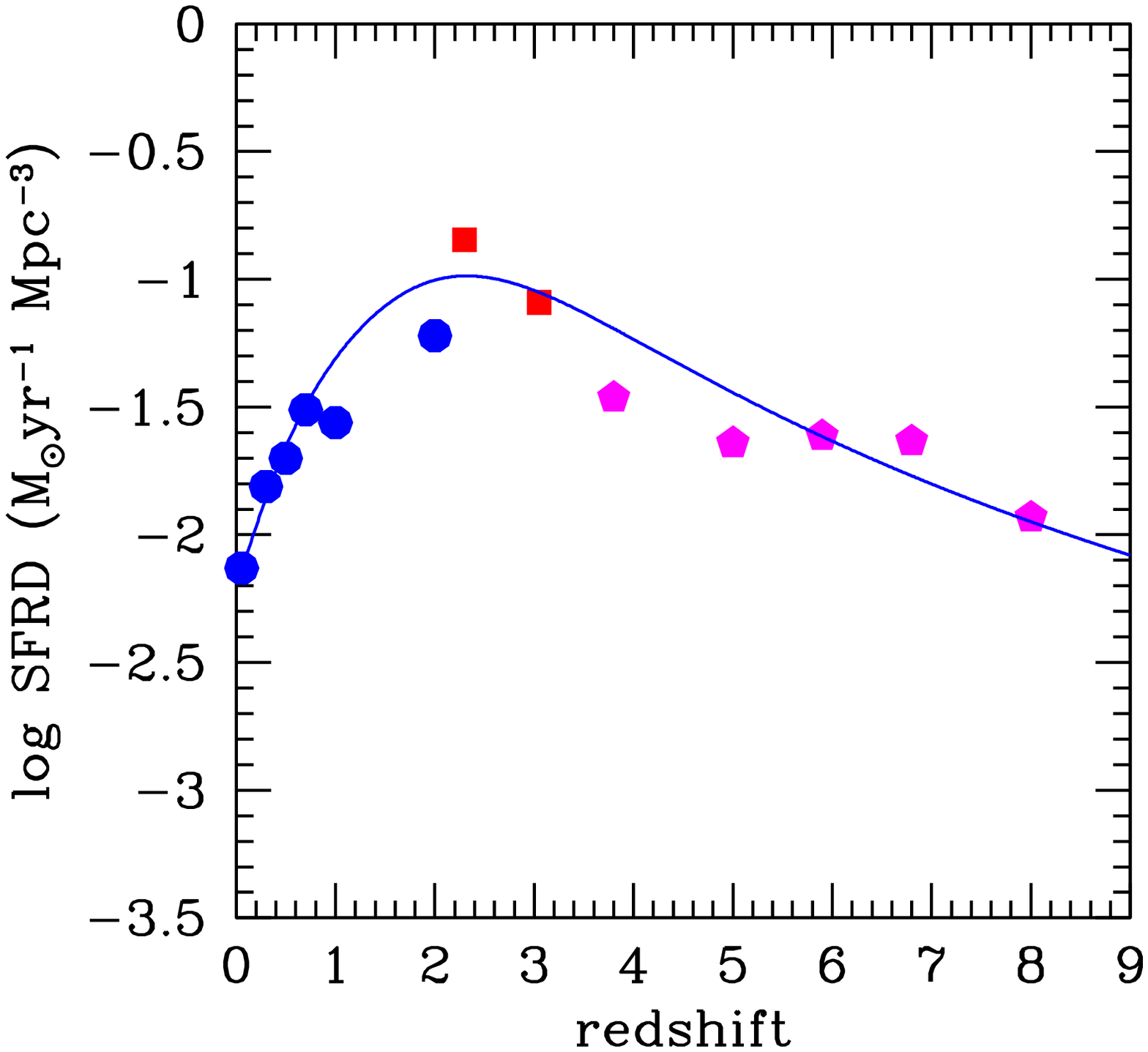}
\includegraphics*[width=0.47\textwidth]{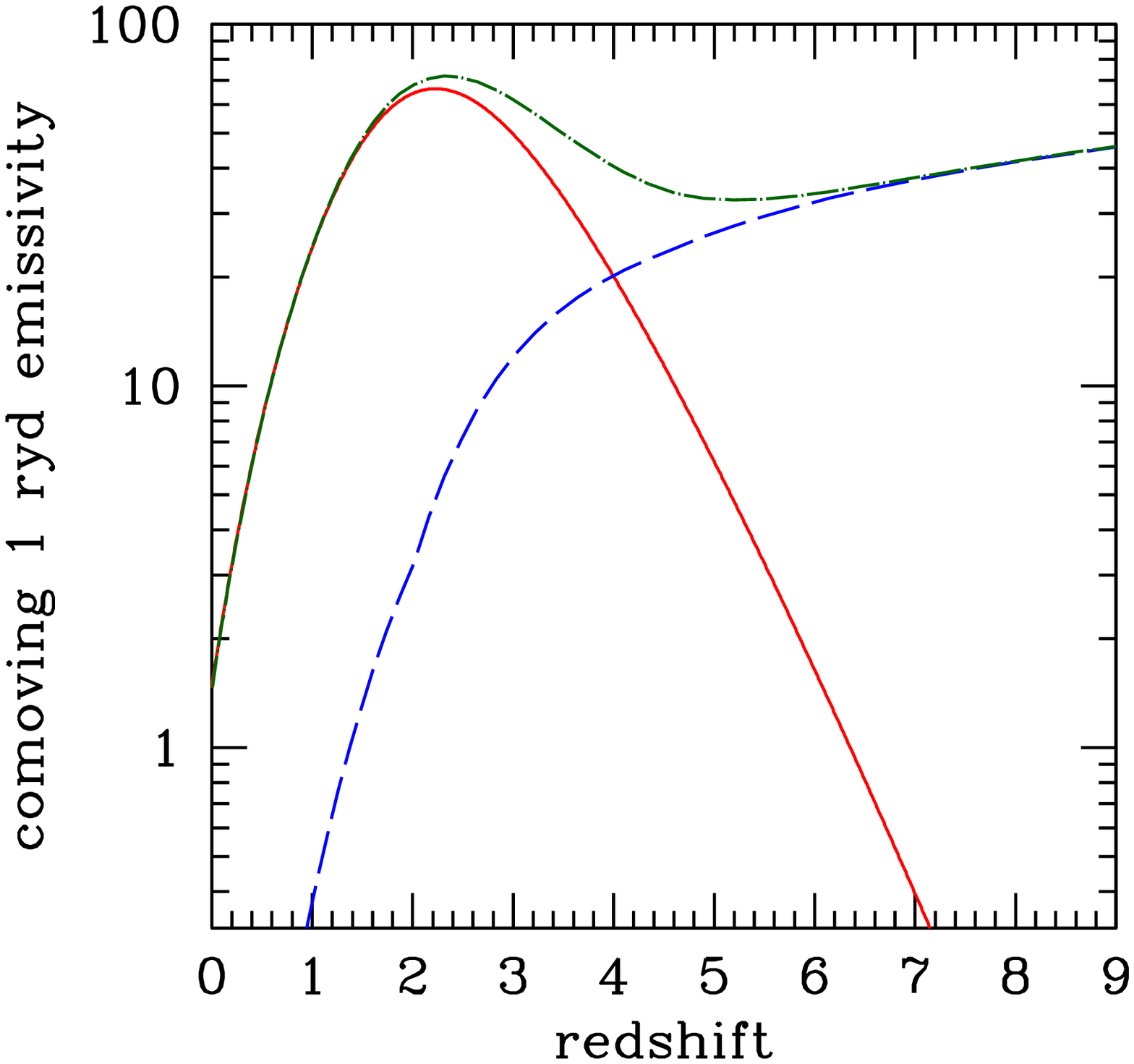}
\caption{\footnotesize {\it Left:} The cosmic history of star formation. The FUV data points from Schiminovich \etal (2005) ({\it blue dots}), 
Reddy \& Steidel (2009) ({\it red squares}) and  Bouwens \etal (2011) ({\it magenta pentagons}) have been converted to instantaneous star 
formation rate density using the conversion factor ${\cal K}=1.05\times 10^{-28}$ (see text for details). The best-fitting star formation 
history, ${\rm SFRD}(z)=[6.9\times 10^{-3}+0.14(z/2.2)^{1.5}]/[1+(z/2.7)^{4.1}]\,\sfrd$, is plotted with the solid blue curve.
{\it Right:} Comoving galaxy emissivity (in units of $10^{23}\,\lumdens$) of 1 ryd photons escaping into the IGM ({\it dashed line}), 
for an escape fraction $\langle f_{\rm esc}\rangle=1.8\times 10^{-4}(1+z)^{3.4}$. The solid line shows the best-fit QSO emissivity of 
eq. (\ref{eqemiss}) for comparison, while the dot-dashed line shows the total quasars $+$ galaxies emissivity.
}
\label{fig7}
\vspace{+0.3cm}
\end{figure}

Dust attenuation was treated using a Calzetti \etal (2000) extinction law, with the function
\begin{equation}
A(\nu,z)=A_{\rm FUV}(z){k(\nu)\over k(1500\,{\rm \AA})}
\label{eq:Anu}
\end{equation}
measuring the magnitudes of attenuation suffered at frequency $\nu$ and redshift $z$. For the luminosity-weighted obscuration at 1500 \AA\
we take
\begin{equation}
A_{\rm FUV}(z)=\begin{cases} 1 & (0\le z\le 2);\\
2.5 \log [(1+1.5/(z-1)] & (z>2).
\end{cases}
\label{eq:AFUV}
\end{equation}
The above expression reproduces at $z\le 2$ the FUV ``minimum dust correction factor" of 2.5 from Schiminovich \etal (2005), the dust 
correction factors of $2.38\pm 0.59$ and $2.0\pm 0.62$ at $z=2.3$ and $z=3.05$ from Reddy \& Steidel (2009), and the decreasing dust 
attenuation at higher redshift from Bouwens \etal (2011). The dust-corrected luminosity densities were smoothed with an approximating 
function and then compared with the results of spectral population synthesis models as follows. The GALAXEV library of Bruzual \& Charlot (2003) 
provides the age-luminosity evolution for a simple stellar population (SSP) at different wavelengths. The FUV luminosity density (before 
dust obscuration) at time $t$ of a ``cosmic stellar population" characterized by a star formation rate density SFRD$(t)$ and a metal-enrichment 
law $Z(t)$ is given by the convolution integral
\begin{equation}
\rho_{\rm FUV}(t)=\int_0^t {\rm SFRD}(t-\tau)l_{\rm FUV}[\tau,Z(t-\tau)]d\tau,
\label{eq:rhofuv}
\end{equation}
where $l_{\rm FUV}[\tau,Z(t-\tau)]$ is specific luminosity radiated at 1500 \AA\ per unit initial stellar mass by an SSP
at age $\tau$ and metallicity $Z(t-\tau)$. We use SSPs of decreasing metallicities with redshift according to 
\begin{equation}
Z(z)=Z_\odot 10^{-0.15z}       
\end{equation}
(Kewley \& Kobulnicky 2007), for a Salpeter IMF between 0.1 and 100 $\msun$. Starting from an initial guess, the function SFRD$(t)$ was adjusted 
in an iterative fashion until the computed FUV luminosity densities as a function of redshift provided a good match to the data. The 
best-fitting star formation history, 
\begin{equation}
{\rm SFRD}(z)={6.9\times 10^{-3}+0.14(z/2.2)^{1.5}\over 1+(z/2.7)^{4.1}}\,\sfrd,
\label{eq:sfrd}
\end{equation}
is shown in Figure \ref{fig7} (left panel), together with the observed luminosity densities adopted in this study. The latter have been 
converted to ongoing star formation rate densities according to 
\begin{equation}
{\rm SFRD}(t)={\cal K}\times \rho_{\rm FUV}(t);~~~~~{\cal K}=1.05\times 10^{-28}, 
\end{equation}
where $\rho_{\rm FUV}$ is expressed in units of $\lumdens$ and SFRD is in units of $\sfrd$. This approximate transformation
makes use of the basic property that the FUV continuum in galaxies is dominated by short-lived massive stars, and is 
therefore a direct measure, for a given IMF and dust content, of the instantaneous star formation rate. The conversion 
factor ${\cal K}$ in the equation above reproduces to within 2\% the results of the synthesis models above redshift 2 
given the adopted star formation and metal enrichment history (cf. Madau, Pozzetti, \& Dickinson 1998). 
At redshift $0<z<1$, ${\cal K}$ decreases from $1.16\times 10^{-28}$ to $1.10 \times 10^{-28}$. Note that these 
newly derived conversion factors are between 21\% and 33\% smaller than the widely used value, ${\cal K}=1.4\times 10^{-28}$,
quoted by Kennicutt (1998) (and based on the calibration by Madau \etal 1998), the differences reflecting 
updated stellar population synthesis models and subsolar metallicities at high redshifts.    

Once the star formation history has been determined, we use stellar synthesis models to compute the dust-reddened frequency-dependent 
UV emissivity as
\begin{equation}
\rho_\nu(t)=C(t)\int_0^t {\rm SFRD}(t-\tau)l_\nu[\tau,Z(t-\tau)]d\tau.
\label{eq:rhoi}
\end{equation}
We take $C(t)\equiv 10^{-0.4A(\nu,t)}$ at all photon energies below 1 ryd, and $C(t) \equiv \langle f_{\rm esc}\rangle$ above the Lyman 
limit. In our treatment, the escape fraction $\langle f_{\rm esc}\rangle$ is a free parameter that incorporates local continuum absorption by hydrogen, helium,
and dust. It is the angle-averaged, absorption cross 
section-weighted, and luminosity-weighted fraction of ionizing radiation that leaks into the IGM from star-forming galaxies: 
the escaping radiation is produced not by sources in a semiopaque medium but by a small fraction of essentially unobscured 
sources (e.g., Gnedin, Kravtsov, \& Chen 2008). In the ``minimal reionization model" discussed in detail in the next section, 
the escape fraction of photons between 1 and 4 ryd is assumed to be a steeply rising function of redshift (see also Inoue, Iwata, \& Deharveng 2006),
\begin{equation}
\langle f_{\rm esc}\rangle=1.8\times 10^{-4}(1+z)^{3.4},
\label{eqno:fesc}
\end{equation}    
and is zero above 4 ryd. The expression above yields an escape fraction at $z=3.3$ of 2.6\%, comparable to the recent upper limit for $L>L_*$ Lyman
break galaxies of Boutsia \etal (2011). The relatively low values of $\langle f_{\rm esc}\rangle$ implied by the above expression in the 
redshift interval from $z=2$ (0.8\%) to $z=5$ (8\%) are dictated in our model by the need to reproduce the hydrogen-ionization rates 
inferred from flux decrement measurements (see 
Fig. \ref{fig8} below). In the same redshift range, the escape fraction of ionizing radiation  from star-forming galaxies hosting 
a $\gamma$-ray burst is measured to be $\langle f_{\rm esc}\rangle\le$ 7.5\% (95\% c.l.) (Chen, Prochaska, \& Gnedin 2007), in agreement
with our expression. The high values predicted by equation (\ref{eqno:fesc}) above redshift 7, in excess of 20\%, are needed to compensate 
for the decline in the star formation rate density and to reionize the IGM at early enough epochs. The resulting galaxy 
emissivity of 1 ryd photons escaping into the IGM is shown in the right panel of Figure \ref{fig7}. Galaxies dominate over QSOs at all redshifts $z>4$, and 
make a negligible contribution to the ionizing background at $z<3$. The total comoving emissivity from quasars $+$ galaxies decreases only weakly
from $z=3$ to $z=5$, and is fairly flat afterwards. This trend is consistent with the conclusions reached by Bolton \& Haehnelt (2007) and 
Faucher-Gigu\`ere \etal (2008a) from empirical measurements of the \Lya\ forest opacity.

\subsection{Ly$\alpha$ emission from galaxies}
Stellar population synthesis codes do not typically include nebular line emission. Here, we provide a simple estimate of the \Lya\ emission from
hydrogen recombinations in the interstellar medium of galaxies. In case B recombination, about 68\% of all the absorbed LyC photons will be converted
locally into \Lya\ (Osterbrock 1989). The \Lya\ proper volume emissivity can then be written as  
\begin{equation}
\epsilon_{\alpha}(z)=h\nu_\alpha \delta(\nu-\nu_\alpha) \dot n_{\alpha}(z), 
\label{eq:lyagal}
\end {equation}
where 
\begin{equation}
\dot n_{\alpha}(z)=0.68 (1-\langle f_{\rm esc}\rangle)\, \int_{\nu_L}^\infty  {d\nu\over h\nu} \epsilon_\nu(z)
\end{equation}
and $\epsilon_\nu$ is the proper volume emissivity from galaxies. We assume here that \Lya\ suffers the same dust extinction as LyC, 
a simple treatment that is unlikely to capture the complex radiative transfer physics of the \Lya\ line as it propagates through the dusty ISM
(see, e.g., Caplan \& Deharveng 1986; Neufeld 1991; Dijkstra 2009; Scarlata et al. 2009; Dayal, Ferrara \& Saro 2010). Inserting equation 
(\ref{eq:lyagal}) into (\ref{Jnu}) yields the additional flux observed at $\nu_o \leq \nu_\alpha$ from galaxy \Lya\ : 
\begin{equation}
J_{\nu_o}(z_o) =
\frac{h}{4\pi}\frac{c}{H(z_\alpha)} \left(\frac{\nu_o}{\nu_\alpha}\right)^3 \dot n_{\alpha}(z_\alpha), 
\end{equation}
where $1+z_\alpha=(\nu_\alpha/\nu_o) (1+z_o)$. We have neglected collisionally excited \Lya\ emission, as this is only about 10-20\% of the 
recombination term (Dayal et al. 2010). A similar contribution is also expected in the emitted spectrum of dense absorbers like the SLLSs and DLAs, 
while collisional excitation is always negligible in lower column density systems.   

\begin{figure}[thb]
\centering
\includegraphics*[width=0.47\textwidth]{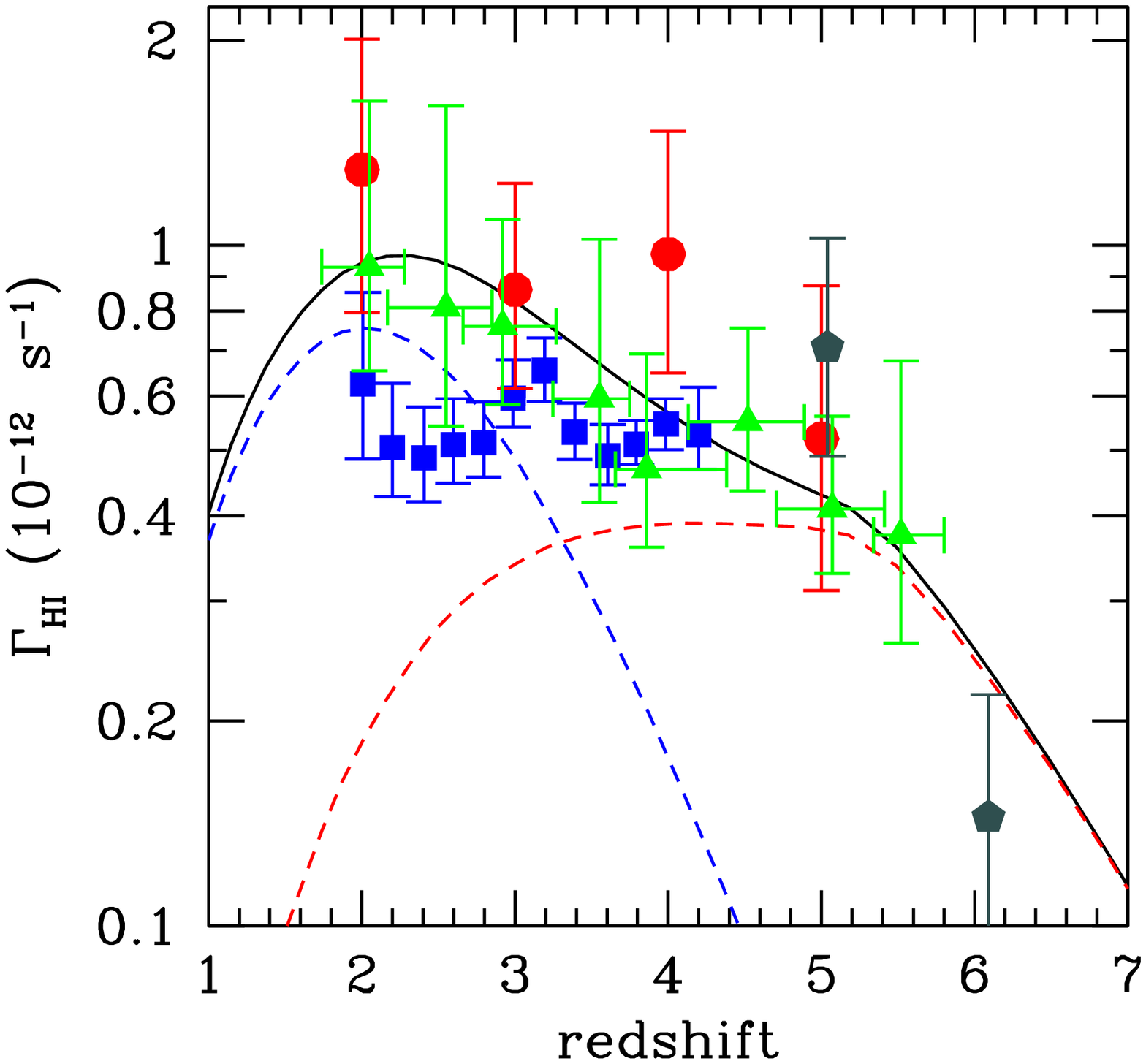}
\includegraphics*[width=0.47\textwidth]{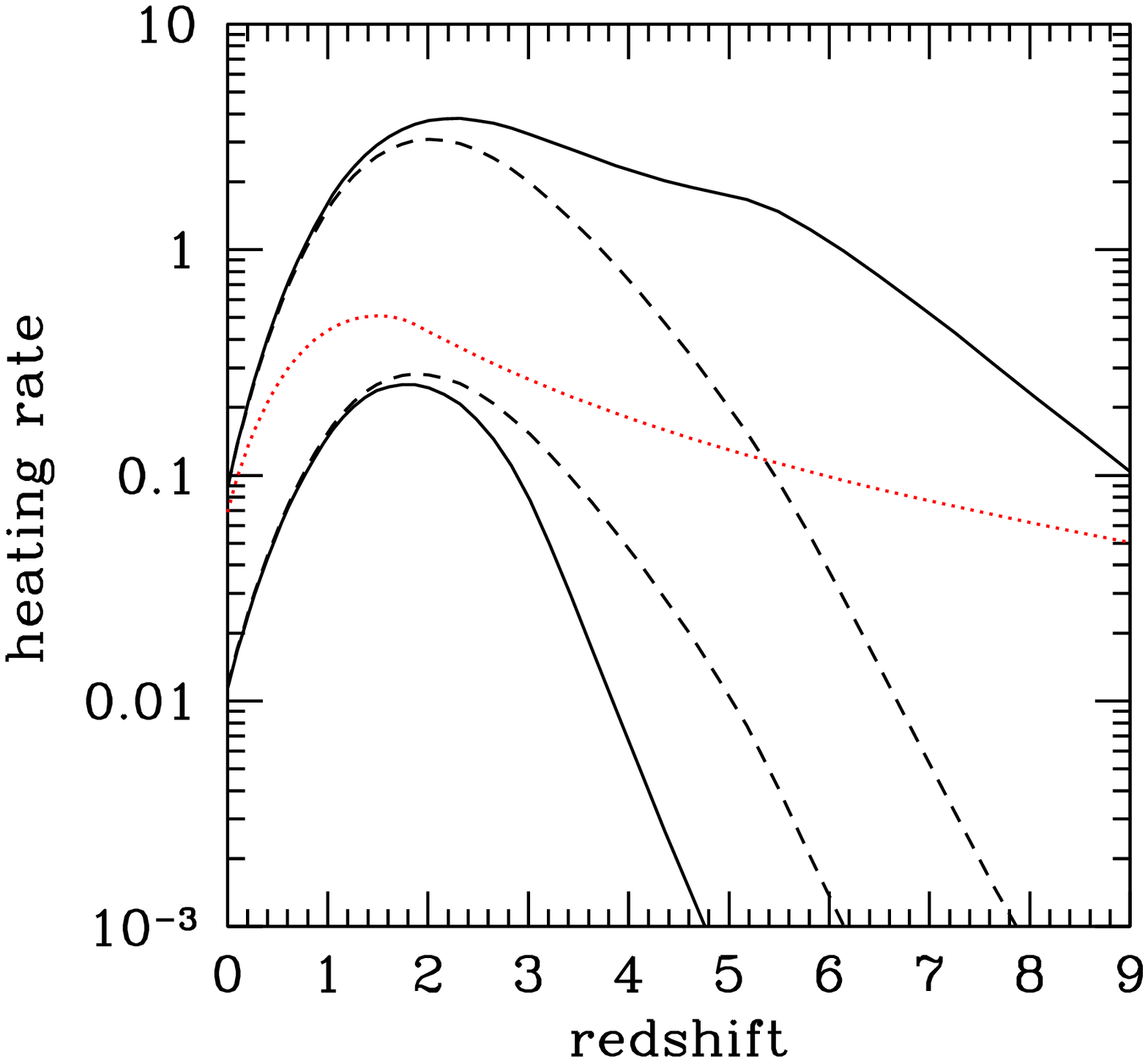}
\vspace{-0.3cm}
\caption{\footnotesize {\it Left:} The hydrogen photoionization rate, $\Gamma_{\rm HI}$, from $z=1$ to $z=7$. {\it Solid curve:} quasars $+$ galaxies
model. The dashed curves depict the individual contributions of the QSO population ({\it blue}) that dominates at low redshift and of the galaxy 
population ({\it red}) that reionize the IGM at early times. {\it Circles:} empirical measurements from the \Lya\ forest effective opacity by 
Bolton \& Haehnelt (2007). {\it Triangles:} same by Becker, Rauch, \& Sargent (2007) (their lognormal model).  
{\it Squares:} same by Faucher-Gigu\`ere \etal (2008a). {\it Pentagons:} same using the quasar proximity effect by 
Calverley \etal (2011).     
{\it Right:} The hydrogen photoheating rate {\it per ion}, ${\cal H}_{\rm HI}$ ({\it upper set of curves}), and the \HeII\ photoheating rate, ${\cal 
H}_{\rm HeII}$ ({\it lower set of curves}), from the present epoch to $z=9$. All photoheating rates are expressed in units of 
$10^{-12}$ eV s$^{-1}$. {\it Solid lines:} quasars $+$ galaxies. 
{\it Dashed lines:} quasar-only. The addition of a galaxy component boosts the \HI\ rate and decreases the \HeII\ rate.
The dotted line shows the Compton heating rate per electron in units of $10^{-18}$ eV s$^{-1}$.
}
\vspace{+0.3cm}
\label{fig8}
\end{figure} 

\section{Basic results} 

This section gives a quick overview of the main results generated by the upgraded CUBA radiative transfer code, 
using the formalism and parameters described above. CUBA solves the radiative transfer equation (\ref{Jnu}) by iteration, as its right-hand term 
implicitly contains $J$ in the recombination emissivity and in the effective helium opacity. 

\subsection{Photoionization and photoheating rates}

The total optically thin photoionization rate of hydrogen, $\Gamma_{\rm HI}$, is shown in Figure \ref{fig8} as a function of redshift (left panel).
For comparison, we have also plotted the individual contributions of the QSO population that dominates at low redshift and of the galaxy 
population that reionize the IGM at early times, together with the empirical measurements from the \Lya\ forest effective opacity by 
Bolton \& Haehnelt (2007), Becker, Rauch, \& Sargent (2007), and Faucher-Gigu\`ere \etal (2008a), and from the quasar proximity effect by 
Calverley \etal (2011). The fractional recombination contribution 
to $\Gamma_{\rm HI}$ increases from 9\% at $z=0$ to 18\% at $z=4$ to up to 37\% at $z\gta 7$: it does so because the mean free path 
of recombination photons decreases with lookback time and a smaller fraction of such photons gets redshifted below the ionization threshold 
before capture (Faucher-Gigu\`ere \etal 2009). While the total \HI\ photoionization rate provides a good match to the data, we note that there are large 
systematic uncertainties in the measurements as these depend on the assumed IGM temperature and gas density distribution.

The right panel of the same figure depicts the optically thin photoheating rates {\it per ion} (see eq. A7 with ${\cal J}_\nu=J_\nu$ for the definition) 
of hydrogen, ${\cal H}_{\rm HI}$, and \HeII, ${\cal H}_{\rm HeII}$, for the quasars $+$ galaxies and quasar-only models. The addition of a galaxy component 
boosts the \HI\ rate as it increases the emissivity of hydrogen-ionizing photons at fixed \HI\ opacity (the latter being determined by the observations). 
The opposite is true for \HeII\ photoheating (as well as \HeII\ photoionization), as galaxies do not contribute to the emissivity above 4 ryd: 
this increases the predicted \HeII\ opacity (again at fixed \HI\ opacity) and causes a large break in the background spectrum at 4 ryd and a smaller 
photoheating rate. While the Compton heating rate {\it per electron} is many orders of magnitude (about 7 dex at redshift 3) below  the \HI\ 
photoheating rate (note the different normalization of the heating rates plotted in Fig. \ref{fig8}), it is a non-negligible source of heating 
for very underdense, highly ionized regions: the Compton heating rate for intergalactic gas at overdensity 0.1, temperature $T=10^4$ K, 
and redshifts $z=(1,2,3)$ is $(53,16,4)$\% of the total photoheating rate. Table 3 tabulates the optically thin photoionization and photoheating 
rates of hydrogen and helium predicted by our ``quasars $+$ galaxies" model for use, e.g., in cosmological hydrodynamics simulations of the \Lya\ forest.  

\begin{table}[h]
\centering
\caption[]{The cosmic background photoionization and photoheating rates.}
\label{photoion}
\begin{tabular}{ccccccc}
\hline\hline\noalign{\smallskip}
$z$ & $\Gamma_\nHI$ & ${\cal H}_\nHI$ & $\Gamma_\nHeI$ & ${\cal H}_\nHeI$ & $\Gamma_\nHeII$ & ${\cal H}_\nHeII$\\
& (s$^{-1})$ & (eV s$^{-1}$) & (s$^{-1})$ & (eV s$^{-1}$) & (s$^{-1})$ & (eV s$^{-1}$)\\
\noalign{\smallskip}\hline\noalign{\smallskip}
     0.00  &   0.228E-13  &   0.889E-13  &   0.124E-13  &   0.112E-12  &   0.555E-15  &   0.114E-13  \\
     0.05  &   0.284E-13  &   0.111E-12  &   0.157E-13  &   0.140E-12  &   0.676E-15  &   0.138E-13  \\
     0.10  &   0.354E-13  &   0.139E-12  &   0.196E-13  &   0.174E-12  &   0.823E-15  &   0.168E-13  \\
     0.16  &   0.440E-13  &   0.173E-12  &   0.246E-13  &   0.216E-12  &   0.100E-14  &   0.203E-13  \\
     0.21  &   0.546E-13  &   0.215E-12  &   0.307E-13  &   0.267E-12  &   0.122E-14  &   0.245E-13  \\
     0.27  &   0.674E-13  &   0.266E-12  &   0.383E-13  &   0.331E-12  &   0.148E-14  &   0.296E-13  \\
     0.33  &   0.831E-13  &   0.329E-12  &   0.475E-13  &   0.408E-12  &   0.180E-14  &   0.357E-13  \\
     0.40  &   0.102E-12  &   0.405E-12  &   0.587E-13  &   0.502E-12  &   0.218E-14  &   0.429E-13  \\
     0.47  &   0.125E-12  &   0.496E-12  &   0.722E-13  &   0.615E-12  &   0.263E-14  &   0.514E-13  \\
     0.54  &   0.152E-12  &   0.605E-12  &   0.884E-13  &   0.751E-12  &   0.317E-14  &   0.615E-13  \\
     0.62  &   0.185E-12  &   0.734E-12  &   0.108E-12  &   0.911E-12  &   0.380E-14  &   0.732E-13  \\
     0.69  &   0.223E-12  &   0.885E-12  &   0.130E-12  &   0.110E-11  &   0.454E-14  &   0.867E-13  \\
     0.78  &   0.267E-12  &   0.106E-11  &   0.157E-12  &   0.132E-11  &   0.538E-14  &   0.102E-12  \\
     0.87  &   0.318E-12  &   0.126E-11  &   0.187E-12  &   0.157E-11  &   0.633E-14  &   0.119E-12  \\
     0.96  &   0.376E-12  &   0.149E-11  &   0.222E-12  &   0.186E-11  &   0.738E-14  &   0.139E-12  \\
     1.05  &   0.440E-12  &   0.175E-11  &   0.261E-12  &   0.217E-11  &   0.852E-14  &   0.159E-12  \\
     1.15  &   0.510E-12  &   0.203E-11  &   0.302E-12  &   0.251E-11  &   0.970E-14  &   0.181E-12  \\
     1.26  &   0.585E-12  &   0.232E-11  &   0.346E-12  &   0.287E-11  &   0.109E-13  &   0.202E-12  \\
     1.37  &   0.660E-12  &   0.262E-11  &   0.391E-12  &   0.323E-11  &   0.119E-13  &   0.221E-12  \\
     1.49  &   0.732E-12  &   0.290E-11  &   0.434E-12  &   0.357E-11  &   0.127E-13  &   0.237E-12  \\
     1.61  &   0.799E-12  &   0.317E-11  &   0.474E-12  &   0.387E-11  &   0.132E-13  &   0.247E-12  \\
     1.74  &   0.859E-12  &   0.341E-11  &   0.509E-12  &   0.413E-11  &   0.134E-13  &   0.253E-12  \\
     1.87  &   0.909E-12  &   0.360E-11  &   0.538E-12  &   0.432E-11  &   0.133E-13  &   0.252E-12  \\
     2.01  &   0.944E-12  &   0.374E-11  &   0.557E-12  &   0.444E-11  &   0.128E-13  &   0.244E-12  \\
     2.16  &   0.963E-12  &   0.381E-11  &   0.567E-12  &   0.446E-11  &   0.119E-13  &   0.229E-12  \\
     2.32  &   0.965E-12  &   0.382E-11  &   0.566E-12  &   0.438E-11  &   0.106E-13  &   0.207E-12  \\
     2.48  &   0.950E-12  &   0.375E-11  &   0.555E-12  &   0.422E-11  &   0.904E-14  &   0.178E-12  \\
     2.65  &   0.919E-12  &   0.363E-11  &   0.535E-12  &   0.398E-11  &   0.722E-14  &   0.145E-12  \\
     2.83  &   0.875E-12  &   0.346E-11  &   0.508E-12  &   0.368E-11  &   0.530E-14  &   0.111E-12  \\
     3.02  &   0.822E-12  &   0.325E-11  &   0.476E-12  &   0.336E-11  &   0.351E-14  &   0.775E-13  \\
     3.21  &   0.765E-12  &   0.302E-11  &   0.441E-12  &   0.304E-11  &   0.208E-14  &   0.497E-13  \\
     3.42  &   0.705E-12  &   0.279E-11  &   0.406E-12  &   0.274E-11  &   0.114E-14  &   0.296E-13  \\
     3.64  &   0.647E-12  &   0.257E-11  &   0.372E-12  &   0.249E-11  &   0.591E-15  &   0.168E-13  \\
     3.87  &   0.594E-12  &   0.236E-11  &   0.341E-12  &   0.227E-11  &   0.302E-15  &   0.925E-14  \\
     4.11  &   0.546E-12  &   0.218E-11  &   0.314E-12  &   0.209E-11  &   0.152E-15  &   0.501E-14  \\
     4.36  &   0.504E-12  &   0.202E-11  &   0.291E-12  &   0.194E-11  &   0.760E-16  &   0.267E-14  \\
     4.62  &   0.469E-12  &   0.189E-11  &   0.271E-12  &   0.181E-11  &   0.375E-16  &   0.141E-14  \\
     4.89  &   0.441E-12  &   0.178E-11  &   0.253E-12  &   0.170E-11  &   0.182E-16  &   0.727E-15  \\
     5.18  &   0.412E-12  &   0.167E-11  &   0.237E-12  &   0.160E-11  &   0.857E-17  &   0.365E-15  \\
     5.49  &   0.360E-12  &   0.148E-11  &   0.214E-12  &   0.146E-11  &   0.323E-17  &   0.156E-15  \\
     5.81  &   0.293E-12  &   0.123E-11  &   0.184E-12  &   0.130E-11  &   0.117E-17  &   0.624E-16  \\
     6.14  &   0.230E-12  &   0.989E-12  &   0.154E-12  &   0.112E-11  &   0.442E-18  &   0.269E-16  \\
     6.49  &   0.175E-12  &   0.771E-12  &   0.125E-12  &   0.952E-12  &   0.173E-18  &   0.128E-16  \\
     6.86  &   0.129E-12  &   0.583E-12  &   0.992E-13  &   0.783E-12  &   0.701E-19  &   0.674E-17  \\
     7.25  &   0.928E-13  &   0.430E-12  &   0.761E-13  &   0.625E-12  &   0.292E-19  &   0.388E-17  \\
     7.65  &   0.655E-13  &   0.310E-12  &   0.568E-13  &   0.483E-12  &   0.125E-19  &   0.240E-17  \\
     8.07  &   0.456E-13  &   0.219E-12  &   0.414E-13  &   0.363E-12  &   0.567E-20  &   0.155E-17  \\
     8.52  &   0.312E-13  &   0.153E-12  &   0.296E-13  &   0.266E-12  &   0.274E-20  &   0.103E-17  \\
     8.99  &   0.212E-13  &   0.105E-12  &   0.207E-13  &   0.191E-12  &   0.144E-20  &   0.698E-18  \\
     9.48  &   0.143E-13  &   0.713E-13  &   0.144E-13  &   0.134E-12  &   0.819E-21  &   0.476E-18  \\
     9.99  &   0.959E-14  &   0.481E-13  &   0.982E-14  &   0.927E-13  &   0.499E-21  &   0.326E-18  \\
    10.50  &   0.640E-14  &   0.323E-13  &   0.667E-14  &   0.636E-13  &   0.325E-21  &   0.224E-18  \\
    11.10  &   0.427E-14  &   0.217E-13  &   0.453E-14  &   0.435E-13  &   0.212E-21  &   0.153E-18  \\
    11.70  &   0.292E-14  &   0.151E-13  &   0.324E-14  &   0.314E-13  &   0.143E-21  &   0.106E-18  \\
    12.30  &   0.173E-14  &   0.915E-14  &   0.202E-14  &   0.198E-13  &   0.984E-22  &   0.752E-19  \\
    13.00  &   0.102E-14  &   0.546E-14  &   0.123E-14  &   0.122E-13  &   0.681E-22  &   0.531E-19  \\
    13.70  &   0.592E-15  &   0.323E-14  &   0.746E-15  &   0.749E-14  &   0.473E-22  &   0.373E-19  \\
    14.40  &   0.341E-15  &   0.189E-14  &   0.446E-15  &   0.455E-14  &   0.330E-22  &   0.257E-19  \\
    15.10  &   0.194E-15  &   0.110E-14  &   0.262E-15  &   0.270E-14  &   0.192E-22  &   0.154E-19 \\
\noalign{\smallskip}\hline\noalign{\smallskip} 
\end{tabular}     
\end{table}       

\subsection{Background spectral energy distribution}

\begin{figure*}[thb]
\centering
\includegraphics*[width=0.9\textwidth]{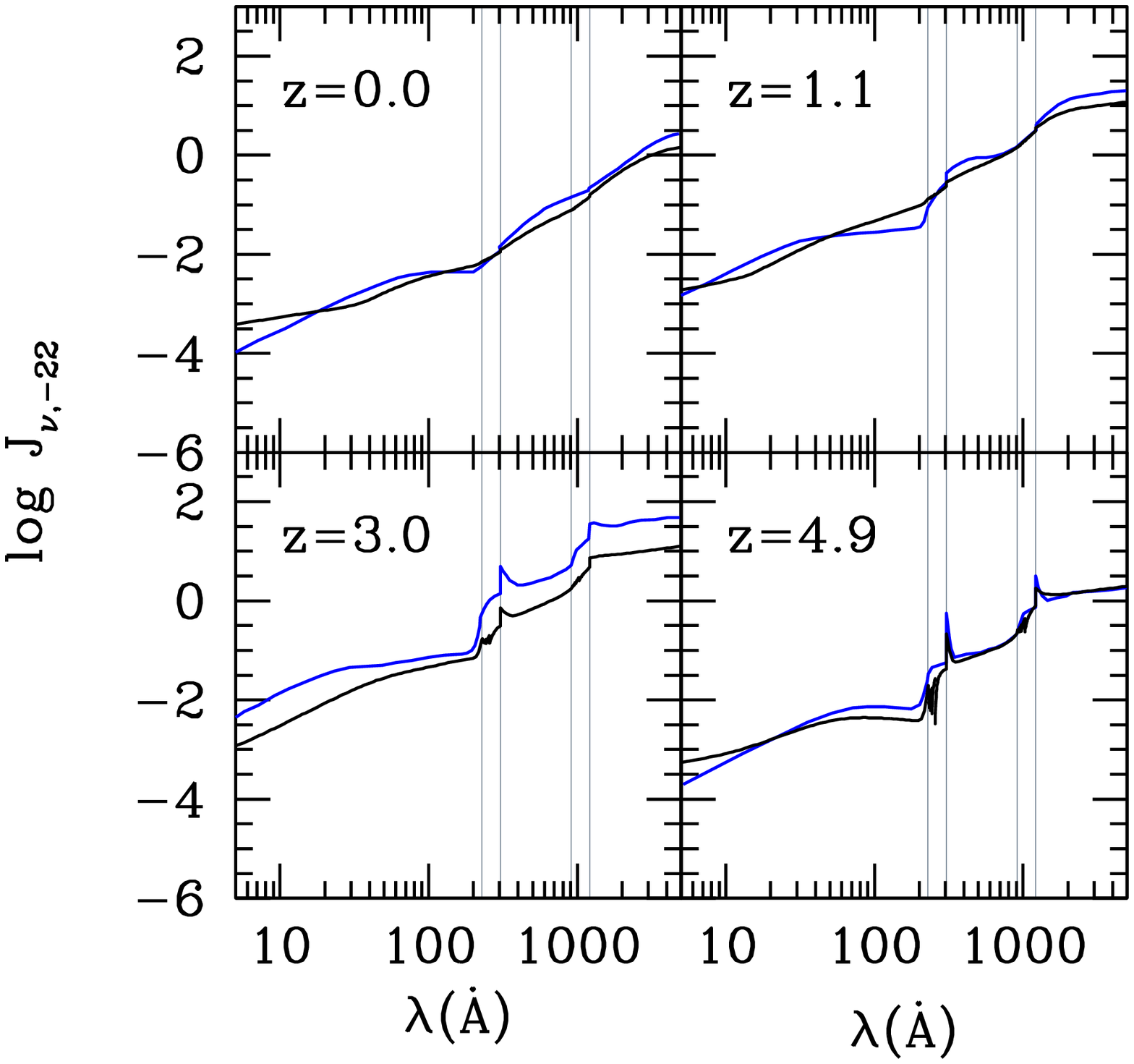}
\vspace{-0.3cm}
\caption{\footnotesize The broadband spectrum of a ``quasar-only" cosmic background between 5 \AA\ and 5,000 \AA\ at epochs $z=0, 1, 3, $and 5. The new 
models ({\it black curves}) are compared with the old results of Paper II ({\it blue curves}). The intensity $J_\nu$ is expressed in 
units of $10^{-22}\,\uvunits$. The vertical thin lines indicate the positions of the \HI\ and \HeII\ \Lya\ and Lyman limit.   
}
\vspace{+0.3cm}
\label{fig9}
\end{figure*}

Figure \ref{fig9} shows the spectrum of the radiation background as a function of redshift for a ``quasar-only" model, together with the old results from Paper 
II. The new spectra are characterized by a lower UV flux (by as much as a factor of 3 at 1 ryd and $z=3$), smaller spectral breaks at 1 and 4 ryd 
because of the reduced \HI\ and \HeII\ LyC absorption, a sawtooth modulation by the Lyman series of \HI\ and \HeII\ that becomes more and more 
pronounced with increasing redshift, and a flatter soft X-ray spectrum. 

\begin{figure*}[thb]
\centering
\includegraphics*[width=0.9\textwidth]{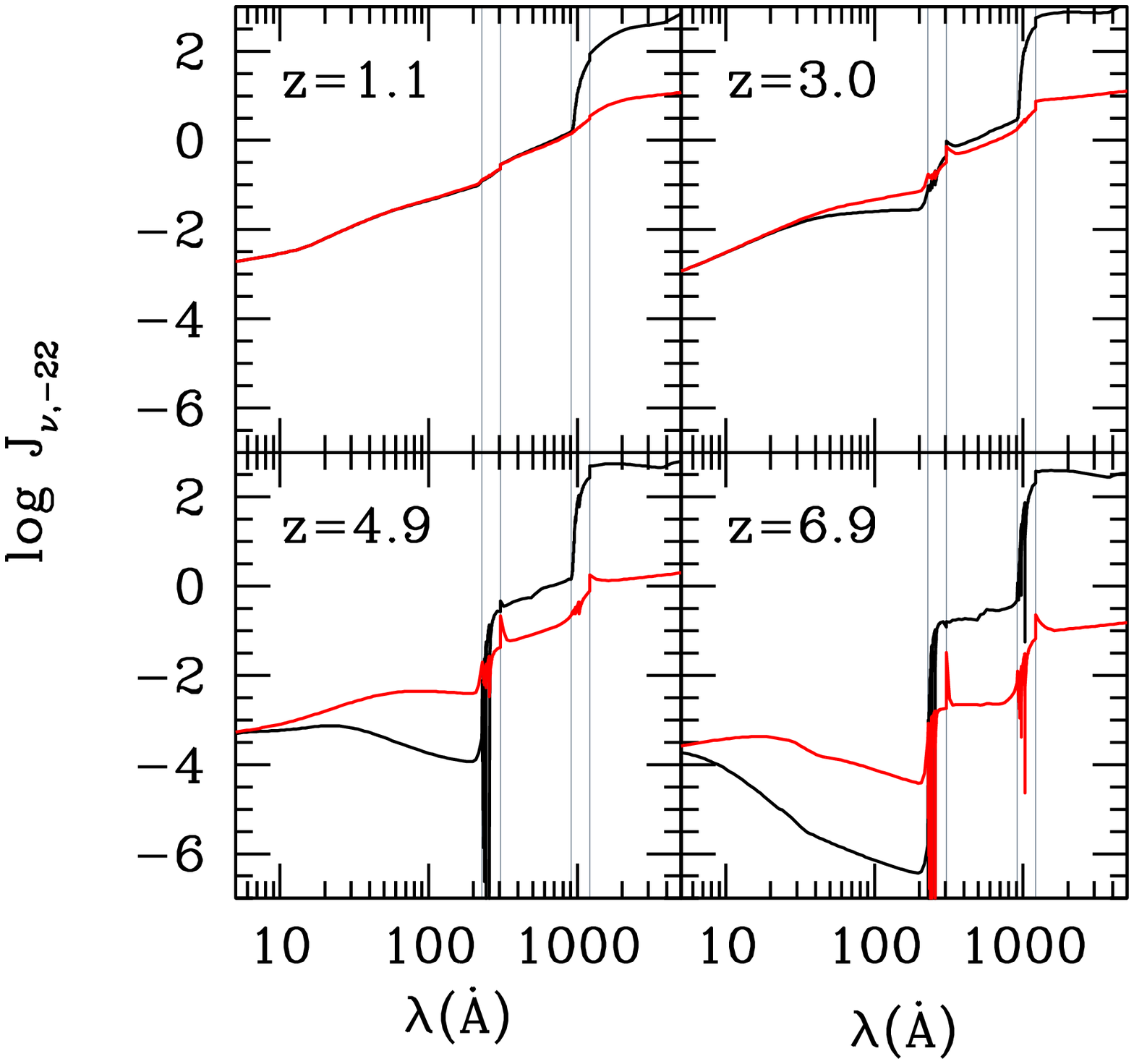}
\vspace{-0.3cm}
\caption{\footnotesize The broadband spectrum of a ``quasars $+$ galaxies" cosmic background at redshifts $z=1.1, 3, 4.9,$ and 5.9 ({\it black curves}). 
The new ``quasar-only" model of Fig. \ref{fig9} is plotted for comparison ({\it red curves}). The intensity $J_\nu$ is expressed in 
units of $10^{-22}\,\uvunits$. The vertical thin lines indicate the positions of the \HI\ and \HeII\ \Lya\ and Lyman limit.   
}
\vspace{+0.3cm}
\label{fig10}
\end{figure*} 

\begin{figure*}[thb]
\centering
\includegraphics*[width=0.9\textwidth]{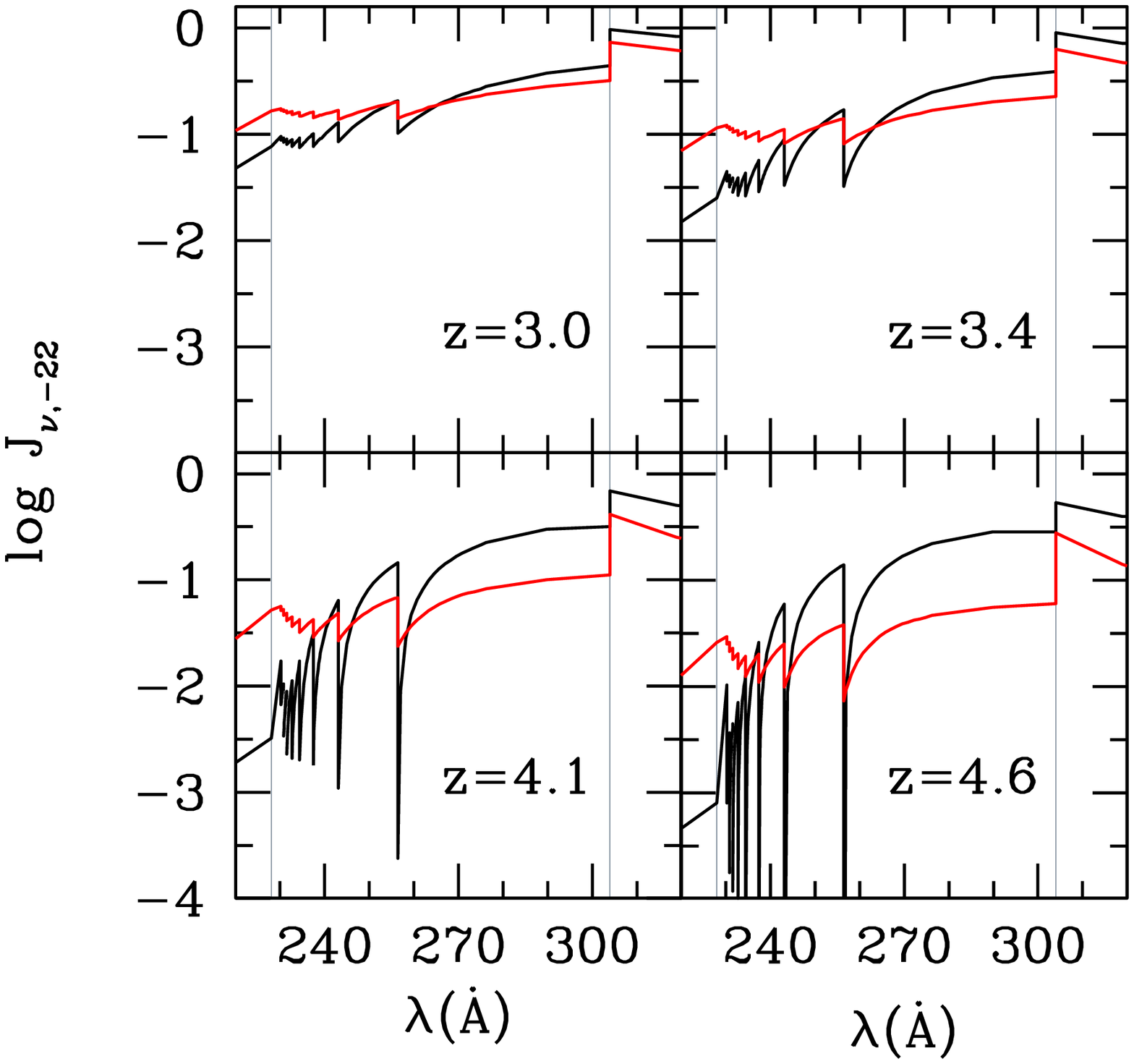}
\vspace{-0.3cm}
\caption{\footnotesize A zoom-in of the ``quasars $+$ galaxies" ({\it black curves}) and ``quasar-only" ({\it red curves}) 
cosmic background spectrum at redshifts $z=3,3.4,4.1,$ and 4.6 showing the sawtooth modulation of the metagalactic 
flux between 220 and 320 \AA\ produced by resonant absorption in the Lyman series of intergalactic \HeII.
The vertical thin lines indicate the positions of the \HeII\ \Lya\ and Lyman limit.
}
\vspace{+0.3cm}
\label{fig11}
\end{figure*}

\begin{figure*}[thb]
\centering
\includegraphics*[width=0.9\textwidth]{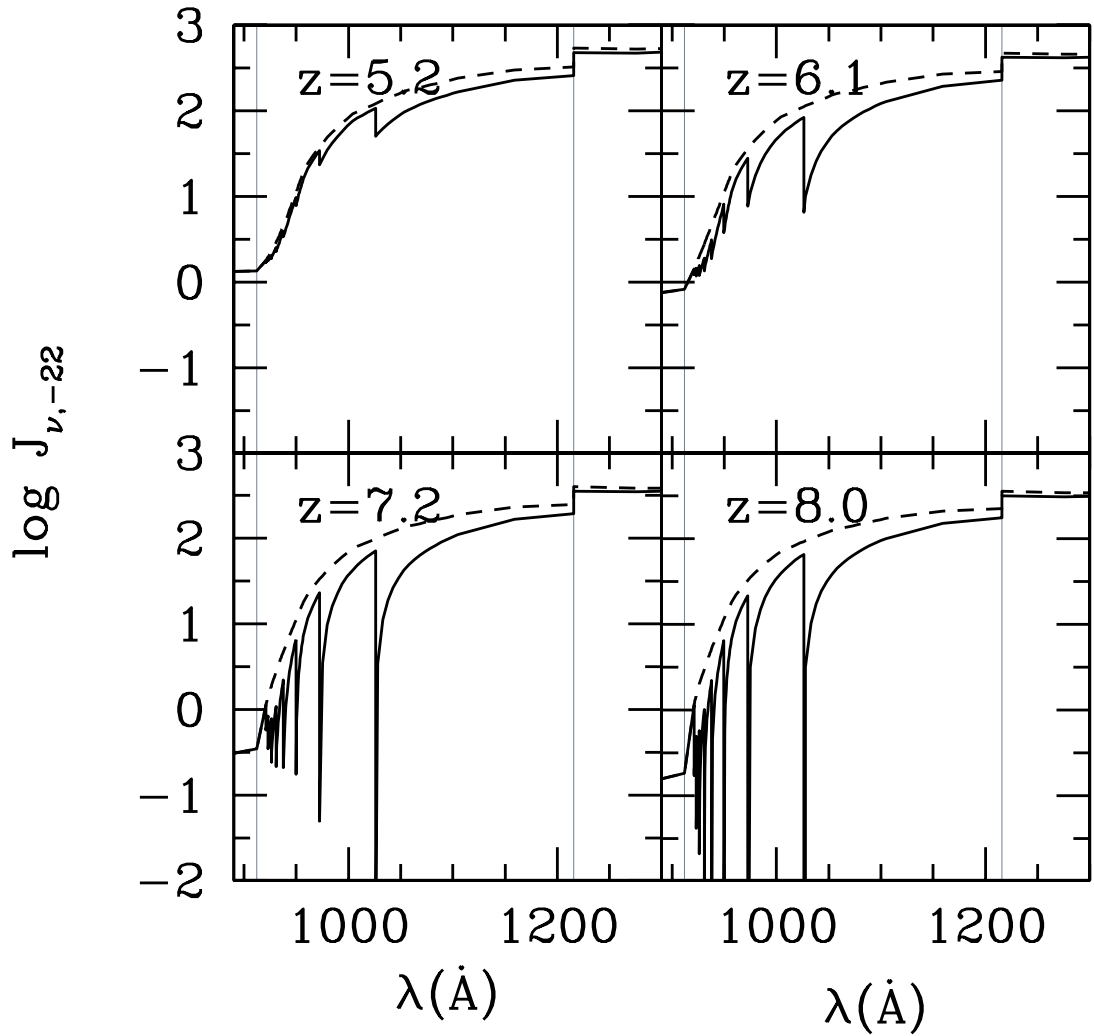}
\vspace{-0.3cm}
\caption{\footnotesize A zoom-in of the ``quasars $+$ galaxies" cosmic background spectrum at redshifts $z=5.2,6.1,7.2,$ and 8.0 showing the sawtooth modulation 
of the metagalactic flux between and 890 and 1300 \AA\ produced by resonant absorption in the Lyman series of intergalactic \HI.
The vertical thin lines indicate the positions of the \HI\ \Lya\ and Lyman limit. The dashed line shows the same spectrum 
without sawtooth for comparison.
}
\vspace{+0.3cm}
\label{fig12}
\end{figure*}

The addition of radiation from galaxies has little effect on the ionizing background at redshifts below 3, as shown in Figure \ref{fig10}, a
consequence of our adopted redshift-dependent escape fraction. At higher redshifts the impact is more dramatic: a large boost at 1 ryd is 
associated with a much sharper \HeII\ sawtooth and \HeII\ absorption edge. As noted above regarding the photoheating rates, 
this arises because galaxy spectra are truncated at 4 ryd, and the large increase in the H-ionizing emissivity from the early galaxy population 
is not accompanied by a similar increase at the \HeII\ edge. The net effect is a larger \HeII\ opacity at fixed \HI\ opacity. 
At $z\gta 5$, the \HeI\ opacity of the IGM also starts building up (it is negligible at lower redshifts), and a small \HeI\ absorption edge  
can be discerned in the spectrum of the background at 24.6 eV. At $z\gta 3$, the sawtooth modulation produced by resonant absorption in the Lyman 
series of intergalactic \HeII\ (see Fig. \ref{fig11}) is clearly a sensitive probe of the nature of the sources that keep the IGM ionized, and may 
be a crucial ingredients in the modelling of the abundances of metal absorption systems (Madau \& Haardt 2009). The analogous sawtooth modulation 
produced by the \HI\ Lyman series becomes significant above redshift 6 (see Fig. \ref{fig12}), and may affect the photodissociation of molecular hydrogen 
during cosmological reionization (Haiman, Rees, \& Loeb 1997).

\begin{figure}[thb]
\centering
\includegraphics*[width=0.7\textwidth]{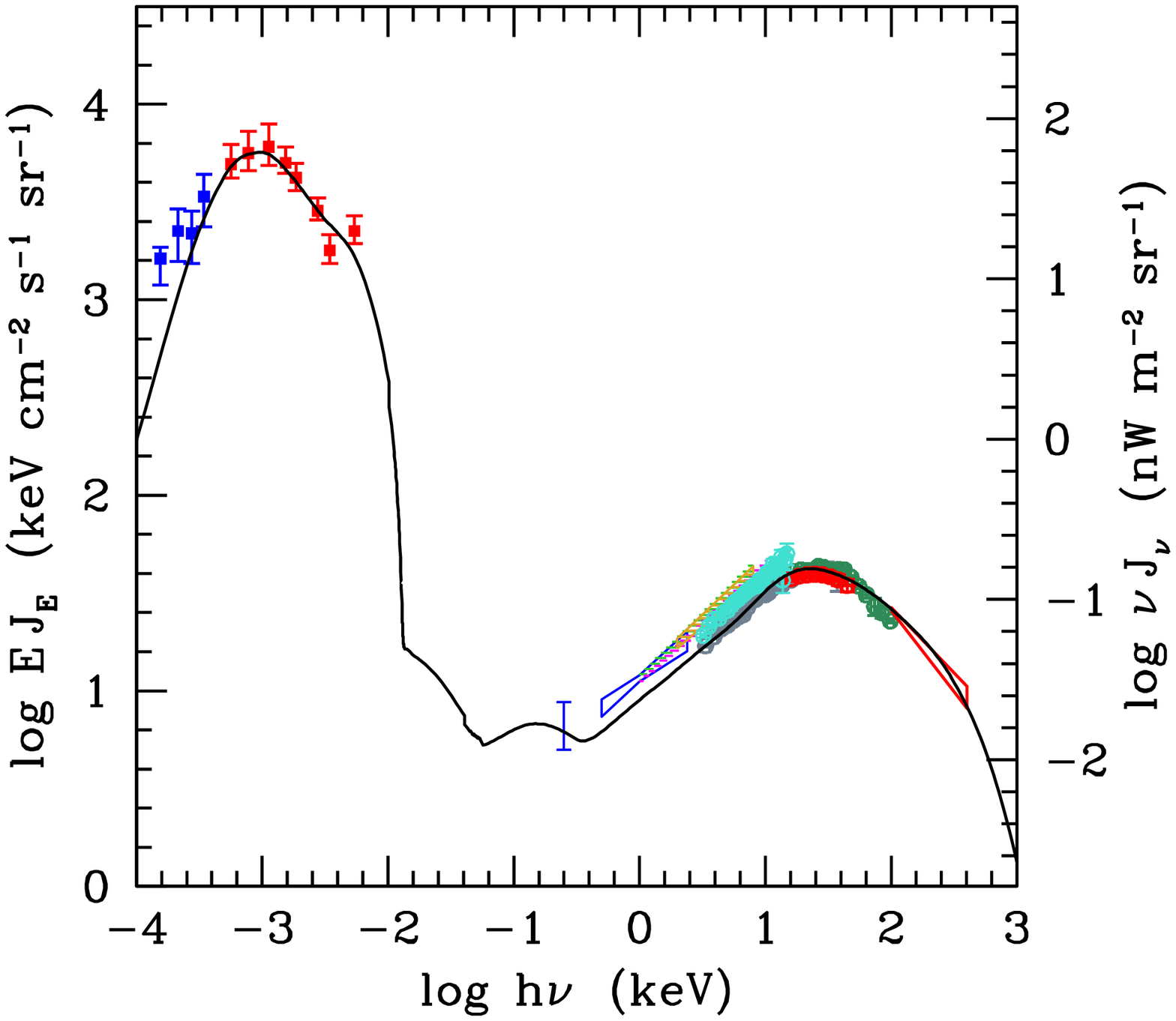}
\vspace{+0.cm}
\caption{\footnotesize 
The predicted broadband extragalactic background light, $\nu J_\nu$, from quasars and galaxies at $z=0$, compared with empirical determinations
at different wavelengths.  {\it Red points:} the optical-near IR EBL from {\it HST} and ground-based galaxy counts (Madau \& Pozzetti 2000). {\it Blue 
points}: the mid-IR EBL from IRAC-{\it Spitzer} galaxy counts (Fazio \etal 2004). The X-ray data points are explained in details in the caption of Fig. 
\ref{fig6}.} 
\vspace{+0.5cm}
\label{fig13}
\end{figure}

Figure \ref{fig13} compares the broadband spectrum of the total extragalactic background light (EBL) from quasars and galaxies, predicted by CUBA at $z=0$, 
with current EBL observations from the mid-IR to the $\gamma$-rays. 

\subsection{A ``minimal reionization model"}

It is interesting at this stage to use the quasar and galaxy ionizing emissivities of \S~6, 7, and 8 
and track the evolution of the volume filling factors of ionized hydrogen and doubly ionized helium regions in the 
universe as a function of cosmic time. As shown in Paper III, the volume filling factor of \HII\ regions, $Q_\nHII$, 
is equal at any given instant $t$ to the integral over cosmic time of the number ionizing photons emitted per hydrogen 
atom by all radiation sources present at earlier epochs,
\begin{equation}
{\cal I}=\int_0^t dt'\,{\dot n_{\rm ion}(t')\over \langle n_\nH(t')\rangle}
\end{equation}
minus the number of radiative recombinations per ionized hydrogen atom, 
\begin{equation}
{\cal R}=\int_0^t {dt'\over \langle t_{\rm rec}(t') \rangle} Q_\nHII(t').
\end{equation}
Here 
\begin{equation}
\dot n_{\rm ion}(t)=\int_{\nu_L}^\infty \langle f_{\rm esc}\rangle {d\nu\over h\nu} \epsilon_\nu(t)
\end{equation}
with $\langle f_{\rm esc}\rangle=1$ in the case of quasars, $\langle n_\nH\rangle=1.9\times 10^{-7}(1+z)^3$ cm$^{-3}$ is 
the mean hydrogen density of the expanding IGM, and 
$\langle t_{\rm rec}\rangle$ is the volume-averaged hydrogen recombination timescale,
\begin{equation}
\langle t_{\rm rec}\rangle=[\chi \langle n_\nH\rangle \alpha_B\,C]^{-1},
\label{eq:trec}
\end{equation}
where $\alpha_B$ is the recombination coefficient to the excited states of hydrogen, $\chi=1.08$ accounts for the presence of 
photoelectrons from singly ionized helium, and $C_{\rm IGM}\equiv \langle n_\nHII^2\rangle/\langle n_\nHII\rangle^2$ is the clumping factor of ionized 
hydrogen. Differentiation yields the \HI\ ``reionization equation" of Paper III, 
\begin{equation}
{dQ_\nHII\over dt}={\dot n_{\rm ion}\over \langle n_\nH \rangle} -{Q_\nHII\over
\langle t_{\rm rec}\rangle}, 
\label{eq:qdot}
\end{equation}
and its equivalent for expanding \HeIII\ regions,
\begin{equation}
{dQ_\nHeIII\over dt}={\dot n_{\rm ion,4}\over \langle n_\nHe \rangle} -{Q_\nHeIII\over
\langle t_{\rm rec,He}\rangle}, 
\label{eq:qHedot}
\end{equation}
where $\dot n_{\rm ion,4}$ now includes only photons above 4 ryd (which are mostly absorbed by \HeII), 
and the recombination timescale of doubly ionized helium, $\langle t_{\rm rec,He}\rangle$, is the about 6 times shorter than the hydrogen recombination 
timescale if \HII\ and \HeIII\ have similar clumping 
factors. We will not attempt here to model the reionization of \HeI, as this occurs nearly simultaneously to and cannot be readily decoupled from that of \HI.
The reionization equation equation: 
1) describes the transition from a neutral universe to a fully ionized one in a statistical way, independently, for a given emissivity, 
of the emission histories of individual radiation sources;  
2) assumes that the mean free path of ionizing photons is much smaller than the horizon, i.e., that they are absorbed before being redshifted 
below the ionization edge; and
3) includes in the source term only those photons above the Lyman limit that escape into the IGM ($\langle f_{\rm esc}\rangle=1$ in the
case of quasars). Photons that are absorbed in loco by dense interstellar gas do not enter in the source term, nor does
the interstellar absorbing material contribute to the recombination rate. The volume-weighted clumping factor reflects only the nonuniformity of the 
ionized low-density IGM, the repository of most of the baryons in the universe, and its use in the recombination timescale is justified 
when the size of the ionized regions is large compared to the scale of the clumping.

When $Q_\nHII\ll 1$ (the ``pre-overlap" stage), individual ionization fronts propagate from star-forming early galaxies into the low-density IGM. The neutral 
phase shrinks as $Q_\nHII$ grows and \HII\ regions start to overlap. The radiation field remains highly inhomogeneous until the reionization process is 
completed at the ``overlap epoch", $Q_\nHII=1$, when all the low-density IGM becomes highly ionized. Pockets of neutral gas remain in collapsed systems 
during the entire ``post-overlap" stage (Gnedin 2000) and may manifest themselves as the SLLSs or DLA systems in quasar absorption spectra. We have 
integrated equation (\ref{eq:qdot}) assuming a gas temperature of $2\times 10^4$ K and a clumping factor for the intergalactic medium of
\begin{equation}
C_{\rm IGM}=1+43\,z^{-1.71}. 
\label{eq:Cigm}
\end{equation}
This is equal to the expression for $C_{100}$ (the clumping factor of gas below a threshold overdensity of 100) found at $z\ge 6$ in a suite of 
cosmological hydrodynamical simulations by Pawlik, Schaye, and van Scherpenzeel (2009). These authors found that photoionization heating by a uniform 
UV background greatly reduces clumping as it smoothes out small-scale density fluctuations, and that the clumping factor at $z=6$ is insensitive to 
the redshift at which the UV background is actually turned on (as long as reheating occurs at $\gta 9$). We use an overdensity of 100 to differentiate between 
dense gas belonging to virialized halos and the diffuse intergalactic gas, and assume that the collapsed mass fraction is small. 
We also extrapolate equation (\ref{eq:Cigm}) down to $z\gta 2$, and assume the same clumping factor for \HII\ and \HeIII.

\begin{figure*}[thb]
\centering
\includegraphics*[width=0.47\textwidth]{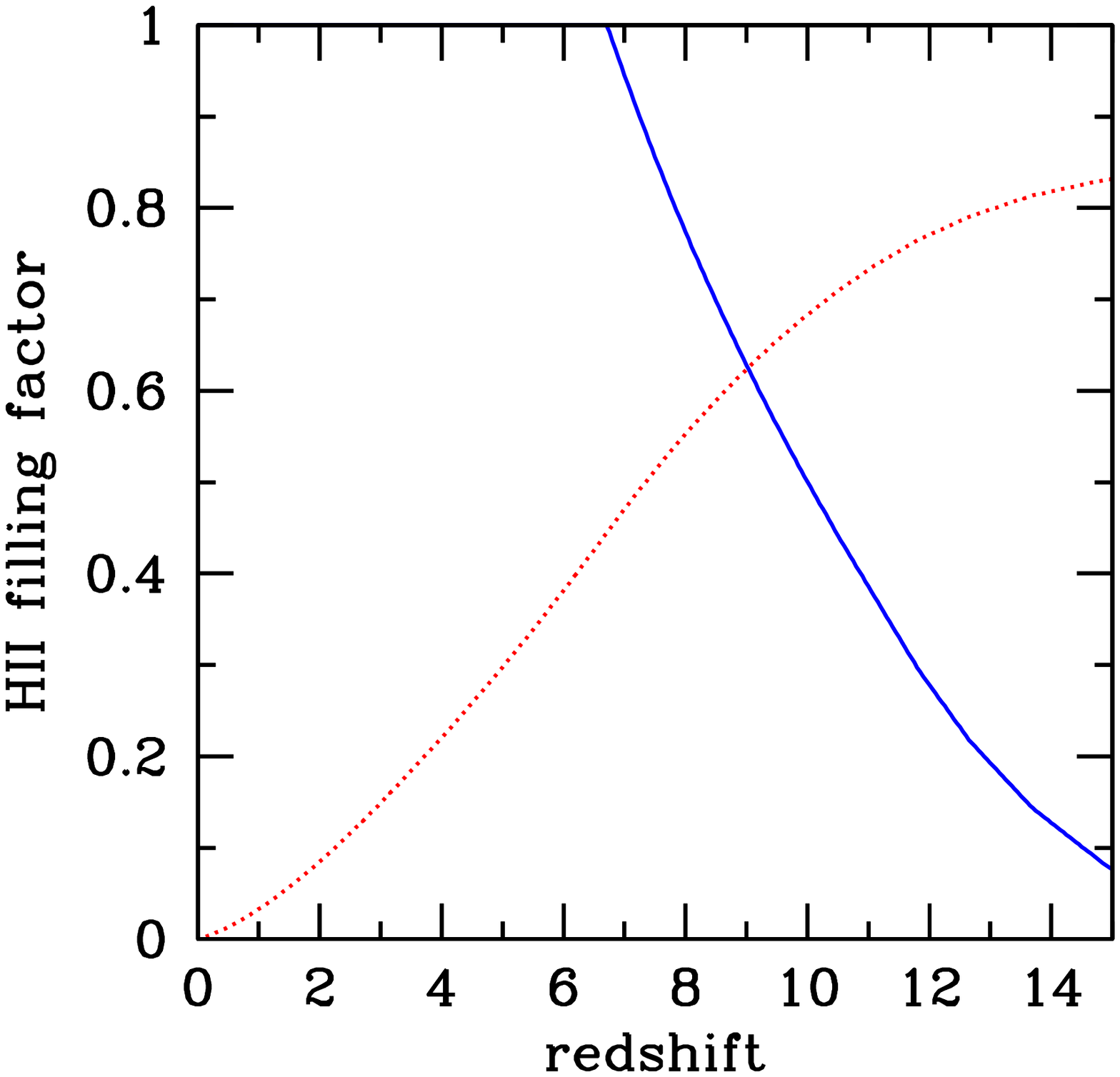}
\includegraphics*[width=0.47\textwidth]{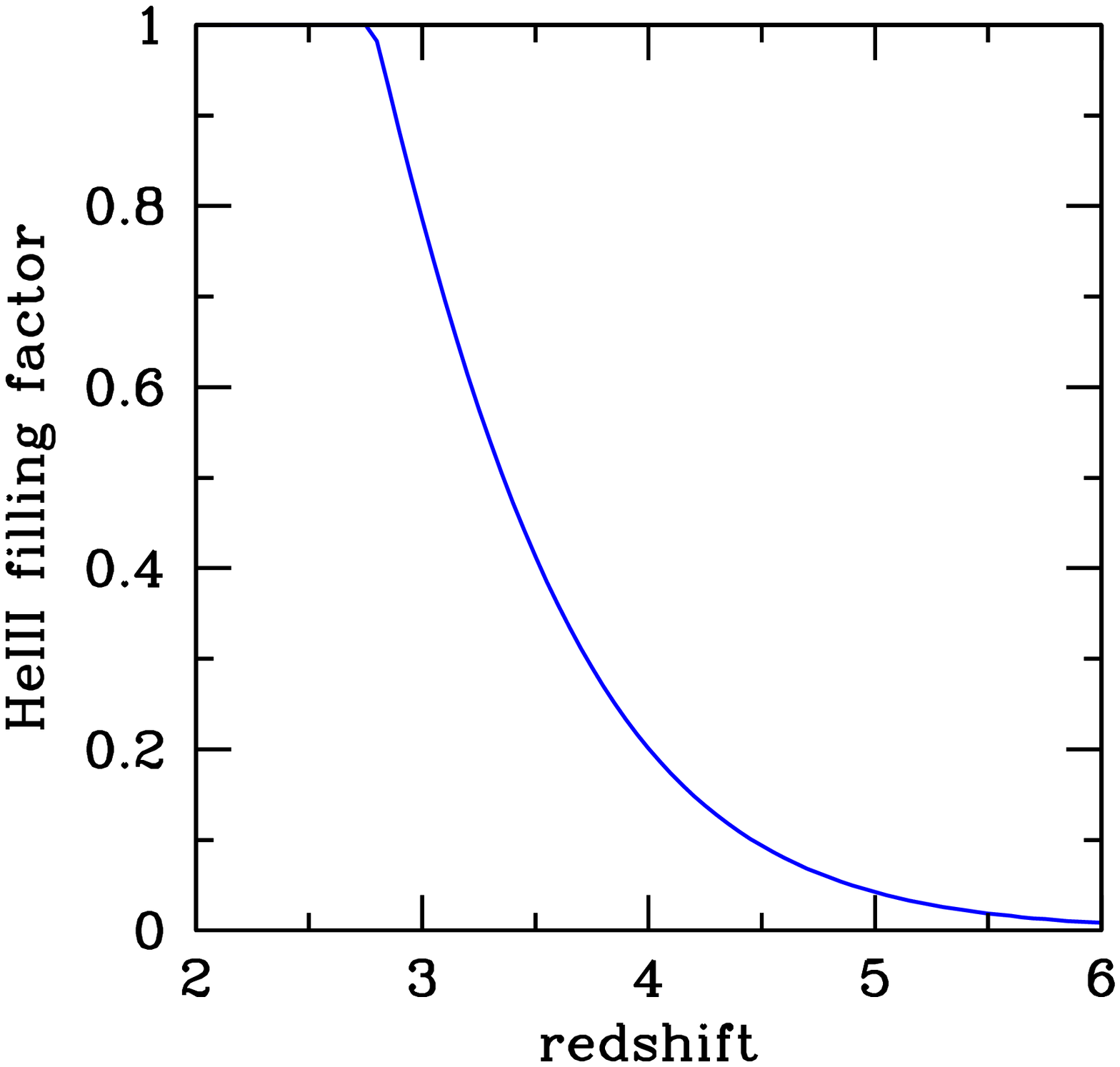}
\vspace{-0.3cm}
\caption{\footnotesize Evolution of the \HII\ ({\it left panel}) and \HeIII\ ({\it right panel}) filling factors as a function of redshft 
for our ``minimal reionization model" (see text for details). The dotted curve in the left panel depicts the cumulative electron scattering optical 
depth of the universe, in units of 10\%, as a function of redshift.
}
\vspace{+0.cm}
\label{fig14}
\end{figure*}

The results of this ``minimal reionization model" are shown in Figure \ref{fig14}. Cosmological \HII\ regions driven by star-forming galaxies
overlap at redshift 6.7, and the hydrogen in the universe is half-ionized (by volume) at redshift 10. \HeIII\ regions driven by 
quasars overlap much later, at redshift 2.8, and their filling factor is only 4\% at redshift 5. These overlap epochs are 
consistent with the SDSS spectra of $z\sim 6$ quasars (Fan \etal 2006b), with numerical simulations of \HI\ reionization (Gnedin \& Fan 2006), and
with observations of the \HeII\ \Lya\ forest at $z\lta 3$ (see, e.g., Worseck \etal 2011; Shull \etal 2010; Fechner \etal 2006; Heap \etal 2000; 
and references therein).      
A simple probe of the reionization history is the integrated optical depth to electron scattering $\tau_{\rm es}$, which depends on the 
path length through ionized gas along the line of sight to the CMB as 
\begin{equation}
\tau_{\rm es}(z)=\sigma_T c \int_0^{z} {dz'\over H(1+z')}[Q_\nHII \langle n_\nH\rangle +Q_\nHeII \langle n_\nHe\rangle +2Q_\nHeIII \langle n_\nHe \rangle]
\label{eq:taues}
\end{equation}
(Wyithe \& Loeb 2003),
where $\sigma_T$ is the Thomson cross section. The seven-year {\it WMAP} results imply $\tau_{\rm es}=0.088\pm 0.015$ (Jarosik \etal 2011). 
Our minimal reionization model assumes $Q_\nHeII=Q_\nHII$ and yields an electron scattering opacity to the epoch of reionization of $\tau_{\rm es}=0.084$, in 
good agreement with the observations. 

The outcome of our minimal reionization model is rather sensitive to the assumed escape fraction of hydrogen-ionizing radiation at early epochs; this
exceeds 50\% at $z\gta 9$ and reaches unity at $z=11.6$. Had we assumed a maximum $f_{\rm esc}$ of 50\% instead, the same model would  
yield $Q_\nHI=1$ at $z=6.2$ and $\tau_{\rm es}=0.06$. We also remark that, in the pre-overlap era, the background spectra shown in Figures \ref{fig9}, 
\ref{fig10}, \ref{fig11}, and \ref{fig12} have only a formal meaning, as they describe a space-averaged radiation field that is in reality highly inhomogeneous. 
Recent spectra taken by the Cosmic Origins Spectrograph on the {\it Hubble Space Telescope} exhibit patchy \HeII\ Gunn-Peterson absorption,
with a mean \HeII/\HI\ abundance ratio that is $47\pm 42$ at $2.4<z<2.73$, and $209\pm 281$ at $z>2.73$ (Shull \etal 2010). In the redshift interval 
$2.4<z<2.73$, our background spectrum yields a \HeII/\HI\ abundance ratio (in the optically thin limit) around 50--70, in good agreement with the 
observations. The predicted 
mean \HeII/\HI\ ratio increases rapidly towards high redshift, to $(280,494,887,1615)$ at $z=(3.42,3.64,3.87,4.1)$ as galaxies start dominating the ionizing 
emissivity and the spectrum of the UVB steepens. The evolution of the \HeII\ abundance and the fluctuating spectrum of the cosmic UVB in the pre-overlap 
era will be the subject of a subsequent paper.

\subsection{The UVB: uncertainties}

\begin{figure}[thb]
\centering
\includegraphics*[width=0.47\textwidth]{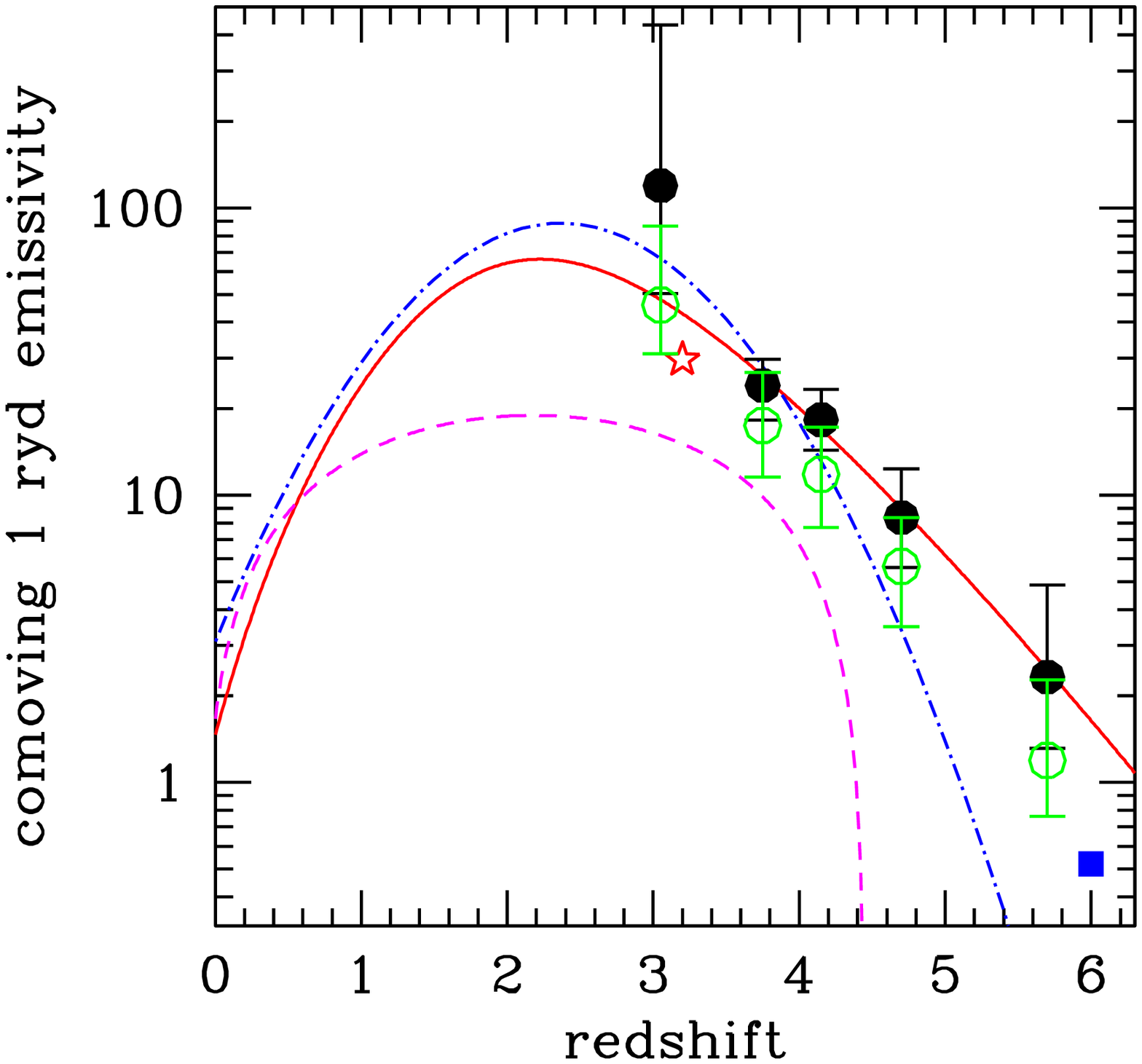}
\includegraphics*[width=0.47\textwidth]{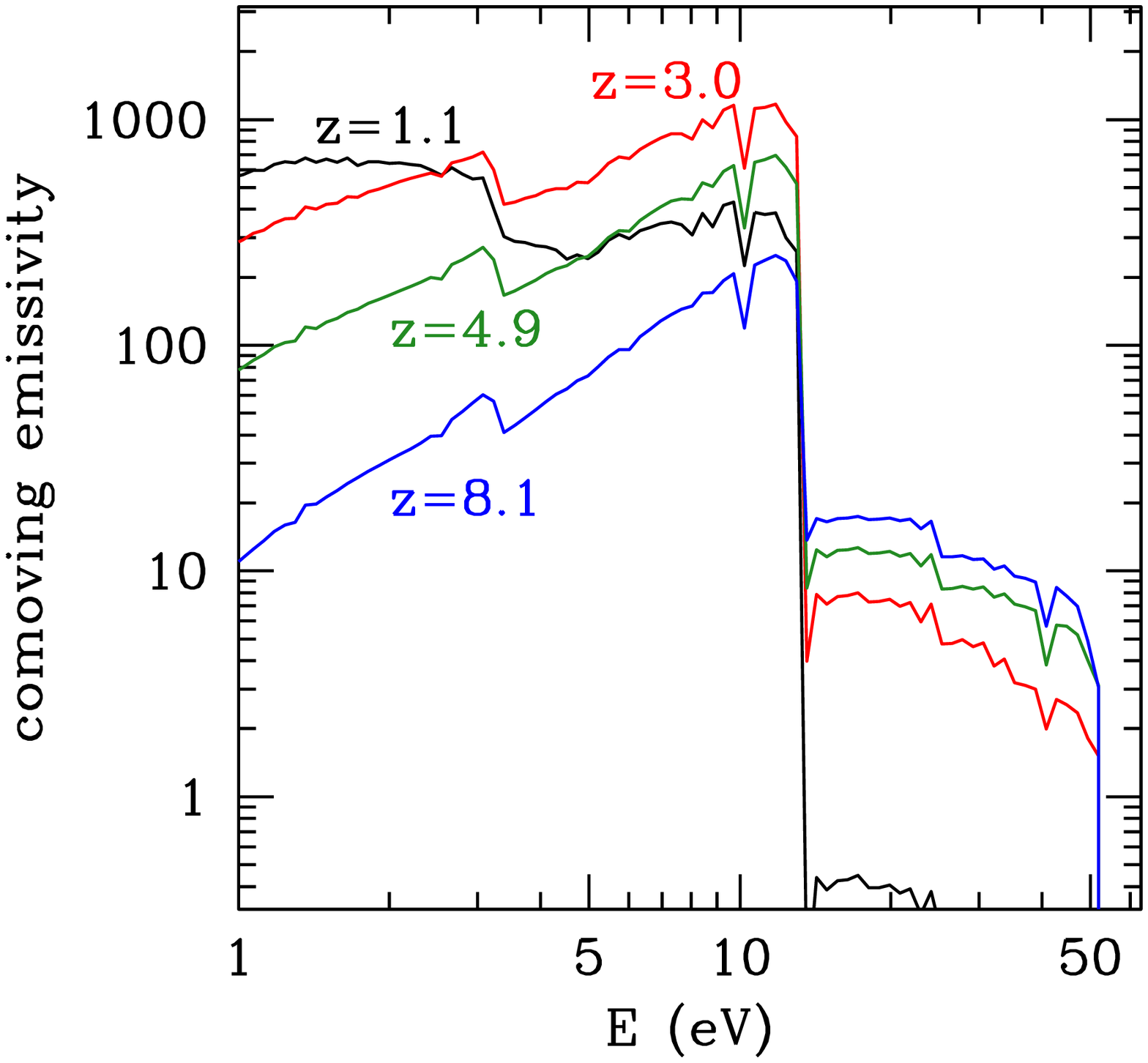}
\caption{\footnotesize {\it Left:} Comoving specific emissivity at 1 ryd (in units of $10^{23}\,\lumdens$)
measured from different quasar surveys. {\it Solid line:} the best-fit function in eq. (\ref{eqemiss}).
{\it Dashed line:} Cowie \etal (2009). {\it Dot-dashed line:} Bongiorno \etal (2007), using a 912 \AA\ to 4400 \AA\ flux ratio of 0.31.
{\it Filled circles:} Meiksin (2005), PLE model. {\it Empy circles:} Meiksin (2005), PDE model. {\it Empty star:} Siana \etal (2008).
{\it Filled square}: Willott \etal (2010) at $z=6$.
{\it Right:} Comoving galaxy emissivity per logarithmic bandwidth (in units of $10^{39}$ ergs s$^{-1}$ Mpc$^{-3}$) escaping into the IGM, 
as a function of photon energy, at four different redshifts. Note the large, time-evolving break at 13.6 eV.
}
\label{fig15}
\vspace{+0.5cm}
\end{figure}

The background spectra computed in the previous section are sensitive to a number of poorly determined input parameters.
In this section we briefly discuss just a few of the uncertainties inherent in our synthesis modelling of the UVB. The left panel of Figure 
\ref{fig15} 
shows the adopted comoving quasar emissivity at 1 Ryd (eq. \ref{eqemiss}), together with the determinations by Meiksin (2005), 
Cowie, Barger, \& Trouille (2009), Bongiorno \etal (2007), Willott \etal (2010), and Siana \etal (2008).
The poorly known faint-end slope of the quasar luminosity function at high redshift, incompleteness corrections,
as well as the uncertain spectral energy distribution (SED) in the UV, all contribute to the large apparent discrepancies between
different measurements.  

The escape fraction of ionizing radiation that leaks into the IGM and its unknown redshift evolution is the major source of uncertainty in the
determination of the galaxy contribution to the UVB. To better gauge the impact of this parameter on our synthesis model, we show 
in the right panel of Figure \ref{fig15} the spectrum of the comoving galaxy emissivity at four different epochs. The relatively large 
leakage of LyC photons assumed at early times is a crucial ingredient of our ``minimal reionization model", which yields 
an optical depth to Thomson scattering in agreement with {\it WMAP} results. A smaller escape fraction at high redshifts would lead to too-late 
reionization, while a significantly larger escape fraction at lower redshifts would produce a hydrogen photoionization rate that appears to 
be too high compared to the observations (see Fig. \ref{fig8}). 

\begin{figure}[thb]
\centering
\includegraphics*[width=0.47\textwidth]{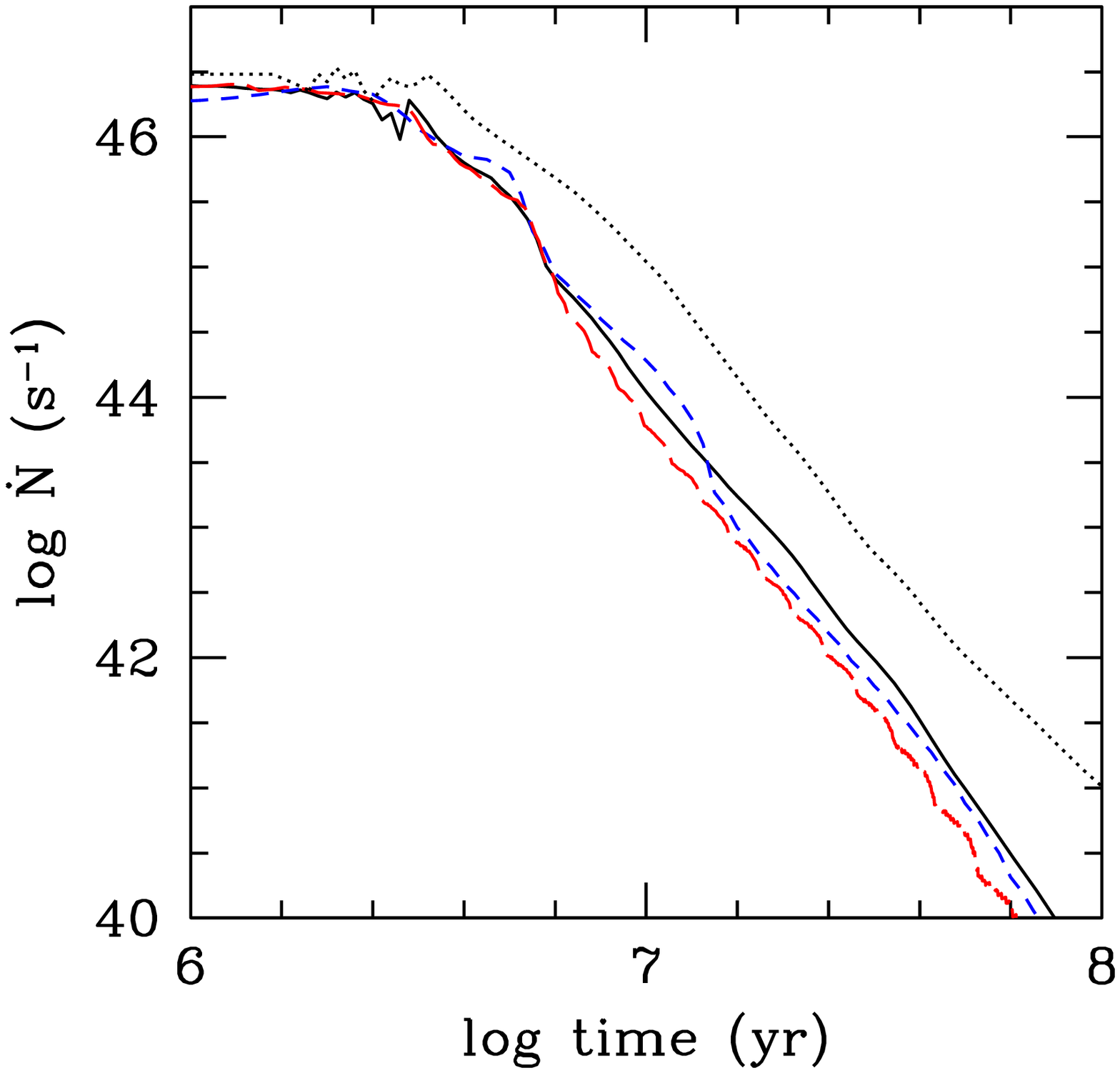}
\includegraphics*[width=0.47\textwidth]{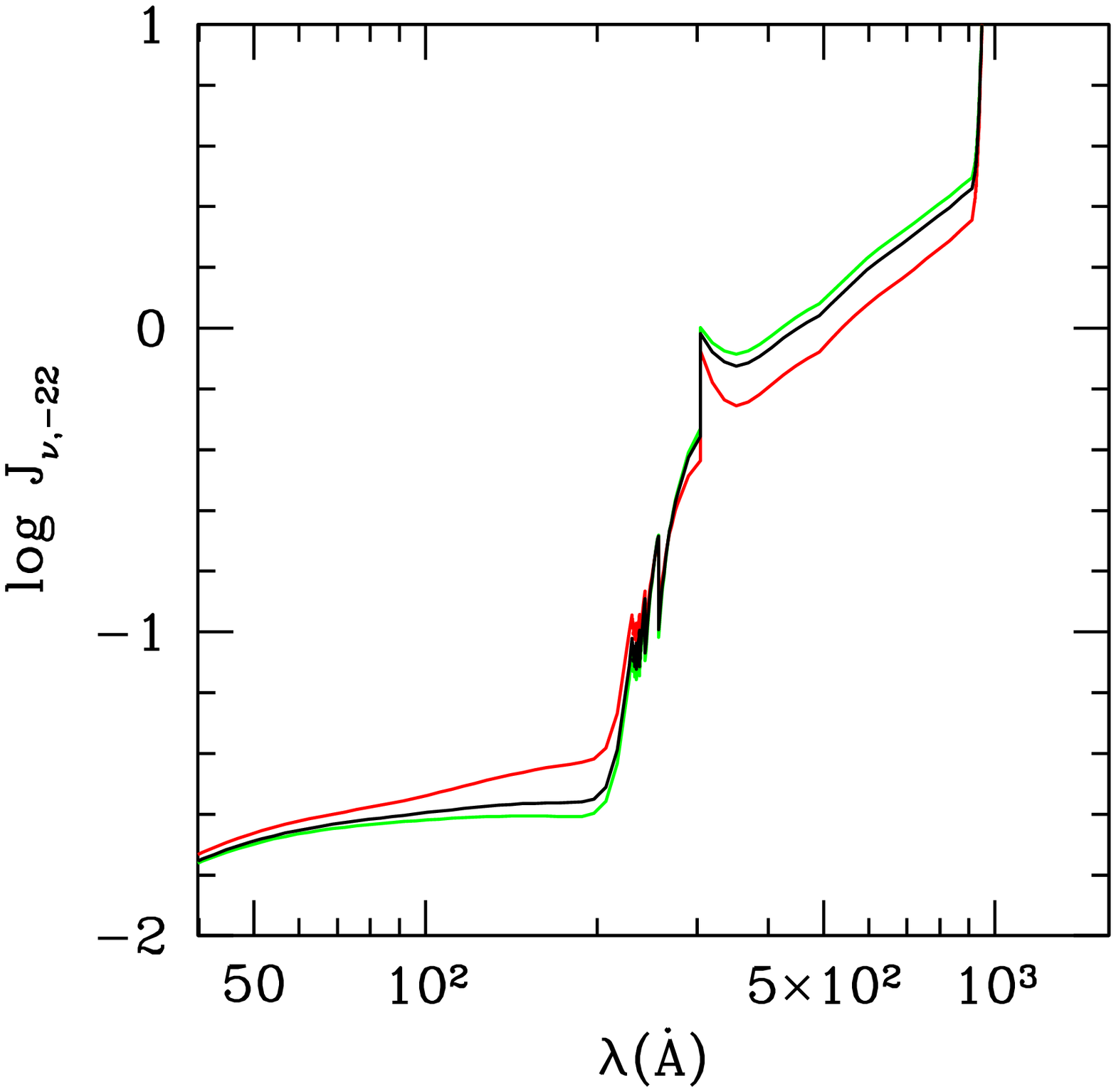}
\caption{\footnotesize {\it Left:} Emission rate of hydrogen-ionizing photons for a SSP of total mass $1\,\msun$, Salpeter IMF, 
and solar metallicity, as a function of age. {\it Solid line:} results from the GALAXEV package of Bruzual \& Charlot (2003). {\it Long-dashed 
line:} same for Starburst99 (Leitherer \etal 1999). {\it Short-dashed line:} same for the FSPS package of Conroy \etal (2009). 
{\it Dotted line:} same as the solid line, but for a metallicity 1/50 of solar.   
{\it Right:} Uncertainties in the broadband spectrum of the ``quasars $+$ galaxies" UVB at redshift 3. {\it Black line:} our fiducial model.
{\it Green line:} a model with a slope of $\beta=0.2$ in the column density distribution of LLSs. 
{\it Red line:} same for $\beta=1.0$. The intensity $J_\nu$ is expressed in units of $10^{-22}\,\uvunits$. 
}
\label{fig16}
\vspace{+0.5cm}
\end{figure}

We have also  checked that, for a given IMF, uncertainties in the stellar population synthesis technique are relatively small.
Figure \ref{fig16} shows the emission rate of hydrogen-ionizing photons for an SSP, calculated as a 
funtion of age with the GALAXEV models of Bruzual \& Charlot (2003), the Starburst99 models of Leitherer \etal (1999), and the 
FSPS models of Conroy, Gunn, \& White (2009). While the three packages use different stellar evolution tracks and
spectral libraries, the total number of ionizing photons emitted agrees to within 10\%. The figure also illustrates the significant
effect of stellar metallicity: an SSP of metallicity 1/50 of solar emits 60\% more hydrogen-ionizing 
photons over its lifetime than a solar metallicity SSP (Salpeter IMF, GALAXEV package).     

Finally, we address the effect of a change in the effective opacity of the IGM. In our parameterization,  
the shape of the $f(N_\nHI,z)$ distribution over the column density range of the LLSs is adjusted at every redshift 
for continuity with the SLLSs. As detailed in \S\,3, this procedure yields the slopes $\beta=0.47,0.61,0.72,0.82$ at 
redshifts $z=2,3,4,5$, respectively. To gauge how an uncertainty in $f(N_\nHI)$ translates into an uncertainty in 
the UVB, we have run CUBA with the fixed values of $\beta=0.2$ and $\beta=1$ in the column density 
interval $10^{17.5}<N_\nHI<10^{19}\,\cmm$. The resulting UVB at $z=3$ is shown in the right panel of Figure \ref{fig16}. 
Compared to our fiducial model, the flat $\beta=0.2$ distribution generates a hydrogen photoionization rate ($\Gamma_\nHI$) that 
is 29\% lower, and a \HeII\ photoionization rate ($\Gamma_\nHeII$) that is 22\% higher. This is because a larger opacity at 1 Ryd  
from the LSSs results in a harder background spectrum, which in turn produces a smaller \HeII\ opacity at 4 Ryd.
Conversely, in the steep $\beta=1$ case, $\Gamma_\nHI$ increases by 8\% and $\Gamma_\nHeII$ decreases by 10\%. 
Notice, however, that the former model would significantly underestimate the 1 Ryd photon mean free path
compared to the measurements of Prochaska \etal (2009).

\section{Summary}

In this paper we have presented improved synthesis models of the evolving spectrum of the UV/X-ray diffuse
background, updating and extending our previous results. Five new main components have been added to our 
cosmological radiative transfer code CUBA and discusses in details: (1) the sawtooth modulation of the background intensity
from resonant line absorption in the Lyman series of cosmic hydrogen and helium; (2) the X-ray
emission from the obscured and unobscured quasars that gives origin to the X-ray background; (3)
a piecewise parameterization of the distribution in redshift and column density of intergalactic
absorbers that fits recent measurements of the mean free path of 1 ryd photons; (4) an accurate
treatment of the absorber photoionization structure, which enters in the calculation of the
helium continuum opacity and recombination emissivity; and (5) the UV emission from star-forming
galaxies at all redshifts. The full implications of our new population synthesis models for the 
thermodynamics and ionization state of the \Lya\ forest and metal absorbers will be addressed in a subsequent paper.   
Here we have provided tables of the predicted \HI\ and \HeII\ photoionization and photoheating rates for use, e.g., 
in cosmological hydrodynamics simulations of the \Lya\ forest, a new metallicity-dependent calibration to the 
UV luminosity density-star formation rate density relation,  
and presented a ``minimal cosmic reionization model" in which the galaxy UV emissivity traces recent determinations of the
cosmic history of star formation, the luminosity-weighted escape fraction of hydrogen-ionizing radiation increases
rapidly with lookback time, the clumping factor of the high-redshift intergalactic medium follows recent determinations
of hydrodynamic simulations that include the effect of photoionization heating,
and Population III stars and miniquasars make a negligible contribution to the metagalactic flux. The model has been shown 
to provide a good fit to the hydrogen-ionization rates inferred from flux decrement and quasar proximity effect measurements, to predict that 
cosmological \HII\ (\HeIII) regions overlap at redshift 6.7 (2.8), and to yield an optical depth to Thomson 
scattering, $\tau_{\rm es}=0.084$ that is agreement with {\it WMAP} results.

Our new background intensities and spectra are sensitive to a
number of poorly determined input parameters and suffer from various degeneracies. Their predictive power should be constantly
tested against new observations. We are therefore making our redshift-dependent UV/X emissivities
and CUBA outputs freely available for public use at \url{http://www.ucolick.org/~pmadau/CUBA}.

\acknowledgments
\ni We have benefited from many informative discussions with A. Boksenberg, S. Charlot, A. Comastri, C.-A. Faucher-Gigu\`ere, M. McQuinn, 
J. Prochaska, C. Scarlata, and G. Worseck. Support for this work was provided by NASA through grant NNX09AJ34G and by the NSF through 
grant AST-0908910 to PM, and by the MIUR, PRIN 2007 to FH.  

{}

\appendix
\section{The ionization and thermal state of intergalactic absorbers} 

In this Appendix we describe the numerical calculations of the ionization structure of individual absorbers outlined in \S~\ref{sec:pho}. 
We approximate each absorber as a semi-infinite slab with uniform hydrogen density $n_{\rm H}$ and thickness equal to its Jeans  length,
\begin{equation}
L=\sqrt{\pi\gamma kT/(G\rho\mu m_p)}=0.92\, {\rm kpc}\, n_\nH^{-1/2}\,T_4^{1/2}\,\left(\frac{f_g}{0.16}\right)^{1/2}
\label{eq:jeans}
\end{equation}
(Schaye 2001; Faucher-Gigu\`ere \etal 2009), where the adiabatic index and mean molecular weight for a monoatomic, fully ionized gas of 
primordial composition are $\gamma=5/3$ and $\mu=0.59$, respectively, and $T_4$ is the gas temperature in units of $10^4$ K. The gas mass 
fraction is set at its universal value, $f_g=\Omega_b/\Omega_M=0.16$. Each slab model provides, given the gas density $n_\nH$ and 
an external isotropic radiation field $J_\nu$, the columns $N_\nHI$, $N_\nHeI$, and $N_\nHeII$.
\begin{figure}[thb]
\centering
\includegraphics*[width=0.47\textwidth]{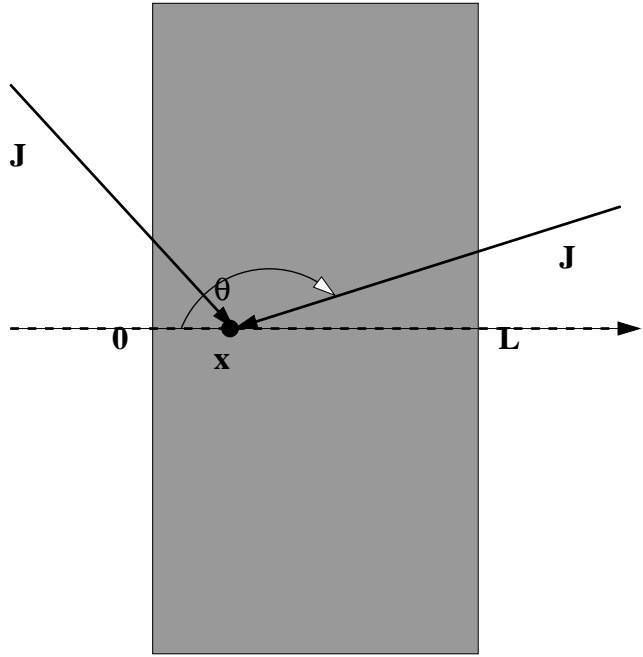}
\vspace{+0.5cm}
\caption{\footnotesize The slab approximation for individual absorbers: a plane-parallel slab of thickness $L$ illuminated by an 
isotropic external radiation field $J$.} 
\vspace{+0.5cm}
\label{fig:slab}
\end{figure} 
The slab geometry reduces the local radiative transfer to a 2-D problem (see Fig. \ref{fig:slab}). Setting $\mu\equiv \cos \theta$, where 
$\theta$ is the angle between the incident ray and the normal at the slab, the specific intensity $I_\nu(x,\mu)$ at
coordinate $x$ within the slab is given by the following implicit solution to the radiative transfer equation:
\begin{equation}
I_\nu(x,\mu)=\begin{cases} I_\nu(0,\mu)\,{\rm e}^{-\tau_\nu(x)/\mu}\,+{1\over 4\pi \mu}\int_0^x dx'\,
j_\nu(x')\,{\rm e}^{-[\tau_\nu(x)-\tau_\nu(x')]/\mu} & (0<\theta<\pi/2);\\
I_\nu(L,\mu)\,{\rm e}^{-[\tau_\nu(L)-\tau_\nu(x)]/|\mu|}\,+{1\over 4 \pi |\mu|} \int_x^L dx'\,
j_\nu(x')\,{\rm e}^{-[\tau_\nu(x')-\tau_\nu(x)]/|\mu|} & (\pi/2<\theta<\pi).
\end{cases}
\label{eq:Imu}
\end{equation}
Here, $j_\nu$ is the recombination radiation emission coefficient along the ray,  
\begin{equation}
\tau_\nu(x)=\sum \sigma_i(\nu)\,\int_0^x{dx' \, n_i(x')},
\label{eq:tau}
\end{equation}
is the optical depth normal to the slab, and the sum is to be taken over all relevant species. 
Integration over solid angle yields the local specific brightness ${\cal J}_\nu(x)$
\begin{equation}
{\cal J}_\nu(x)=\frac{1}{4\pi} \int d\Omega\, I_\nu\,=\frac{J_\nu}{2}\,\{E_2[\tau_\nu(x)]+E_2[\tau_\nu(L)-\tau_\nu(x)]\}+\frac{1}{2}
\int_0^L dx'\, \frac{j_\nu(x')}{4\pi}\,E_1(|\tau_\nu(x)-\tau_\nu(x')|),
\label{eq:J}
\end{equation}
where $E_1$ and $E_2$ are the first and second order exponential integral functions, respectively. 
The symmetry of the problem allows us to study just half of the slab, as for all ions $n(x)=n(L-x)$. With this in mind, we can rewrite the 
external contribution to ${\cal J}_\nu(x)$ (the first term on the right hand side of eq. \ref{eq:J}) as 
\begin{equation}
\frac{J_\nu}{2}\,\{E_2[\tau_\nu(x)]+E_2[2\tau_\nu (L/2)-\tau_\nu(x)]\},
\end{equation}
($0<x<L/2$), and the contribution of recombinations (the second term on the right hand side of eq. \ref{eq:J}) as 
\begin{equation}
\frac{1}{2}
\int_0^{L/2} dx'\, \frac{j_\nu(x')}{4\pi}\,\{E_1(|\tau_\nu(x)-\tau_\nu(x')|)\,+E_1(2\tau_\nu(L/2)-\tau_\nu(x')-\tau_\nu(x))\}.
\end{equation}

Thermal equilibrium in the absorbers is obtained by balancing photoionization heating with a number of different energy loss mechanisms:  
free-free, collisional excitations and ionizations, recombinations, Compton cooling against the cosmic microwave backgroud, and adiabatic 
cooling from cosmic expansion. The photoheating rate (per ion) is given by
\begin{equation}
{\cal H}_i=\int{d\nu \, \frac{4\pi {\cal J}_\nu}{h\nu}h(\nu-\nu_i)\sigma_i(\nu)},
\label{eq:thermal}
\end{equation}
where $h\nu_i$ is the ionization potential of species $i$. The equation of thermal equilibrium can then be written as
\begin{equation}
n_\nHI {\cal H}_\nHI+n_\nHeI {\cal H}_\nHeI+n_\nHeII {\cal H}_\nHeII=n_e\Lambda_c\,+\,3kT\,H\,(n_\nH+4n_\nHe)/\mu,
\end{equation}
where $\Lambda_c$ is the atomic cooling rate and the last term accounts for adiabatic cooling. Our treatment of the thermal 
state of cosmological absorbers is over-simplified in two aspects: 1) the lowest density IGM is not in thermal equilibrium;
and 2) dense quasar absorbers do not expand with the Hubble flow. We find, however, that: 1) adiabatic cooling is never important 
for absorbers with $N_\nHI \gta10^{15}$ cm$^{-2}$, so the inclusion of this process does not alter the thermal balance in this 
column density regime; and 2) in low the density, optically thin IGM, the functions $\eta=N_\nHeII/N_\nHI$ and 
(to a lesser extent) $\zeta=N_\nHeI/N_\nHI$ are independent on gas temperature, and hence the assumption of 
thermal equilbrium should not affect our results. 

The ionization and thermal structure of the slab, for a given $J_\nu$ and $n_\nH$, are solved by iteration. We start by assuming an 
almost fully ionized slab at $T=10^4$ K, and compute the ionization rates $\Gamma_i$ setting the recombination emissivity to zero.
We then solve for the ionization structure within the slab, compute updated ion fractions and opacities, and evaluate new ionization rates 
including recombination radiation. The process is iterated until convergence to better than 0.1\% is obtained 
at every point of the slab. At each iteration the size of the slab is changed accordingly to the Jeans criterion (eq.~\ref{eq:jeans}).  

\end{document}